\documentclass[a4paper, amsfonts, amssymb, amsmath, reprint, showkeys, nofootinbib, twoside]{revtex4-1}
\usepackage[english]{babel}
\usepackage[utf8]{inputenc}
\usepackage[colorinlistoftodos, color=green!40, prependcaption]{todonotes}
\usepackage{amsthm}
\usepackage{mathtools}
\usepackage{physics}
\usepackage{xcolor}
\usepackage{graphicx}
\usepackage[left=23mm,right=13mm,top=35mm,columnsep=15pt]{geometry} 
\usepackage{adjustbox}
\usepackage{placeins}
\usepackage[T1]{fontenc}
\usepackage{lipsum}
\usepackage{csquotes}

\newcommand{\figref}{Fig.\,\ref}
\newcommand{\secref}{Sec.\,\ref}
\newcommand{\tabref}{Tab.\,\ref}
\newcommand{\Eqref}{Eq.\,\ref}

\begin{document}
\title{The threshold of semiconductor nanolasers}

\author{Marco Saldutti, Yi Yu, and Jesper Mørk}
    \email[Correspondence email address: ]{jesm@dtu.dk}% Your name
    \affiliation{DTU Electro, Technical University of Denmark, DK-2800 Kgs. Lyngby, Denmark}
    \affiliation{NanoPhoton - Center for Nanophotonics, Technical University of Denmark, DK-2800 Kgs. Lyngby, Denmark}

\date{\today} % Leave empty to omit a date

\begin{abstract}
Nanolasers based on emerging dielectric cavities with deep sub-wavelength confinement of light offer a large light-matter coupling rate and a near-unity spontaneous emission factor, $\beta$. These features call for reconsidering the standard approach to identifying the lasing threshold. Here, we suggest a new threshold definition, taking into account the recycling process of photons when $\beta$ is large. This threshold with photon recycling reduces to the classical balance between gain and loss in the limit of macroscopic lasers, but qualitative as well as quantitative differences emerge as $\beta$ approaches unity. We analyze the evolution of the photon statistics with increasing current by utilizing a standard Langevin approach and a more fundamental stochastic simulation scheme. We show that the threshold with photon recycling consistently marks the onset of the change in the second-order intensity correlation, $g^{(2)}(0)$, toward coherent laser light, irrespective of the laser size and down to the case of a single emitter. In contrast, other threshold definitions may well predict lasing in light-emitting diodes. These results address the fundamental question of the transition to lasing all the way from the macro- to the nanoscale and provide a unified overview of the long-lasting debate on the lasing threshold. 
\end{abstract}

%\keywords{nanolasers, threshold, extreme dielectric confinement}

\maketitle

\section{Introduction} \label{sec:Intro}

  %Lasers exist in many forms and are essential for numerous scientific and technical applications. Currently, 
  Researchers are currently writing a new fascinating chapter in the exciting history \cite{Hecht_ApplOpt_2010} of laser development: semiconductor nanolasers \cite{Hill_NatPhot_2014,Ning-SPIE-2019,Ma_NatNanotech_2019,Deng_AdvOptMat_2021}. Applications range from chip-scale optical communications \cite{Nozaki_NatPhot_2019,Takeda_OptExpress_2021,Dimopoulos_Optica_2023}, biochemical sensing \cite{Zhang_PhotRes_2021} and quantum technologies \cite{Kreinberg_LightScAppl_2018,Zhang_Nature_2022}, to artificial intelligence and neuromorphic computing \cite{Shastri_NatPhot_2021}, to name a few. 

  \textcolor{black}{With increasing pumping, conventional macroscopic lasers abruptly transition from incoherent to coherent light. Over a pumping range that is negligibly small, the output power increases and the linewidth of the emission spectrum decreases by orders of magnitude. Thus, the pump rate marking the kink of the light-current characteristic uniquely identifies the laser threshold \cite{Coldren-book2ndEd-2012}.} 
  
  \textcolor{black}{As lasers shrink to the micro- and nanoscale, the fraction of spontaneous emission funneled into the lasing mode - the so-called $\beta$-factor - approaches unity, and the phase transition vanishes \cite{Rice-PRA-1994}. In this operation regime, neither the power increase nor linewidth reduction is a sufficient indicator of lasing \cite{Ning_JSTQE_2013,Kreinberg_LightScAppl_2017}. The absence of a phase transition has spawned and fueled the long-standing discussion about properly defining the lasing threshold – a debate that recent advances in nanolaser technology are only igniting further \cite{Bjork_JQE_1991,Jin_PRA_1994,Bjork_PRA_1994,Rice-PRA-1994,Ning_JSTQE_2013,Kreinberg_LightScAppl_2017,Lohof_PhysRevApplied_2018,Takemura_PRA_2019,Kreinberg_LPR_2020,Carroll_PhysRevLett_2021,Khurgin_LPR_2021,Lippi_ChaosSolitonsFractals_2022,Yacomotti-LPR-2022,Carroll_PRA_2023}. Not only is the question of fundamental nature - essentially, what defines the onset of lasing - but also of practical relevance - in order to design and optimize lasers with low threshold, e.g. for applications in on-chip interconnects with ultra-low power consumption \cite{Miller_JLWT_2017}.}

  \textcolor{black}{The second-order intensity correlation, $g^{(2)}(0)$, is an important measure of the light statistics and is often employed to distinguish nanolasers from nanoLEDs \cite{Kreinberg_LightScAppl_2017,Deng_AdvOptMat_2021}. Yet, a unified and clear understanding of the transition to lasing, from the macro- to the nanoscale, is still missing \cite{Yacomotti-LPR-2022}. For instance, expressions reported in standard textbooks \cite{Coldren-book2ndEd-2012} predict a threshold current equal to zero in the case of negligible nonradiative recombination and unity $\beta$-factor - a so-called "zero-threshold" laser \cite{DeMartini_PRL_1988}. On the other hand, it was pointed out by Björk and Yamamoto \cite{Bjork_PRA_1994} that transitions usually associated with lasing, such as the increase of quantum efficiency or linewidth reduction, are smooth when plotted versus the average intracavity photon number, irrespective of the $\beta$-factor. Along these lines, it has been argued that the onset of lasing is better characterized by reaching a given average photon number larger than one \cite{Kreinberg_LightScAppl_2017}. Yet, how large this photon number should be is still unclear, with different values being reported depending on the laser scale \cite{Kreinberg_LightScAppl_2017,Lohof_PhysRevApplied_2018}, and without a simple and unique rationale.} 
  
  \textcolor{black}{In this article, we argue that the threshold can be uniquely defined even in the case of a near-unity $\beta$-factor by the condition of gain-loss balance, but with the inclusion of photon recycling. We name this generalized condition \textit{threshold with photon recycling}. While conventional rate equations \cite{Bjork_JQE_1991} already account for photon recycling, a mechanism identified in \cite{Yamamoto_JJAP_1991}, including the effect when defining the lasing threshold is this paper's original and central contribution. We shall show that this definition marks the onset of the asymptotic transition towards coherent light, $g^{(2)}(0)=1$, and robustly distinguishes between lasers and LEDs. A comprehensive comparison is made to previous definitions.} 
  %We utilize rate equations of discrete two-level emitters with adequate inclusion of quantum noise \cite{Mørk_APL_2018,Bundgaard_PRL_2023}.    
  
  \textcolor{black}{In the past decades, microscopic theories \cite{Gies_PRA_2007,Wiersig_Nature_2009} have been developed which successfully explain cavity quantum electrodynamics (QED) effects in semiconductor nanolasers, such as Rabi oscillations \cite{Nomura_NatPhys_2010,Gies_PRA_2017} and superradiance \cite{Leymann_PRA_2015,Jahnke_NatCom_2016,Kreinberg_LightScAppl_2017}. Yet, these effects are mostly relevant if the decoherence time of the emitters is long enough \cite{Auffèves_NewJournPhys_2011}, and indeed experimental demonstrations are usually restricted to cryogenic temperatures \cite{Wiersig_Nature_2009,Nomura_NatPhys_2010,Jahnke_NatCom_2016,Gies_PRA_2017,Kreinberg_LightScAppl_2017}. Furthermore, cavity QED models may disagree with experiments - to a qualitative level - on such a fundamental quantity as the second-order intensity correlation \cite{Kreinberg_LightScAppl_2017}. This indicates that even semiconductor microscopic models - despite the significant level of complexity - need to be further developed.} 

  \textcolor{black}{The scope of this article is to analyze the onset of lasing in the practically relevant case of room-temperature semiconductor nanolasers. For this purpose, we utilize rate equations for an ensemble of two-level emitters with the inclusion of quantum noise \cite{Mørk_APL_2018,Bundgaard_PRL_2023}. 
  %using rate equations of discrete two-level emitters suffices, as long as the quantum noise is properly described \cite{Mørk_APL_2018,Bundgaard_PRL_2023}. 
  The use of rate equations of discrete two-level emitters is in line with previous works investigating nanolasers with extended active media by rate equations \cite{Takemura_PRA_2019} or microscopic models \cite{Lohof_PhysRevApplied_2018}. In those works, the transition to lasing is analyzed by mapping the electron-hole pairs to discrete emitters. The total number of emitters reflects the active region volume and a given transparency carrier density.}

  From the application point of view, this work is further motivated by emerging dielectric cavities with extreme dielectric confinement (EDC) \cite{Hu_ACSPhot_2016,Choi_PRL_2017,Wang_APL_2018,Albrechtsen_NatComm_2022,Albrechtsen_OptExpress_2022,Kountouris-OptExress-2022}. These cavities confine light to an effective mode volume \cite{Kristensen_OptLett_2012,Sauvan_PRL_2013} deep below the so-called diffraction limit \cite{Coccioli_1998,Khurgin_NatNanotech_2015,Bozhevolnyi_Optica_2016} while keeping the quality factor (Q-factor) sufficiently high. 
  % As a result, these cavities with extreme dielectric confinement (EDC) can achieve strong light-matter interaction. 
  In EDC cavities, the optical density of states is largely dominated by a single mode \cite{Kountouris-OptExress-2022}. 
  % thereby limiting the fraction of undesired spontaneous emission coupled into other modes. Therefore, EDC cavities are 
  Therefore, EDC cavities may lead to nanolasers - from now on EDC lasers - with a near-unity $\beta$-factor. 

   \begin{figure}[ht]
	  \centering\includegraphics[width=1\linewidth]{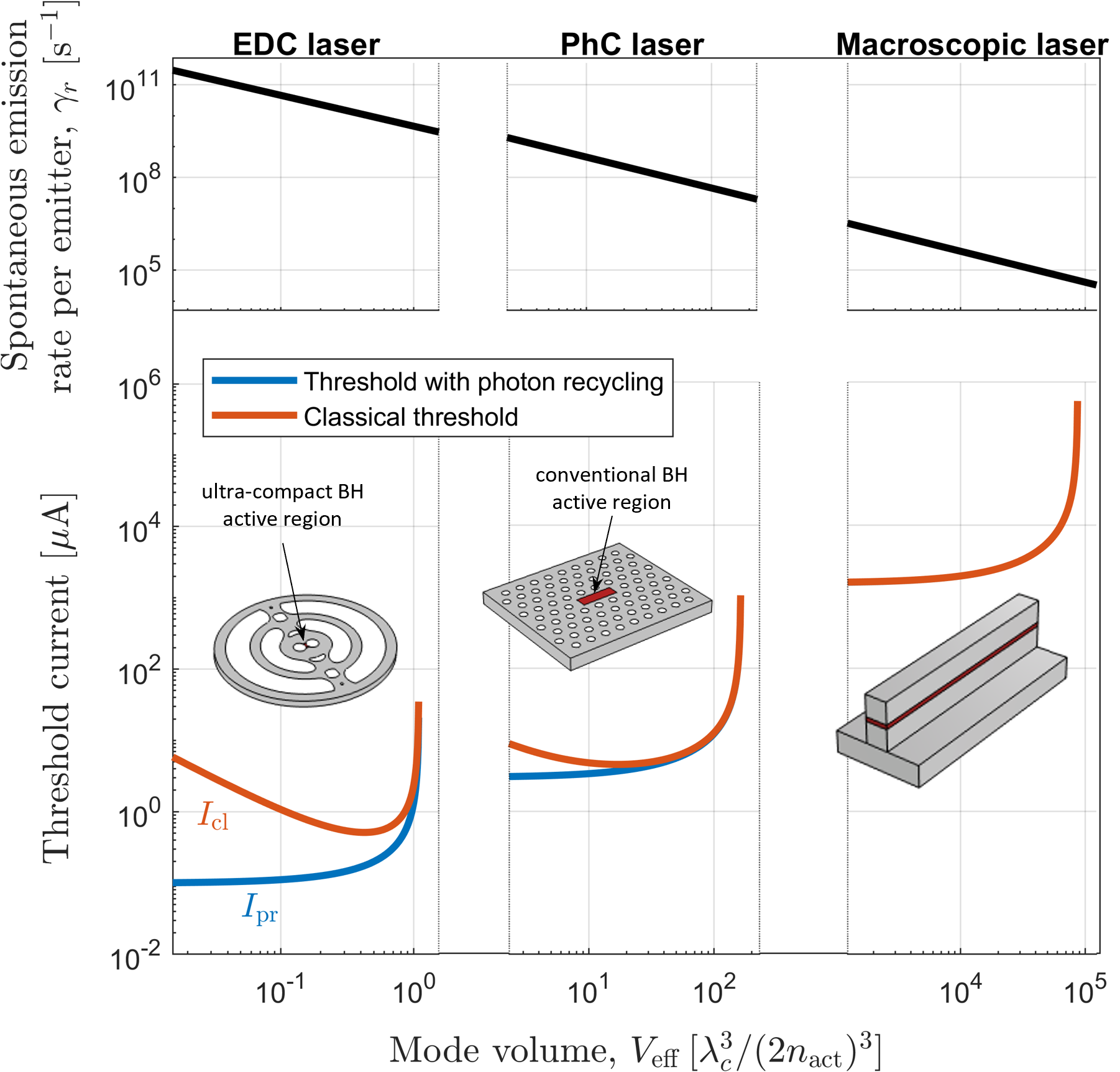}
	  \caption{\label{fig:sec:INTRO:FIG1} Spontaneous emission rate per emitter (top) and threshold current (bottom) versus mode volume for a nanolaser with extreme dielectric confinement (EDC laser, left), a photonic crystal laser (PhC laser, center) with a buried heterostructure (BH) active region and a conventional macroscopic laser (right). Both the threshold with photon recycling (blue) and the classical threshold (red) are shown. The number of emitters is fixed, while the mode volume varies with the optical confinement factor. Apart from the mode volume, the parameters are listed in \tabref{tab:nano-devices}, with the EDC laser being EDC laser 1. %The EDC laser is inspired by recently demonstrated cavity designs \cite{Wang_APL_2018,Albrechtsen_NatComm_2022,Kountouris-OptExress-2022}.
      }
   \end{figure} 
  %In this article, we propose a simple expression for the threshold current, valid all the way from the macro to the nanoscale. By incorporating the physics of photon recycling \cite{Yamamoto_JJAP_1991}, this threshold extends to semiconductor nanolasers with a high $\beta$-factor, such as EDC lasers, the classical definition of balance between gain and loss \cite{Bjork_JQE_1991,Coldren-book2ndEd-2012}. We name this generalized condition \textit{threshold with photon recycling}. Importantly, the expression accounts for the finite number of electronic states available to the lasing process, which must be considered in the case of nanolasers.
  \textcolor{black}{\figref{fig:sec:INTRO:FIG1} highlights the importance of using a proper threshold definition. In the example shown, the classical threshold and the threshold with photon recycling thus give qualitatively different results when varying the mode volume.} The figure illustrates the spontaneous emission rate per emitter, $\gamma_r$ (top), and the threshold current (bottom) versus mode volume for an EDC laser (left), a photonic crystal (PhC) laser (center) and a macroscopic laser (right). The carrier lifetime is $(\gamma_r+\gamma_\mathrm{bg})^{-1}$, where $\gamma_\mathrm{bg}$ reflects nonradiative recombination and recombination into nonlasing modes. Importantly, $\gamma_r$ is inversely proportional to the mode volume. In macroscopic lasers, the threshold current with photon recycling, $I_{\mathrm{pr}}$ (blue), reduces to the classical balance between gain and loss, $I_{\mathrm{cl}}$ (red). However, with decreasing mode volume, qualitative and quantitative differences between the two thresholds gradually emerge. In EDC lasers, $I_{\mathrm{cl}}$ strongly increases when the mode volume becomes very small, due to the strong carrier lifetime reduction. Conversely, $I_{\mathrm{pr}}$ decreases and eventually saturates, due to an effective saturation of the carrier lifetime induced by photon recycling.      
  
  Notably, the threshold with photon recycling consistently marks the onset of the transition toward coherent laser light. 
  % and it clarifies the role \cite{Khurgin_LPR_2021} of the $\beta$-factor. 
  We demonstrate this property by studying the photon statistics \cite{Rice-PRA-1994} from below to above threshold. We consider both a conventional Langevin approach \cite{Coldren-book2ndEd-2012,Mørk_APL_2018} and a more fundamental stochastic simulation scheme \cite{Andre_OptExpress_2020,Bundgaard_PRL_2023}. 
  
  Besides the threshold with photon recycling, we scrutinize several other threshold definitions \cite{Bjork_PRA_1994,Jin_PRA_1994,Takemura_PRA_2019,Carroll_PhysRevLett_2021,Lippi_ChaosSolitonsFractals_2022,Yacomotti-LPR-2022} and explicitly show that some of them are not reliable indicators of lasing. These definitions, such as the quantum threshold \cite{Bjork_PRA_1994}, may well predict the possibility of lasing in structures that remain in the LED regime of thermal light at all pumping levels.  

  The article is organized as follows. In \secref{sec:EDC}, we review the physics of extreme dielectric confinement and clarify the mode volume definitions relevant to nanolasers. In \secref{sec:rate-equation-model}, we illustrate the laser model and discuss the two approaches utilized in this article to describe the quantum noise. In \secref{sec:lasing-threshold-intro}, we derive and analyze the threshold with photon recycling and thoroughly discuss the laser input-output characteristics, from the macro- to the nanoscale. In \secref{sec:beta-factor}, we elucidate the role of the spontaneous emission factor, often misunderstood. In \secref{sec:other-threshold-definitions} and \secref{Sec: Photon statistics: stochastic simulations}, we review other threshold definitions and discuss the photon probability distributions. In \secref{sec:discussion}, we finally summarize the discussion and draw the main conclusions. In the Supplementary Material, we derive the spontaneous emission rate per emitter, as well as the variance of the photon number as obtained from the Langevin approach.

\section{Extreme dielectric confinement}\label{sec:EDC} 

  Generally speaking, concentrating the light spatially and for a sufficiently long time is key to achieving strong light-matter interaction. The figures of merit reflecting the spatial and temporal confinement of light are the mode volume and the quality factor (Q-factor), respectively \cite{Notomi_RepProgrPhys_2010}. Plasmonics \cite{Maier_book_2007} may achieve deep sub-wavelength optical confinement, but the absorption losses in metals severely limit the quality factor \cite{Wang_PRL_2006}. Photonic crystal (PhC) cavities \cite{Notomi_RepProgrPhys_2010,Saldutti_Nanomat_2021} can easily offer Q-factors of several thousands if not millions \cite{Asano_OptExpress_2017}, but often have mode volumes exceeding $\lambda_c^3/(2n)^3$ by a factor of at least $2$ or $3$ \cite{Zhang_OE_2004}, where $\lambda_c^3/(2n)^3$ sometimes is referred to as the diffraction-limited mode volume \cite{Coccioli_1998,Khurgin_NatNanotech_2015,Bozhevolnyi_Optica_2016}. Here, $\lambda_c$ is the vacuum wavelength and $n$ is the material refractive index. 
  
  Emerging cavities with extreme dielectric confinement (EDC) \cite{Hu_ACSPhot_2016,Choi_PRL_2017,Wang_APL_2018,Albrechtsen_NatComm_2022,Albrechtsen_OptExpress_2022,Kountouris-OptExress-2022} promise the perfect blend of an ultra-small mode volume and a sufficiently high quality factor. Without trading energy efficiency for speed, EDC cavities are excellent candidates for numerous applications, including few-photon nonlinearities \cite{Choi_PRL_2017}, optical switches \cite{Saldutti_IEEE_2022} and nanoscale light-emitting sources with squeezed intensity noise \cite{Mørk_Optica_2020}. 

  %In particular, EDC cavities may achieve strong light-matter interaction without needing extremely high Q-factors. Without trading energy efficiency for speed, EDC cavities are excellent candidates for numerous applications where high energy efficiency and wide bandwidths are simultaneously desirable. These include few-photon nonlinearities \cite{Choi_PRL_2017} and quantum technologies \cite{Nomura_NatPhys_2010,Lodahl_RevModPhys_2015}, photodetectors \cite{Takiguchi_ACSPhot_2020} and optical switches \cite{Ono_NatPhot_2020,Saldutti_IEEE_2022}, as well as nanoscale light-emitting diodes and nanolasers with squeezed intensity noise \cite{Mørk_Optica_2020}.

   \begin{figure}[ht]
	  \centering\includegraphics[width=0.95\linewidth]{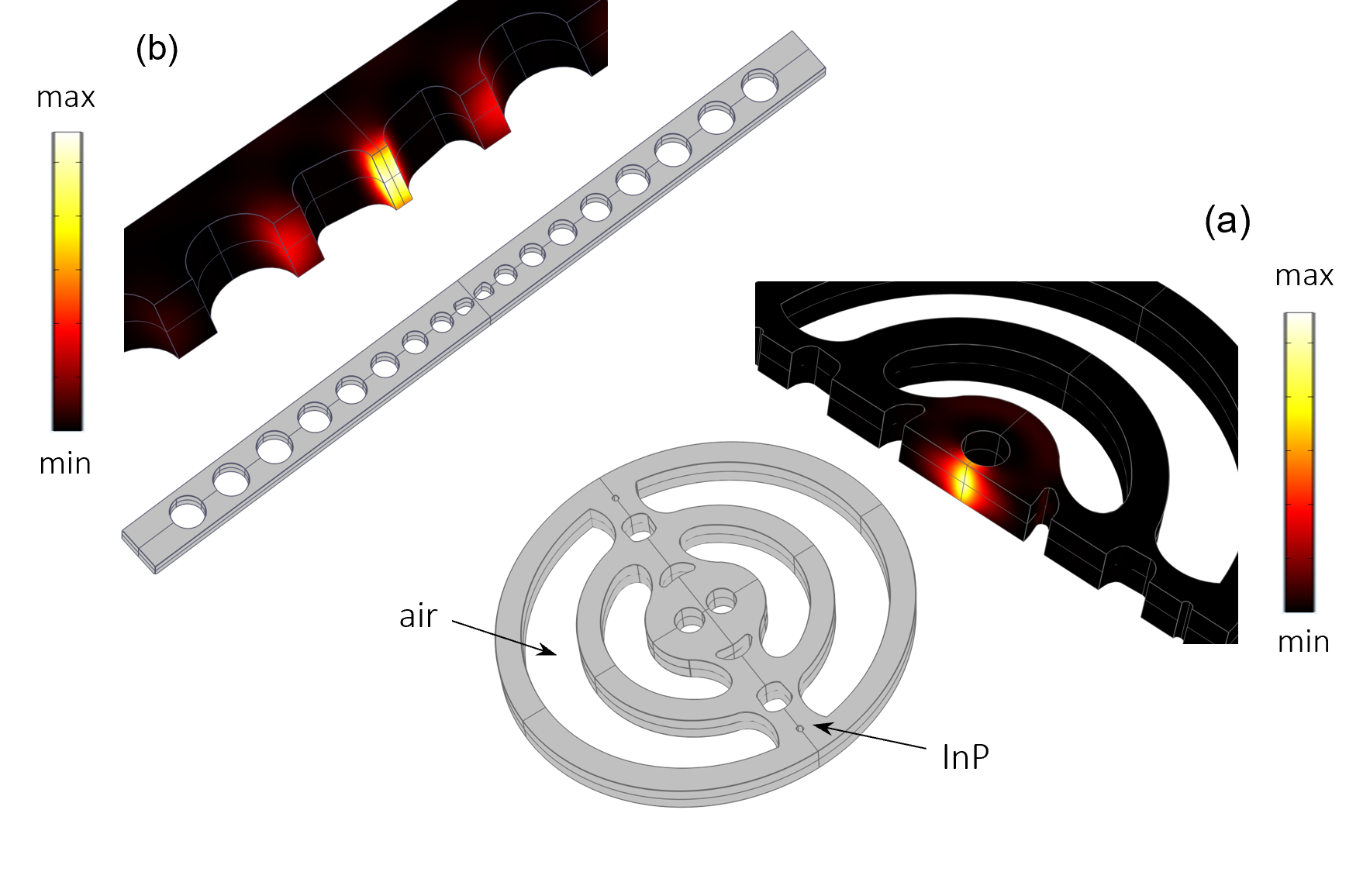}
	  \caption{\label{fig:sec:EDC:FIG1} Optical cavities with extreme dielectric confinement (EDC) inspired by recently demonstrated designs 
        \cite{Hu_ACSPhot_2016,Choi_PRL_2017,Wang_APL_2018,Albrechtsen_NatComm_2022,Kountouris-OptExress-2022}. The color plots show the squared magnitude of the electric field, $\left|\mathbf{E}(\mathbf{r})\right|^2$, on a cross-section of the bowtie.} 
   \end{figure} 
  The boundary conditions of Maxwell's equations require the normal component of the electric displacement field and the tangential component of the electric field to be continuous across a dielectric interface \cite{Jackson-book3rdEd-1999}. A bowtie geometry, as present in EDC cavities, conveniently leverages these two requirements, increasing the electric energy density around a given point. For example, \figref{fig:sec:EDC:FIG1} illustrates two EDC cavities inspired by recent designs \cite{Hu_ACSPhot_2016,Choi_PRL_2017,Wang_APL_2018,Albrechtsen_NatComm_2022,Kountouris-OptExress-2022}, with a bowtie located at the center. The rings (cf. \figref{fig:sec:EDC:FIG1}a) or the holes (cf. \figref{fig:sec:EDC:FIG1}b) surrounding the bowtie effectively act as a distributed Bragg grating, which suppresses the in-plane radiation loss and increases the Q-factor. The distance between the holes forming the bowtie is of the order of a few nanometers \cite{Albrechtsen_NatComm_2022,Kountouris-OptExress-2022}.   

  The relevant definition of mode volume depends on the specific type of light-matter interaction that is considered \cite{Saldutti_Nanomat_2021}. In lasers, stimulated and spontaneous emission scale with $\gamma_r$, the spontaneous emission rate per emitter. For lasers with discrete and identical two-level emitters, such as quantum dots with a dominant electronic transition and negligible inhomogeneous broadening, $\gamma_r$ may be expressed as \cite{Mørk_APL_2018} 
	\begin{equation}
		\label{eq:gamma_r}
		\gamma_r = \frac{2d^2}{\hbar\epsilon_0n_{\mathrm{act}}^2}\frac{\omega_c}{\gamma_2}\frac{1}{V_p}
	\end{equation}
  %In \secref{Sec: Appendix A}, we provide a detailed derivation of \Eqref{eq:gamma_r} on the basis of Fermi's golden rule. 
  In \Eqref{eq:gamma_r}, $V_p$ is the mode volume and $\omega_c$ is the angular frequency of the lasing mode. The refractive index at the position of the emitters, $n_{\mathrm{act}}$, and the emitter dipole moment, $d$, are taken to be the same for all the emitters. The dephasing rate, $\gamma_2 = 2/T_2$, reflects the emitter homogeneous broadening, with $T_2$ being the dephasing time. The emitters are assumed to be identical and in resonance with the lasing mode. Any detuning would limit $\gamma_r$, thereby effectively reducing the emitter dipole moment (see Sec. I in the Supplementary Material). 
  
  To a first approximation (see \secref{sec:rate-equation-model}), \Eqref{eq:gamma_r} is also valid for lasers with an extended active region. In this case, $\gamma_2$ may also reflect the inhomogeneous broadening of the gain material (see Sec. I in the Supplementary Material). We note that \Eqref{eq:gamma_r} considers the so-called good-cavity limit \cite{Mørk_APL_2018}, where the emitter broadening is larger than the cavity linewidth. This is the typical scenario of semiconductor lasers working at room temperature \cite{Romeira_JQE_2018}. We also note that the light-matter coupling rate \cite{Bundgaard_PRL_2023}, $g$, is given by $\sqrt{\gamma_2\gamma_r}/2$.
 
    For nanolasers with a single emitter \cite{Nomura_OptExpress_2009,Bundgaard_PRL_2023}, the mode volume appearing in \Eqref{eq:gamma_r} is expressed as \cite{Gérard_Springer_2003}
	\begin{equation}
		\label{eq:Vp-single-emitter-main-text}
		\frac{1}{V_p} \rightarrow \frac{1}{V_{\mathrm{opt}}} = \frac{\epsilon_0 n^2(\mathbf{r}_e) \left|\hat{\mathbf{d}}_e\cdot\mathbf{E}(\mathbf{r}_e)\right|^2}{\int_V \epsilon_0 n^2(\mathbf{r})\left|\mathbf{E}(\mathbf{r})\right|^2d^3\mathbf{r}} 
	\end{equation}
	where $\mathbf{r}_e$ is the position of the emitter, $\hat{\mathbf{d}}_e$ is the unit vector which represents the orientation of the emitter dipole and $\mathbf{E}$ is the electric field of the lasing mode. We note that a rigorous definition should be based on the theory of quasi-normal modes \cite{Lalanne_LPR_2018,Kristensen_AdvOptPhot_2020}. However, if the Q-factor is high enough, generalized definitions \cite{Kristensen_OptLett_2012,Sauvan_PRL_2013} reduce to \Eqref{eq:Vp-single-emitter-main-text}. The integration volume, $V$, should suitably enclose the physical volume of the cavity with a margin of a few wavelengths. 
    % \Eqref{eq:Vp-single-emitter-main-text} is a reasonable approximation for cavity designs based on extreme dielectric confinement (EDC) \cite{Hu_ACSPhot_2016,Choi_PRL_2017,Wang_APL_2018,Albrechtsen_NatComm_2022,Kountouris-OptExress-2022}, with a quality factor larger than one thousand. 
    
    For nanolasers with an extended active region, averaging over the active region (see Sec. I in the Supplementary Material) leads to an \textit{effective} mode volume
	\begin{equation}
		\label{eq:Vp-extended-medium-main-text}
		\frac{1}{V_p} \rightarrow \frac{1}{V_{\mathrm{eff}}} = \frac{\Gamma}{V_\mathrm{act}} 
	\end{equation}
	Here, $V_\mathrm{act}$ is the physical volume of the active region. The optical confinement factor, $\Gamma$, is the fraction of electric energy stored within the active region volume:
 	\begin{equation}
		\label{eq:Gamma}
		\Gamma = \frac{ \int_{V_\mathrm{act}} \epsilon_0 n^2_{\mathrm{act}} \left|\mathbf{E}(\mathbf{r}_e)\right|^2 d^3\mathbf{r}_e}{ \int_V \epsilon_0 n^2(\mathbf{r})\left|\mathbf{E}(\mathbf{r})\right|^2d^3\mathbf{r} }
	\end{equation}
    The active region refractive index, $n_{\mathrm{act}}$, is assumed to be uniform throughout the active region. \Eqref{eq:Vp-extended-medium-main-text} is the usual definition of mode volume as encountered in the conventional theory of semiconductor lasers \cite{Coldren-book2ndEd-2012}. We note that \Eqref{eq:Vp-extended-medium-main-text} and \Eqref{eq:Gamma} reduce to \Eqref{eq:Vp-single-emitter-main-text} if the active region volume is much smaller than the scale on which the electric field amplitude varies. Any non-perfect alignment between the electric field and the dipoles is considered by effectively reducing the emitter dipole moment in \Eqref{eq:gamma_r} (see Sec. I in the Supplementary Material).
        
    %While offering Q-factors much larger than plasmonic cavities, EDC cavities can squeeze the mode volume of a single emitter well below what was previously thought to be \cite{Coccioli_1998,Khurgin_NatNanotech_2015,Bozhevolnyi_Optica_2016} the diffraction limit, roughly $\lambda_c^3/(2n)^3$. 
    Notably, a passive EDC cavity with a record mode volume, $V_{\mathrm{opt}}$, around $0.08\,\lambda_c^3/(2n)^3$ has been demonstrated experimentally \cite{Albrechtsen_NatComm_2022}. The cavity, designed by topology optimization \cite{Jensen-Sigmnud_LPR_2011,Wang_APL_2018}, is similar to that in \figref{fig:sec:EDC:FIG1}a. The mode volume is evaluated at the center of the cavity by assuming perfect alignment between the emitter dipole and the cavity field. 
    % As a comparison, the smallest single-emitter mode volume reported for photonic crystal cavities is around $2.30\,\lambda_c^3/(2n)^3$ \cite{Zhang_OE_2004}. 
    
    Future designs may specifically target nanolaser applications. In particular, designs optimizing the optical confinement factor in a small active region, as enabled by the buried heterostructure technology \cite{Kuramochi_OptExpress_2018,Dimopoulos_LPR_2022}, may lead to EDC lasers with an ultra-low threshold current.  

\section{Laser model}\label{sec:rate-equation-model}

    \textcolor{black}{Rate equations are derived from microscopic \cite{Lorke_PRB_2013} or master equation models \cite{Moelbjerg_JQE_2013} by adiabatic elimination of the photon-assisted polarization and neglecting emitter-emitter correlations. These approximations are well-grounded in the case of semiconductor lasers working at room temperature - even at the nanoscale - due to the large dephasing rate \cite{Auffèves_NewJournPhys_2011}.} 
    
    \textcolor{black}{In particular, rate equations that only track the number of photons and the \textit{total} number of carriers are obtained when the intraband dynamics is sufficiently fast \cite{Lorke_PRB_2013,Moelbjerg_JQE_2013}. Many-body Coulomb effects \cite{Coldren-book2ndEd-2012} are neglected \cite{Gies_PRA_2007}. The light-current characteristic and modulation response of two-level rate equations are in good agreement with microscopic theories of semiconductor nanolasers \cite{Lorke_PRB_2013} when treating on the same footing the stimulated emission and spontaneous emission into the lasing mode \cite{Gregersen_APL_2012,Romeira_JQE_2018} and accounting for the energy distribution of the carriers via the electronic density of states (DOS) and Fermi-Dirac occupation probabilities \cite{Coldren-book2ndEd-2012}.} 

  \textcolor{black}{When considering nanolasers operating at very low power levels, proper consideration of the quantum noise is of utmost importance. For a single-emitter laser, it was shown that the intensity quantum noise calculated using a stochastic interpretation of the rate equations, to be explained later, quantitatively agrees with a full quantum mechanical approach \cite{Bundgaard_PRL_2023}. Thus, higher-order correlation terms required in, e.g., the cluster expansion approach \cite{Gies_PRA_2007} to break the mean-field approximation and capture the photon statistics, are accounted for by the stochastic rate equations \cite{Bundgaard_PRL_2023}. For multiple emitters, collective effects, i.e. super-radiance and sub-radiance due to coherent interactions between the emitters, may play a role in the presence of small dephasing rates and small inhomogeneous broadening \cite{Leymann_PRA_2015}. However, as we limit ourselves to room-temperature operation, these effects will not be important here.}  
  %However, these rate equations \cite{Gregersen_APL_2012,Romeira_JQE_2018} still constitute a purely numerical approach, and conventional strategies to include the quantum noise \cite{Coldren-book2ndEd-2012} are inapplicable. Thus, approximations expressing the stimulated and carrier recombination rates as simple functions of the carrier number \cite{Coldren-book2ndEd-2012} are highly valuable and have been pivotal for developing the current understanding and technology of semiconductor lasers.
  
  \textcolor{black}{While the rate equations we employ are derived for a set of identical two-level emitters, they are also expected to be a good representation for a nanolaser with a spatially-confined quantum well active region, e.g. the buried heterostructure laser considered in \cite{Dimopoulos_Optica_2023}. Thus, approximations expressing the stimulated and carrier recombination rates as simple functions of the carrier number \cite{Coldren-book2ndEd-2012} are highly valuable and have been pivotal for developing the current understanding and technology of semiconductor lasers.}
  %In this framework, rate equations of discrete two-level emitters - where spontaneous and stimulated emission rates are \textit{linear} in the number of carriers - represent the crudest approximation, and they are inapplicable, strictly speaking, to semiconductor lasers \cite{Gies_PRA_2007}. Yet, the approximation is widely employed and has proved successful in reproducing measurements on quantum dot (QD) \cite{Ota_OptExp_2017,Kaganskiy_Optica_2019} and even quantum well (QW) \cite{Pan_Optica_2016,Takemura_PRA_2019} semiconductor nanolasers. The caveat is that accurate estimates of sensitive parameters, such as the $\beta$-factor, are not feasible \cite{Gies_PRA_2007}, and one must resort, for such a purpose, to more rigorous practices.

  \textcolor{black}{In this framework, rate equations where spontaneous and stimulated emission rates are linear in the number of carriers represent the simplest approximation. Yet, the approximation has proved successful in reproducing measurements on quantum dot (QD) \cite{Ota_OptExp_2017,Kaganskiy_Optica_2019} and even quantum well (QW) \cite{Pan_Optica_2016,Takemura_PRA_2019} semiconductor nanolasers. The caveat is that important parameters, such as the $\beta$-factor, must be treated with care and will, in general, depend on the pumping level \cite{Gies_PRA_2007}.}  
  %Importantly, rate equations of discrete two-level emitters enable a direct comparison with full-scale master equation models \cite{Moelbjerg_JQE_2013},and a simple description of quantum noise \cite{Mørk_APL_2018}. We point out that modeling the quantum noise is necessary to calculate the photon statistics and describe the second-order intensity correlation. In particular, it has been shown \cite{Bundgaard_PRL_2023} that a stochastic interpretation of the rate equations accurately captures the RIN and second-order intensity correlation, $g^{(2)}(0)$, of single-emitter nanolasers, including the photon antibunching at low values of current. Interestingly, the agreement holds even when the master equation model displays vacuum Rabi oscillations.
    
 With these considerations in mind, we model the laser by rate equations for the average number of excited emitters, $n_e$, and the average number of photons in the lasing mode, $n_p$ \cite{Mørk_APL_2018}: 
    \begin{subequations}
	\begin{align}
		& \frac{dn_e}{dt} = \frac{I}{q}\left(1-\frac{n_e}{n_0}\right) - \gamma_r(2n_e-n_0)n_p - (\gamma_r+\gamma_\mathrm{bg})n_e  \label{eq:rate-eq-1}\\
		& \frac{dn_p}{dt} = \gamma_r(2n_e-n_0)n_p + \gamma_rn_e - \gamma_cn_p  \label{eq:rate-eq-2}
	\end{align}
    \end{subequations}
    \textcolor{black}{Terms like excited emitters, carriers, and electron-hole pairs are used interchangeably in this article, while differences exist when using more sophisticated models \cite{Gies_PRA_2007,Kreinberg_LightScAppl_2017}}. 
    
    In \Eqref{eq:rate-eq-1} and \Eqref{eq:rate-eq-2}, $I$ is the injection current, $n_0$ is the total number of emitters and $q$ is the electron charge. The net stimulated emission rate, $\gamma_r(2n_e-n_0)n_p$, is the difference between the stimulated emission rate, $\gamma_rn_en_p$, and the stimulated absorption rate, $\gamma_r(n_0-n_e)n_p$. The spontaneous emission rate into the lasing mode, $\gamma_rn_e$, coincides with the stimulated emission rate with one photon in the mode, as following from Einstein's relation \cite{Bjork_JQE_1991,Romeira_JQE_2018,Khurgin_LPR_2021}. Nonradiative recombination and radiative recombination into non-lasing modes are embedded into the background recombination rate, $\gamma_\mathrm{bg}$. The photon decay rate, accounting for intrinsic and mirror losses, is denoted by $\gamma_c$ and determines the Q-factor, $Q_c = \omega_c/\gamma_c$. The total decay rate per emitter is $\gamma_t = \gamma_r + \gamma_\mathrm{bg}$, whereas the pump rate per emitter is $\gamma_p = I/(qn_0)$. 

    We define the spontaneous emission factor ($\beta$-factor) as the ratio between the spontaneous emission rate into the lasing mode and the total carrier recombination rate \cite{Rice-PRA-1994,Mørk_APL_2018,Jagsch_NatComm_2018}: 
   \begin{equation}
	\label{eq:beta-factor}
	\beta = \frac{\gamma_rn_e}{\gamma_rn_e + \gamma_\mathrm{bg}n_e} = \frac{\gamma_r}{\gamma_r + \gamma_\mathrm{bg}}
   \end{equation}
   Therefore, the $\beta$-factor herein defined includes the effect of nonradiative recombination and quantifies the overall efficiency of carrier recombination into the lasing mode.

   %\Eqref{eq:rate-eq-1} and \Eqref{eq:rate-eq-2} are derived from cavity quantum electrodynamics models of discrete, two-level emitters \cite{Gies_PRA_2007,Auffeves_PRB_2010,Moelbjerg_JQE_2013} by adiabatic elimination of the medium polarization and by neglecting correlations between emitters. 
   As for conventional \cite{Coldren-book2ndEd-2012} rate equations, coherent effects such as Rabi oscillations \cite{Nomura_NatPhys_2010,Gies_PRA_2017} and superradiance \cite{Leymann_PRA_2015,Jahnke_NatCom_2016} are not included. \textcolor{black}{Furthermore, the model only applies in the weak-coupling regime \cite{Xu-Yariv_PRA_2000,Gérard_Springer_2003}, where the emitter dephasing rate, $\gamma_2$, is much larger than the light-matter coupling rate, $g = \sqrt{\gamma_2\gamma_r}/2$.} As an extension to conventional rate equations, we include the pump blocking term, $(1-n_e/n_0)$, which represents Pauli-blocking and reflects the finite number of emitters \cite{Mørk_APL_2018}.           
   %To a first approximation, the equations also apply to semiconductor lasers with extended active media, such as layers of quantum wells \cite{Matsuo_JSTQE_2013,Jagsch_NatComm_2018,Dimopoulos_LPR_2022} or quantum dots \cite{Ota_OptExp_2017,Zhou_JSTQE_2022}. In this case, the inhomogeneous broadening of the electronic transitions, as described by the electronic density of states \cite{Coldren-book2ndEd-2012}, generally makes the stimulated and spontaneous emission rates nonlinear functions of the number of carriers. Therefore, more accurate modeling requires purely numerical approaches \cite{Gregersen_APL_2012,Romeira_JQE_2018,Jagsch_NatComm_2018}. However, if the injection level is not much higher than the transparency condition, the spontaneous and stimulated emission rates are approximately linear in the number of carriers, as entailed by \Eqref{eq:rate-eq-1} and \Eqref{eq:rate-eq-2} and often assumed in the literature \cite{Bjork_JQE_1991,Matsuo_JSTQE_2013,Ota_OptExp_2017,Dimopoulos_LPR_2022}. 
   
   In the case of extended active media, the pump blocking term effectively reflects the saturation \cite{Gregersen_APL_2012} with increasing current of the number of electron-hole pairs interacting with the cavity field. These electronic states are located in a bandwidth typically determined by the homogeneous broadening and centered around the cavity mode frequency \cite{Mark_APL_1992} (see Sec. I in the Supplementary Material). The occupation of the electronic states that contribute to the gain of the optical mode being considered will thus approach unity as the quasi-Fermi levels of electrons and holes move past the cavity mode frequency. % and they are pushed deeper into the conduction and valence band. 
   We note that the consideration of a single optical mode is particularly appropriate in EDC cavities \cite{Kountouris-OptExress-2022}. %In macroscopic lasers, on the contrary, the presence of several cavity modes and the possibility of mode hopping nullify, in practice, the effect of pump blocking. 

   The rate equations govern the average number of photons and carriers, but the intrinsic quantum noise induces fluctuations around these averages. As a result, laser light possesses characteristic statistical properties, depending on the pumping level. 
   % The standard approach to modeling the quantum noise is complementing the rate equations with Langevin noise sources. Hence, one may perform a small-signal analysis \cite{Coldren-book2ndEd-2012} or integrate the equations numerically \cite{Ahmed_JQE_2001}. %The two approaches are found to be in good agreement if the correlations between photons and carriers are properly taken into account \cite{Ahmed_JQE_2001}.  
   In the following, we briefly illustrate two techniques to model the quantum noise: the Langevin approach \cite{Coldren-book2ndEd-2012} and a stochastic approach \cite{Andre_OptExpress_2020}. 
   
   By following the Langevin approach, we add Langevin noise forces to \Eqref{eq:rate-eq-1} and \Eqref{eq:rate-eq-2} and carry out a small-signal analysis. Hence, we compute the fluctuation in the photon number and the corresponding power spectral density \cite{Coldren-book2ndEd-2012}. By integrating the spectral density over all frequencies, we calculate the variance of the photon number, $\langle\Delta n_p^2\rangle$. The expression of this variance depends on the small-signal coefficients of the rate equations, correlation strengths of the Langevin noise forces, and the pumping level (see Sec. II in the Supplementary Material). From the variance, we obtain the relative intensity noise (RIN), the Fano factor \cite{Rice-PRA-1994}, $F_F$, and the second-order intensity correlation, $g^{(2)}(0)$ \cite{Fox-book-2006}:
       \begin{equation}
       \label{eq:statistical-measures}
          \mathrm{RIN} = \frac{\langle\Delta n_p^2\rangle}{\langle n_p\rangle^2}, \ \: F_F = \frac{\langle\Delta n_p^2\rangle}{\langle n_p\rangle}, \ \: g^{(2)}(0) = \frac{\langle n_p(n_p-1)\rangle}{\langle n_p\rangle^2}
       \end{equation}
   The average of the photon number, $\langle n_p\rangle$, is calculated analytically from the carrier and photon rate equations. Since the variance may be expressed as $\langle\Delta n_p^2\rangle = \langle n_p^2\rangle - \langle n_p\rangle^2$, one finds $g^{(2)}(0) = 1 + \mathrm{RIN} - 1/\langle n_p\rangle$. 
   
   It should be stressed that here we focus on the statistical properties of the intracavity photon number. The output optical power is affected by an additional noise contribution, the partition noise \cite{Coldren-book2ndEd-2012} at the laser output. As a result, the variance of the photon number and the variance of the output power generally differ \cite{Yamamoto_PRA_1986,Coldren-book2ndEd-2012,Mørk_Optica_2020,Bundgaard_PRL_2023}.

   The Langevin approach is the standard technique to model the quantum noise. However, describing the quantum noise with Langevin noise sources is questionable in the case of very few emitters \cite{Roy-Choudhury_PRA_2010,Bundgaard_PRL_2023}. Indeed, by definition, the small-signal analysis assumes the noise fluctuations to be a small perturbation of the state of the laser. In nanolasers, variations on the order of a single photon or a single emitter excitation may constitute a non-negligible perturbation, due to the small number of photons and carriers. Hence, a small-signal analysis is not guaranteed to properly describe the noise properties.    
 
    %However, the numerical approach becomes increasingly questionable as one moves from above to below threshold. Indeed, by definition, the small-signal analysis assumes the quantum fluctuations to be a small perturbation of the state of the laser. In nanolasers, variations on the order of a single photon or a single emitter may constitute a non-negligible perturbation, due to the small number of photons and carriers. Furthermore, the necessity of enforcing a non-negative photon number to ensure physically meaningful solutions may lead to incorrect statistics. Hence, a small-signal analysis is not guaranteed to properly describe the noise properties.   
	
    Instead of modeling the \textit{collective} effect of the photon and carrier fluctuations with Langevin noise forces, the stochastic approach more fundamentally describes the laser dynamics as a stochastic Markov process in the space of the photon and carrier numbers \cite{Andre_OptExpress_2020}. More specifically, the Langevin approach describes the number of photons and carriers as \textit{continuous} time variables. Conversely, the stochastic approach reflects the shot noise affecting photons and carriers by considering the latter as \textit{discrete} stochastic time variables. The shot noise originates from the finite values of the electronic charge and photon energy, imposing a granularity on the electron current and optical power. 
    % intrinsic granularity of the electron charge and photon energy and from the probabilistic nature of the various emission and recombination events. 
    The rates of the various events involving emitters and photons (stimulated emission, spontaneous emission, etc.) are regarded as probabilities per unit of time that a given event would occur, based on the current number of particles. How long it takes for the next event to occur and which event it occurs is determined via a Monte Carlo simulation framework, the so-called Gillespie's first reaction method \cite{Gillespie_JCP_1976}. The number of photons and carriers is resolved versus time with a nonuniform time step. Compared to similar stochastic approaches with fixed time increment \cite{Puccioni_OptExpress_2015,Mørk_APL_2018,Mørk_Optica_2020}, Gillespie's first reaction method is numerically exact, in the sense that it does not require any convergence study on the width of the time step \cite{Andre_OptExpress_2020}. The average photon number and the photon variance are obtained from the time-resolved number of photons. Subsequently, one finds the RIN, the Fano factor and the intensity correlation.
    
    In general, good agreement is found between the Langevin approach and the stochastic approach when considering the average photon number and RIN, even in the presence of very few emitters \cite{Mørk_APL_2018,Andre_OptExpress_2020} and down to the case of a single emitter \cite{Bundgaard_PRL_2023}. Qualitative discrepancies, though, may emerge in the intensity correlation \cite{Bundgaard_PRL_2023} and the Fano factor, as illustrated in \secref{Sec: Photon statistics: stochastic simulations}. For a single emitter, for instance, the stochastic approach, contrary to the Langevin approach, well reproduces the change in the intensity correlation with increasing current as obtained from full quantum simulations \cite{Bundgaard_PRL_2023}. % Furthermore, stochastic simulations give access to the probability distribution of the photon number, irrespective of the number of emitters. On the contrary, scaling up full quantum simulations beyond the case of a single emitter is either computationally demanding or requires specific assumptions \cite{Yacomotti-LPR-2022}. 

    Finally, a comment is due on the impact of pump broadening. The incoherent pumping leads to pump-induced dephasing, effectively increasing the emitter dephasing rate \cite{Gartner_PhysRevA_2011,Moelbjerg_JQE_2013}. As a result, the spontaneous emission rate per emitter may be quenched at high current values. However, the effect is negligible if the pumping rate per emitter, $\gamma_p$, is much smaller than the nominal value of $\gamma_2$. Therefore, one can safely ignore pump broadening if considerations are limited to values of current much smaller than $I_{pb} = qn_0\gamma_2$. 
    
    Throughout this article, we assume $d = 10^{-28}\,\mathrm{Cm}$ \cite{Yamada_book_2014,Mørk_APL_2018}, $n_{\mathrm{act}} = 3.3$, $\lambda_c = 1.55\,\mu\mathrm{m}$ and $T_2 = 50\,\mathrm{fs}$. The value of the dephasing time, in the range of a few tens of femtoseconds, reflects operation at room temperature \cite{Borri_APL_2000}. % For example, one finds $I_{pb} \approx 6.4\,\mu\mathrm{A}$ for $n_0 = 1$ and $I_{pb} \approx 769\,\mu\mathrm{A}$ for $n_0 = 120$. 
    Unless explicitly noted, pump broadening would not affect the results herein discussed and is therefore neglected. 

   % \textcolor{black}{[Discussion on $\gamma_\mathrm{bg}$?]}

    \section{Threshold with photon recycling}\label{sec:lasing-threshold-intro}

    %Despite decades of research, the concept of lasing threshold in semiconductor micro- and nanolasers is still the center of a vivid debate \cite{Bjork_JQE_1991,Jin_PRA_1994,Bjork_PRA_1994,Rice-PRA-1994,Ning_JSTQE_2013,Ma_LPR_2013,Lohof_PhysRevApplied_2018,Takemura_PRA_2019,Khurgin_LPR_2021,Carroll_PhysRevLett_2021,Lippi_ChaosSolitonsFractals_2022,Vyshnevyy_PhysRevLett_2022,Carroll_PhysRevLett_2022,Yacomotti-LPR-2022,Carroll_PRA_2023}. As we shall see, the large spontaneous emission rate per emitter achievable in EDC lasers questions the standard approach \cite{Coldren-book2ndEd-2012} for calculating the threshold current. In the following, we generalize the classical definition of balance between gain and loss by building on the concept of photon recycling \cite{Yamamoto_JJAP_1991}.

    Classically, the lasing threshold is reached when the \textit{net} stimulated emission, $\gamma_r(2n_e-n_0)n_p$, balances the optical loss, $\gamma_cn_p$. This balance requires the number of carriers to be
	\begin{equation}
		\label{eq:ne-gain-loss-balance}
		n_{e_\mathrm{th}} = \frac{n_0}{2} + \frac{\gamma_c}{2\gamma_r} 
	\end{equation}
    Here, the first term is the number of carriers at transparency and the second is the additional fraction one needs to excite to compensate for the cavity loss. In the presence of lasing, the number of carriers is clamped at $n_{e_\mathrm{th}}$. 

    The transparency current is defined unambiguously. By substituting the number of carriers at transparency into the rate equations, one finds the number of photons at transparency
	\begin{equation}
            \label{eq:xi-definition}
		\xi = \frac{n_0\gamma_r}{2\gamma_c} = N_\mathrm{tr}\Gamma \frac{2d^2}{\hbar\epsilon_0n_\mathrm{act}^2} \frac{Q_c}{\gamma_2}
	\end{equation}
	%This is the ratio between the rate of spontaneous emission into the lasing mode at transparency, $n_0\gamma_r/2$, and the photon decay rate, $\gamma_cn_p$, with one photon. 
    and the transparency current 
    \begin{equation}
        \label{eq:Itr-definition}
	I_\mathrm{tr} = q\frac{2\xi\gamma_c}{\beta} \quad \text{(transparency)}
    \end{equation}
    In \Eqref{eq:Itr-definition}, the factor of $2$ is absent if pump blocking is neglected. In the last step of \Eqref{eq:xi-definition}, we have assumed an extended active region and expressed the total number of emitters as $n_0 = 2N_\mathrm{tr}V_\mathrm{act}$, with $N_\mathrm{tr}$ being the transparency carrier density. 

    \begin{figure}[ht]
		\centering\includegraphics[width=1\linewidth]{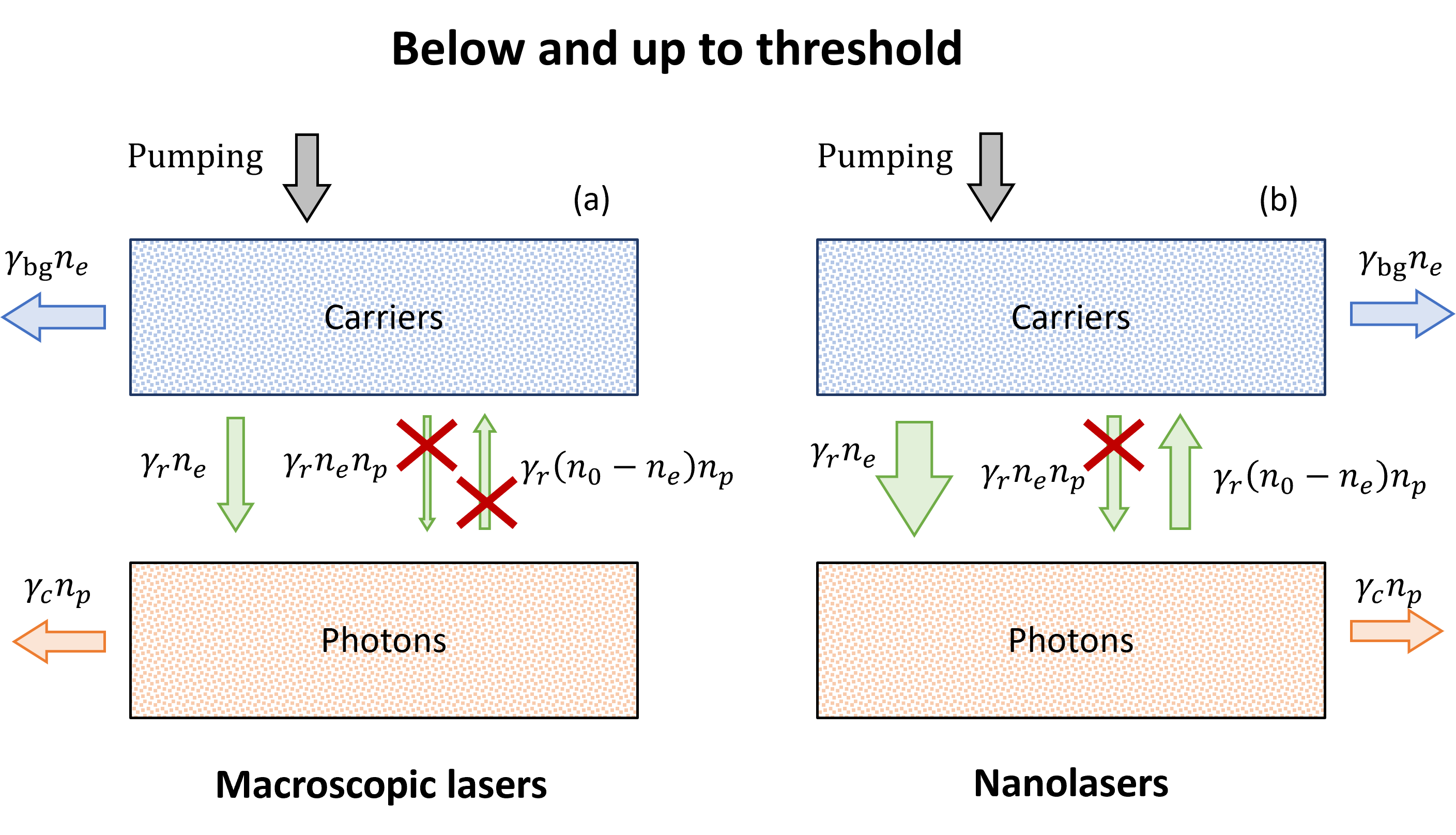}
		\caption{\label{fig:Sec4:FIGphot-res} Major contributions to the carrier and photon dynamics below and up to threshold in (a) macroscopic lasers and (b) nanolasers. The processes marked with a cross have a minor impact. In nanolasers, for $\beta$ approaching one and $\xi$ sufficiently larger than one-half, a large fraction of the photons funneled into the lasing mode are re-absorbed before decaying, leading to photon recycling.}
    \end{figure} 
    As opposed to the transparency current, the threshold current is not necessarily a well-defined quantity. Indeed, a quick inspection of the photon rate equation shows that the balance between gain and loss is only achieved, strictly speaking, in the limit of an infinite number of photons, due to spontaneous emission into the lasing mode. 
	%This inconsistency is not an issue if the $\beta$-factor is much smaller than one. One may then argue that the concept of threshold itself is meaningless for lasers operating far from the thermodynamic limit of $\beta\rightarrow0$ \cite{Rice-PRA-1994}. The relevance of this argument, however, does not solve the equally relevant problem \cite{Ning_JSTQE_2013,Lippi_ChaosSolitonsFractals_2022} of a definition at least functional to applications. 
    The standard approach \cite{Coldren-book2ndEd-2012} is to estimate the number of carriers below threshold by wholly neglecting the net stimulated emission term in the carrier rate equation, as illustrated by \figref{fig:Sec4:FIGphot-res}a. Thus, by requiring the number of carriers to fulfill \Eqref{eq:ne-gain-loss-balance}, one finds the classical threshold current, $I_\mathrm{cl} = q(1-n_{e_\mathrm{th}}/n_0)\gamma_tn_{e_\mathrm{th}}$, with $\gamma_t = \gamma_r + \gamma_\mathrm{bg}$. The term in brackets is due to pump blocking, often neglected \cite{Bjork_JQE_1991,Coldren-book2ndEd-2012}, and $\gamma_tn_{e_\mathrm{th}}$ is the total loss rate of carriers at threshold. 
    
    By introducing the number of photons at transparency, one finds
	\begin{equation}
		\label{eq:Ith1}
		I_\mathrm{cl} = \frac{q}{\eta}\frac{\gamma_c}{2\beta}\left(2\xi + 1\right) \quad \text{(classical threshold)}
	\end{equation}
    where the pump injection efficiency, $\eta$, is
	\begin{equation}
		\label{eq:eta}
		\eta = \frac{1 - 1/(2\xi)}{2}
	\end{equation} 
    We emphasize that $\eta$ reflects the leakage due to pump blocking. If pump blocking is neglected, then $\eta=1$. Any additional leakage would further degrade the injection efficiency by a factor $\eta_i$ \cite{Coldren-book2ndEd-2012}, amounting to quantitative, but not qualitative changes. For simplicity, we assume $\eta_i=1$ throughout this article. 
    
    Importantly, in the presence of pump blocking, the maximum available gain is finite ($n_0\gamma_r$). Therefore, according to the classical definition, lasing is unattainable if $n_0\gamma_r\leq\gamma_c$, namely
    \begin{equation}
	\label{eq:LED-regime}
	\xi \leq \frac{1}{2} \quad \text{(LED regime)}
    \end{equation} 
    In this case, the laser operates as a light-emitting diode (LED) and the photon number saturates at sufficiently high currents. This saturation is a characteristic feature of LEDs, though in practice other detrimental effects herein neglected, such as the temperature increase, may cause a similar saturation even in lasers.
    
    The standard approach to estimate the number of carriers below threshold is invalid if the $\beta$-factor is close to one. In this case, due to the large spontaneous emission into the lasing mode, the number of photons below transparency may be non-negligible. Furthermore, if $\xi$ is sufficiently larger than one-half, a generated photon has a higher chance of being re-absorbed than decaying. Indeed, $\xi$ is one-half of the ratio between the stimulated absorption rate well-below inversion, $\gamma_r n_0n_p$, and the photon decay rate, $\gamma_c n_p$. This process of large spontaneous emission into the lasing mode and significant stimulated re-absorption recycles the generated photons without major energy loss and effectively increases the carrier lifetime \cite{Yamamoto_JJAP_1991}. 
    
    We have found that a better estimate of the number of carriers below threshold is obtained by neglecting the stimulated emission, but retaining the stimulated absorption, as illustrated by \figref{fig:Sec4:FIGphot-res}b. In other terms, we assume $\gamma_r(2n_e-n_0)n_p\approx-\gamma_r n_0n_p$ in both the carrier and photon rate equations, leading to 
	\begin{equation}
		\label{eq:ne-approx-PR}
		n_e \approx \frac{(I/q)(2\xi+1)}{\gamma_r/\beta_{\mathrm{eff}} + (2\xi+1)\gamma_p} \quad \text{(below threshold)}
	\end{equation}
	Here, we define an effective $\beta$-factor: 
	\begin{equation}
		\label{eq:beta-tilde}
		\beta_{\mathrm{eff}} = \frac{\beta}{1+ 2\xi(1-\beta)}
	\end{equation}
        % If pump blocking is neglected, the denominator of \Eqref{eq:ne-approx-PR} is simply given by $\gamma_r/\beta_{\mathrm{eff}}$. 
        % Our approach differs from the standard procedure of fully neglecting the net stimulated emission term in the carrier rate equation. 
        Then, we require the approximate number of carriers from \Eqref{eq:ne-approx-PR} to reach the level set by the gain-loss balance condition, $n_{e_\mathrm{th}}$. This requirement leads to an expression for the threshold current including the effect of photon recycling:     
	\begin{equation}
		\label{eq:Ith3}
		I_\mathrm{pr} = \frac{q}{\eta}\frac{\gamma_c}{2\beta_{\mathrm{eff}}} \quad \text{(threshold with photon recycling)}
	\end{equation}  
    
    \textcolor{black}{The corresponding average photon number is} 
      \begin{equation}
     \label{eq:npth3}
     n_{p,{\mathrm{pr}}} = \frac{1}{2} \sqrt{ \frac{1}{\eta\beta_{\mathrm{eff}}} }
     \end{equation}
     \textcolor{black}{This photon number depends on the pump injection efficiency, $\eta$, and \textit{effective} $\beta$-factor, $\beta_{\mathrm{eff}}$. We emphasize that the Q-factor and active region volume affect $\eta$ and $\beta_{\mathrm{eff}}$ via the number of photons at transparency, $\xi$ (cf. \Eqref{eq:xi-definition}). Therefore, for a given value of $\beta$, the average intracavity photon number needed for reaching the threshold with photon recycling may differ significantly.} 
    
    As illustrated by \figref{fig:sec:INTRO:FIG1}, the threshold with photon recycling, \Eqref{eq:Ith3}, and the classical expression, \Eqref{eq:Ith1}, coincide for $\beta\ll1$ (macroscopic lasers), but differ qualitatively for $\beta$ approaching $1$ (EDC lasers). At the end of this section, we will return to this point in more detail.    
	
         \begin{figure}[ht]
		\centering\includegraphics[width=0.9\linewidth]{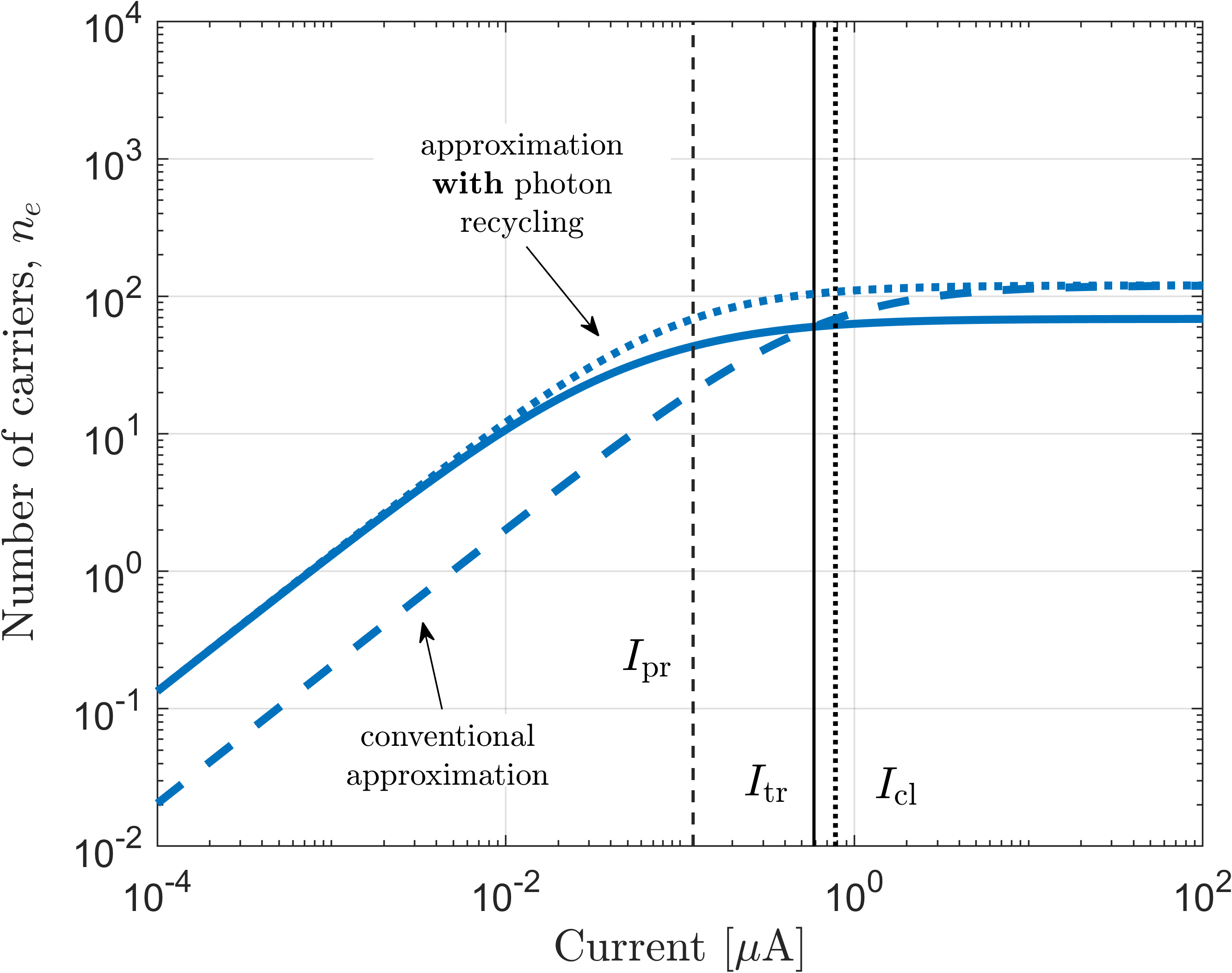}
		\caption{\label{fig:SecIth:FIG1} Number of excited carriers of EDC laser 1 in \tabref{tab:nano-devices} versus current (solid). The conventional approximation used for calculating the threshold current (dashed) and our approximation with the inclusion of photon recycling (dotted, cf. \Eqref{eq:ne-approx-PR}) are also shown. The carriers start clamping at the threshold current with photon recycling, $I_\mathrm{pr}$, even before reaching the transparency current, $I_\mathrm{tr}$. The classical threshold current, $I_\mathrm{cl}$, is slightly larger than $I_\mathrm{tr}$.}
	\end{figure} 
\begin{table*}[ht]
\footnotesize	
\centering
\begin{tabular}{p{0.19\linewidth}p{0.19\linewidth}p{0.19\linewidth}p{0.19\linewidth}p{0.19\linewidth}}
\hline
Device & $n_0$ & $V_p\,[\lambda_c^3/(2n_{\mathrm{act}})^3]$ & $Q_{c}\,[2.43\times10^{3}]$ & $\gamma_\mathrm{bg}\,[10^9\,\mathrm{s^{-1}}]$ \\
\hline
EDC laser 1 & $120$ & $0.154$ & $1$ & $1$\\
PhC laser & $1.8\times10^4$ & $23.16$ & $1$ & $1$\\
macroscopic laser & $9.6\times10^6$ & $1.23\times10^4$ & $1$ & $1$\\
EDC laser 2 & $120$ & $0.154$ & $10$ & $1$\\
EDC laser 3 & $120$ & $0.154$ & $10$ & $100$\\
EDC laser 4 & $120$ & $0.154$ & $1$ & $100$\\
EDC laser 5 & $120$ & $1$ & $1$ & $1$\\
EDC laser 6 & $1$ & $0.01$ & $10$ & $1$\\
nanoLED A & $120$ & $1.54$ & $1$ & $1$\\
\hline
\end{tabular}
\caption{\label{tab:nano-devices} Parameters of the main devices considered in this article: total number of emitters ($n_0$), mode volume ($V_p$), quality factor ($Q_c$) and background recombination rate ($\gamma_\mathrm{bg}$).}  
\end{table*}

    To illustrate the role of photon recycling, we consider an EDC laser, namely EDC laser 1 in \tabref{tab:nano-devices}.
    % with $n_0 = 120$, $V_p \approx 0.154\,\lambda_c^3/(2n)^3$, $\gamma_c = 5\times10^{11}\,\mathrm{s^{-1}}$ (i.e. $Q_c\approx2.43\times10^3$) and $\gamma_\mathrm{bg} = 10^9\,\mathrm{s^{-1}}$, namely EDC laser 1 in \tabref{tab:nano-devices}. 
    The number of emitters, $n_0$, reflects a BH active region having a typical transparency carrier density of $10^{18}\,\mathrm{cm^{-3}}$, an area of $50\times50\,\mathrm{nm}^2$ and $3$ active layers, each being $8\,\mathrm{nm}$ thick. The mode volume follows from the active region volume and a total optical confinement factor equal to $3\,\%$, according to \Eqref{eq:Vp-extended-medium-main-text}.  
    % As mentioned in relation to \figref{fig:SecIthIntro:FIG1}, this value of $n_0$ may reflect an EDC laser with a BH active region having a transparency carrier density of $10^{18}\,\mathrm{cm^{-3}}$, an area of $50\times50\,\mathrm{nm}^2$ and $3$ active layers, each being $8\,\mathrm{nm}$ thick. The mode volume, $V_p$, follows from the active region volume and a total optical confinement factor equal to $3\,\%$. Furthermore, we assume $\gamma_\mathrm{bg} = 10^9\,\mathrm{s^{-1}}$ and $\gamma_c = 5\times10^{11}\,\mathrm{s^{-1}}$. 
    These parameters lead to $\beta\approx0.97$ and $\xi\approx3.54$. \figref{fig:SecIth:FIG1} shows the number of carriers (solid) versus current, as well as the estimates with (dotted, \Eqref{eq:ne-approx-PR}) and without (dashed) photon recycling. The vertical lines pinpoint the threshold current with photon recycling (dashed), the transparency current (solid) and the classical threshold current (dotted). 
    
    \Eqref{eq:ne-approx-PR} well describes the number of carriers below the threshold with photon recycling. Importantly, it is seen that the number of carriers at a given current below transparency is higher than expected from the conventional approximation. This discrepancy is due to photon recycling, which effectively increases the carrier lifetime. As a result, a lower current suffices to reach the number of carriers required by the gain-loss balance condition. 
    
    We emphasize that the large fraction of spontaneous emission into the lasing mode plays a key role, unlike the usual regime encountered in macroscopic lasers. As further discussed in \secref{sec:beta-factor} and shown in \figref{fig:SecIth:FIG1}, the threshold with photon recycling may well be attained below transparency. In such a case, the net stimulated emission term is negative. Therefore, the growth of the photon number is largely sustained by the spontaneous emission into the lasing mode. 
    % It is the spontaneous emission that balances the photon loss resulting from out-coupling loss, internal loss and stimulated absorption.

    % In this section, we discuss the fundamental changes in the photon statistics that typically characterize the onset of lasing, all the way from the macro to the nanoscale. Generally speaking, the transition to lasing is abrupt in macroscopic lasers, which approximate well the thermodynamic limit of $\beta\rightarrow0$ \cite{Rice-PRA-1994}. In this case, the lasing threshold is well-defined (see \secref{sec:other-threshold-definitions}). On the contrary, the process becomes increasingly smooth as the $\beta$-factor increases and a grey area appears, where multiple definitions have been proposed. In the following, we focus on the role of the threshold current with photon recycling.

       \begin{figure}[ht]
	    \centering\includegraphics[width=1\linewidth]{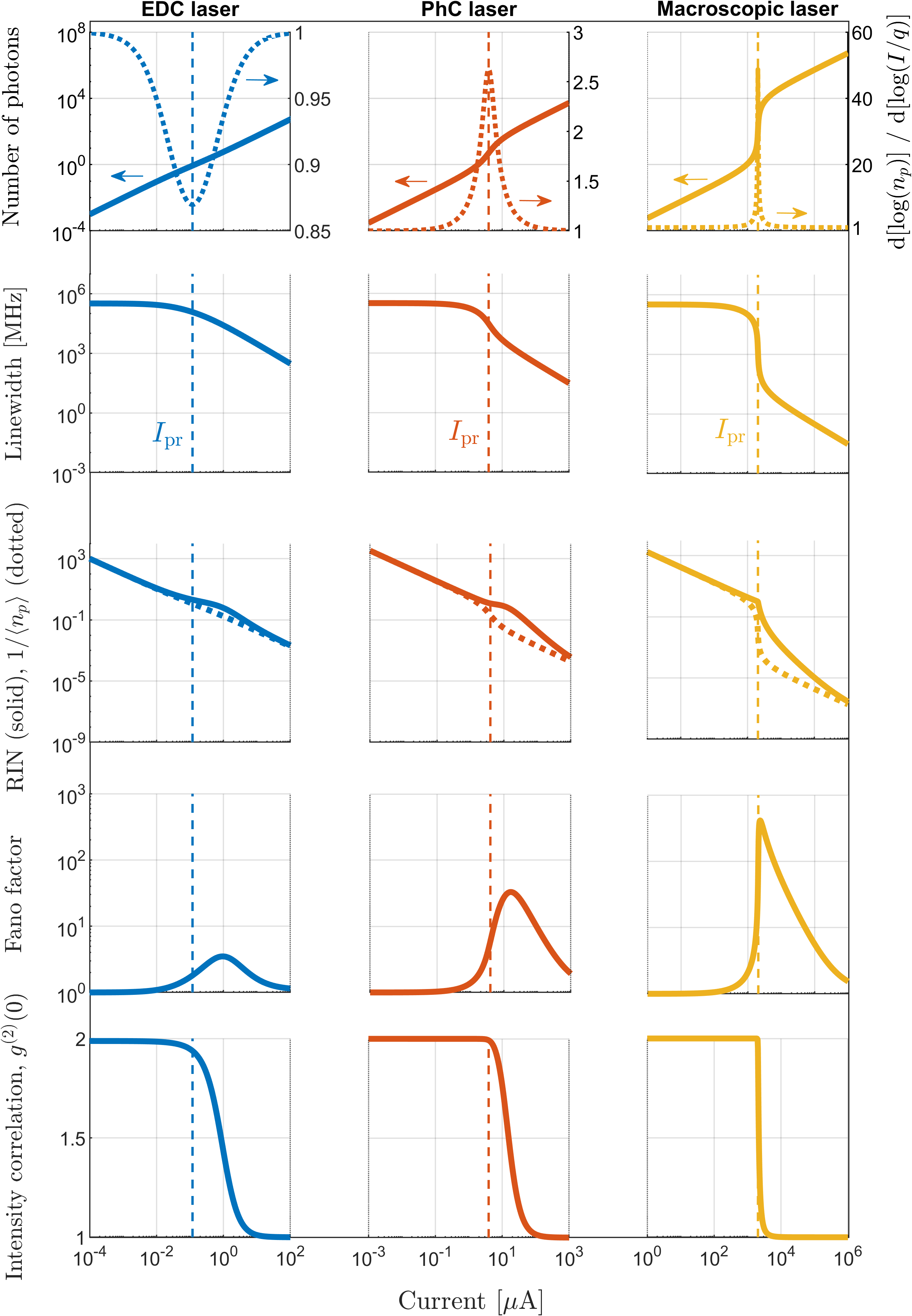}
	      \caption{\label{fig:SecIth:FIG3} Input-output characteristics of a nanolaser with extreme dielectric confinement (EDC laser, left), a photonic crystal laser (PhC laser, center) and a conventional macroscopic laser (right). The parameters for the EDC laser (EDC laser 1) as well as the PhC and macroscopic laser are given in \tabref{tab:nano-devices}. The vertical, dashed lines indicate the threshold with photon recycling.}
       \end{figure} 
   We shall now examine the variation with current of the average photon number, the laser linewidth and the photon statistics, including key figures of merit such as the RIN, the Fano factor, and the intensity correlation, introduced in \secref{sec:rate-equation-model}. Besides EDC laser 1, we consider a PhC laser and a macroscopic laser, with parameters summarized in \tabref{tab:nano-devices}. The number of emitters reflects an active region with a typical transparency carrier density of $10^{18}\,\mathrm{cm^{-3}}$ and consisting of $3$ active layers, each being $8\,\mathrm{nm}$ thick and having an area of $0.3\times1.25\,\mu\mathrm{m}^2$ \cite{Takemura_PRA_2019} (PhC laser) and $2\times100\,\mu\mathrm{m}^2$ \cite{Coldren-book2ndEd-2012} (macroscopic laser). The mode volume corresponds to a total optical confinement factor equal to $3\%$, leading, in all cases, to the same value of $\xi$ (cf. \Eqref{eq:xi-definition}). 
   
   \figref{fig:SecIth:FIG3} illustrates the input-output characteristics of the EDC laser (left), the PhC laser (center) and the macroscopic laser (right) versus current. The vertical, dashed lines indicate the threshold with photon recycling. The quantum noise (see \secref{sec:rate-equation-model}) is modeled by the Langevin approach, but we note that the stochastic approach would provide similar results. Deviations emerge when further scaling down the number of emitters, as illustrated explicitly in \secref{Sec: Photon statistics: stochastic simulations}.    
       
   In the first row of \figref{fig:SecIth:FIG3}, the light-current (LI) characteristic is displayed (left axis). The macroscopic laser features a marked intensity jump, with a typical S-shape and a steep increase in the photon number at the threshold with photon recycling. Conversely, the intensity jump reduces significantly for the PhC laser, and the EDC laser features an almost linear characteristic. 
   
   The shape of the LI characteristic and role of the $\beta$-factor are discussed in detail in \secref{sec:beta-factor}. \textcolor{black}{In particular, it is shown therein that a linear LI characteristic - referred to as "thresholdless" \cite{Yokoyama_JAP_1989,Yokohama_Science_1992} with an unfortunate yet widespread terminology - does \textit{not} require $\beta=1$. Instead, $\beta$ and $\xi$ must obey \Eqref{eq:linear-LI-xi-cond}. Here, we note that \textit{if} an intensity jump is present, \textit{then} $I_{\mathrm{pr}}$ coincides with the inflection point of the LI characteristic on a double logarithmic scale:}
	\begin{equation}
		\label{eq:LIchar-2ndder-Ith2}
		\left.\frac{d^2[\log(n_p)]}{d[\log(I/q)]^2}\right|_{I_\mathrm{pr}} = 0
	\end{equation}
   % This key property makes the threshold with photon recycling a useful operational definition rather than a mere theoretical construction. % Furthermore, this property allows us to define precisely the intensity jump as the ratio between the asymptotes of the upper and lower branches of the LI characteristic at the threshold current with photon recycling.
   In the first row of \figref{fig:SecIth:FIG3}, the first derivative, $d[\log(n_p)]/d[\log(I/q)]$, is also shown (right axis). Consistently with \Eqref{eq:LIchar-2ndder-Ith2}, the threshold with photon recycling coincides with the maximum or minimum of the first derivative, depending on whether the LI characteristic features a typical or inverse \cite{Jagsch_NatComm_2018} S-shape, respectively. 
   
   \textcolor{black}{We emphasize, though, that \Eqref{eq:LIchar-2ndder-Ith2} is \textit{not} a definition, but a mere corollary. If the LI characteristic is linear (cf. \Eqref{eq:linear-LI-xi-cond}), there is no intensity jump and \Eqref{eq:LIchar-2ndder-Ith2} is invalid. Yet, the threshold with photon recycling, \Eqref{eq:Ith3}, remains well-defined. In contrast, \textit{defining} the lasing threshold as the inflection point \cite{Ning_JSTQE_2013} shows its limits in the case of linear LI characteristics and fuels the confusion around "zero-threshold" \cite{DeMartini_PRL_1988} or "thresholdless" \cite{Yokoyama_JAP_1989,Yokohama_Science_1992} lasers. Instead, it is a central conclusion of this article that a threshold \textit{does} exist even in lasers with a linear LI characteristic.} 
   
   \textcolor{black}{Furthermore, previous works \cite{Kreinberg_LightScAppl_2017} have shown that nanoLEDs at cryogenic temperatures could also feature LI characteristics with an inflection point. 
   %This may be due to heating or saturation effects not accounted for by the rate equations used here. 
   This emphasizes that, experimentally, \Eqref{eq:LIchar-2ndder-Ith2} should be complemented by other measures for the achievement of lasing, such as the second-order intensity correlation.}  
       
       The second row of \figref{fig:SecIth:FIG3} shows the laser linewidth, which we calculate by the modified Schawlow–Townes formula \cite{Coldren-book2ndEd-2012,Henry_JLWT_1986} 
       \begin{equation}
       \label{eq:laser-linewidth}
       \Delta\nu = \frac{\gamma_rn_e}{4\pi n_p} = \frac{\gamma_c}{4\pi}\frac{\xi+\frac{1}{2}}{n_p + \frac{1}{2}}
       \end{equation}
       The Schawlow–Townes linewidth determines the intrinsic coherence time of the laser light, limited by fluctuations induced by spontaneous emission on the phase of the electric field. The original formula derived by Schawlow and Townes \cite{Schawlow_PR_1958} applies well below threshold and it is a factor of two larger than \Eqref{eq:laser-linewidth}. Well above threshold, instead, the intensity fluctuations are quenched and only phase fluctuations are important \cite{Henry_JLWT_1986}, leading to \Eqref{eq:laser-linewidth}. We note that in the presence of gain-induced refractive index modulation, herein neglected, the Schawlow–Townes linewidth should be multiplied by an additional factor of $1+\alpha^2$, with $\alpha$ being the linewidth enhancement factor \cite{Henry_JQE_1982,Coldren-book2ndEd-2012}. Furthermore, the linewidth may generally reflect additional effects, such as non-orthogonality of modes and bad-cavity effects \cite{Pick_PRA_2015,Cerjan_OptExpress_2015}. Nonetheless, \Eqref{eq:laser-linewidth} retains the fundamental dependence on the number of photons. 
       %Well below threshold, where the photon number is negligible, the linewidth is the same irrespective of the laser size (all lasers in \figref{fig:SecIth:FIG3} share the same value of $\xi$). 
       At threshold, the linewidth of the macroscopic laser abruptly decreases, reflecting the marked intensity jump. The linewidth reduction, instead, becomes gradually smoother as one moves from the PhC laser to the EDC laser. 
       % In this case, the narrowing sets in around the threshold with photon recycling, as also found by quantum simulations of single-emitter nanolasers \cite{Bundgaard_ArXiV_2023}. 

       %The Schawlow–Townes linewidth depends on the average number of photons. To access information such as the variance of the photon number and the RIN (RIN) \cite{Coldren-book2ndEd-2012}, the presence of quantum noise should be explicitly taken into account. In \secref{Sec: Photon statistics: stochastic simulations}, we utilize a more fundamental stochastic simulation scheme \cite{Andre_OptExpress_2020,Bundgaard_ArXiV_2023}, which provides further information on the photon statistics.    

       The third row of \figref{fig:SecIth:FIG3} displays the RIN (solid) and the inverse of the average photon number (dotted). Well below threshold, the lasers operate as LEDs, hence generating thermal light. The variance is $\langle\Delta n_p^2\rangle = \langle n_p\rangle^2 + \langle n_p\rangle$ \cite{Fox-book-2006} and the RIN reduces to $\mathrm{RIN} = 1 + 1/\langle n_p\rangle \approx 1/\langle n_p\rangle$. The approximation holds due to the low average photon number. Well above threshold, the photon statistics asymptotically approaches a Poisson process, corresponding to coherent laser light. For a Poisson process, the variance equals the average \cite{Fox-book-2006}, $\langle\Delta n_p^2\rangle = \langle n_p\rangle$, leading to $\mathrm{RIN} = 1/\langle n_p\rangle$. At intermediate pumping levels, the photon statistics is more blurred and the RIN is enhanced above the level of $1/\langle n_p\rangle$. Interestingly, a weaker enhancement is observed in the EDC laser.

       In the fourth row of \figref{fig:SecIth:FIG3}, the Fano factor is shown. Irrespective of the laser size, the Fano factor increases around threshold and decreases at larger current values. For the macroscopic laser, the maximum of the Fano factor coincides with the lasing threshold \cite{Rice-PRA-1994}. This motivates an alternative definition of threshold current, corresponding to the maximum of the Fano factor \cite{Jin_PRA_1994} (see \secref{sec:other-threshold-definitions}). For the EDC and PhC lasers, instead, a much smoother increase is observed, and the Fano factor peaks at a pumping level above the threshold with photon recycling. 
       %Interestingly, the photon statistics in the macroscopic laser remains super-Poissonian for a significant current range above threshold, not differently from the case of the EDC laser. This observation is consistent with stochastic simulations of semiconductor lasers with different values of the $\beta$-factor \cite{Andre_OptExpress_2020}. 
       In addition, the maximum value of the Fano factor is seen to scale with the laser size. The EDC laser features the Fano factor with the lowest peak, signifying fluctuations in the photon number, at this current value, with the smallest amplitude relative to the average. This is due to the larger damping rate of EDC lasers (not shown in the figure). The damping rate determines the lifetime of a given noise fluctuation and, consequently, the level of noise accumulation. Therefore, in spite of the larger RIN above threshold, EDC lasers are expected to be more dynamically stable, as generally argued for nanolasers with a high $\beta$-factor \cite{Wang_OptExpress_2021,Lippi_ChaosSolitonsFractals_2022}.

       Finally, we consider the second-order intensity correlation, in the last row of \figref{fig:SecIth:FIG3}. As the current grows from below to above threshold, the intensity correlation changes from $g^{(2)}(0)\approx2$, well below threshold, to $g^{(2)}(0)\approx1$, well above threshold, reflecting the change from thermal to Poissonian statistics. For the macroscopic laser, the change in the intensity correlation is steep and sharply localized at threshold. 
       % In this case, all threshold definitions coincide, apart from the quantum threshold (cf. \figref{fig:SecIthIntro:FIG2}) \cite{Rice-PRA-1994,Lippi_ChaosSolitonsFractals_2022}. 
       A more gradual variation is observed for the EDC and PhC lasers.  
       %In this case, the threshold with photon recycling marks the onset of the change in the intensity correlation. %It should be stressed that the limit of $g^{(2)}(0)=1$ is only reached asymptotically, thus preventing a unique definition of lasing threshold based on the intensity correlation.

    \begin{figure}[ht]
		\centering\includegraphics[width=1\linewidth]{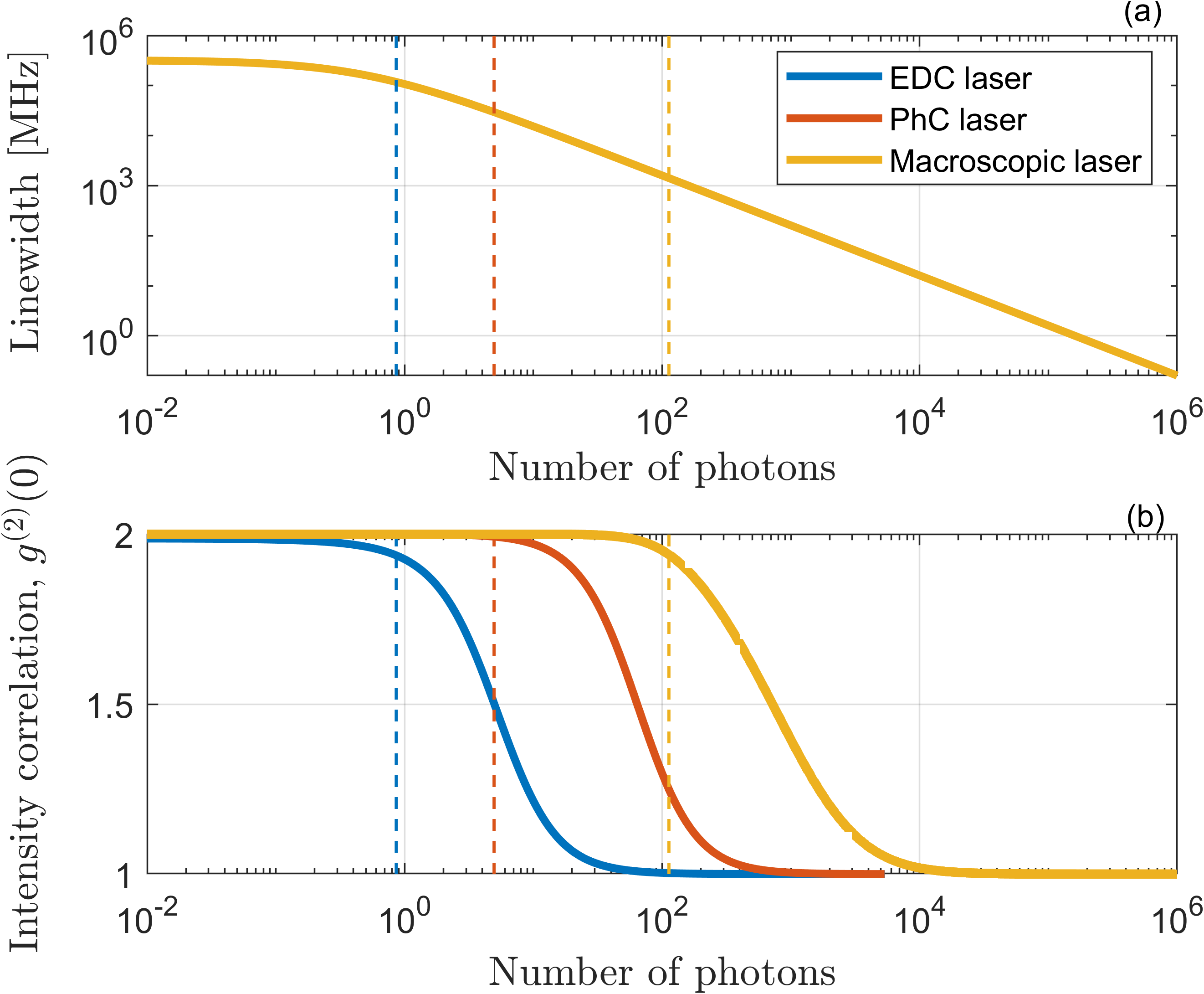}
		\caption{\label{fig:SecIth:Deltanu-g20-vs-np} \textcolor{black}{(a) Modified Schawlow–Townes linewidth and (b) intensity correlation versus photon number of the EDC laser (blue), PhC laser (red), and macroscopic laser (yellow) in \figref{fig:SecIth:FIG3}. The vertical lines indicate the threshold with photon recycling.}}
    \end{figure}
    \textcolor{black}{We point out that the input-output characteristics are smooth when plotted versus the photon number, irrespective of the laser scale \cite{Bjork_PRA_1994,Kreinberg_LightScAppl_2017,Lohof_PhysRevApplied_2018}. This is exemplified by \figref{fig:SecIth:Deltanu-g20-vs-np}, illustrating (a) the linewidth and (b) intensity correlation of the same lasers as in \figref{fig:SecIth:FIG3}. The lasers share the same value of $\xi$, and thus their linewidths coincide for all values of the photon number (cf. \Eqref{eq:laser-linewidth}). The photon number at threshold (dashed lines) significantly varies, due to the different values of the effective $\beta$-factor (cf. \Eqref{eq:npth3}). Hence, the linewidth at threshold substantially differs among the three lasers. Yet, the threshold with photon recycling consistently marks the onset of the change in the second-order intensity correlation, $g^{(2)}(0)$, toward coherent laser light, with small differences in the absolute value of $g^{(2)}(0)$ from one laser to another.}

    \begin{figure}[ht]
		\centering\includegraphics[width=1\linewidth]{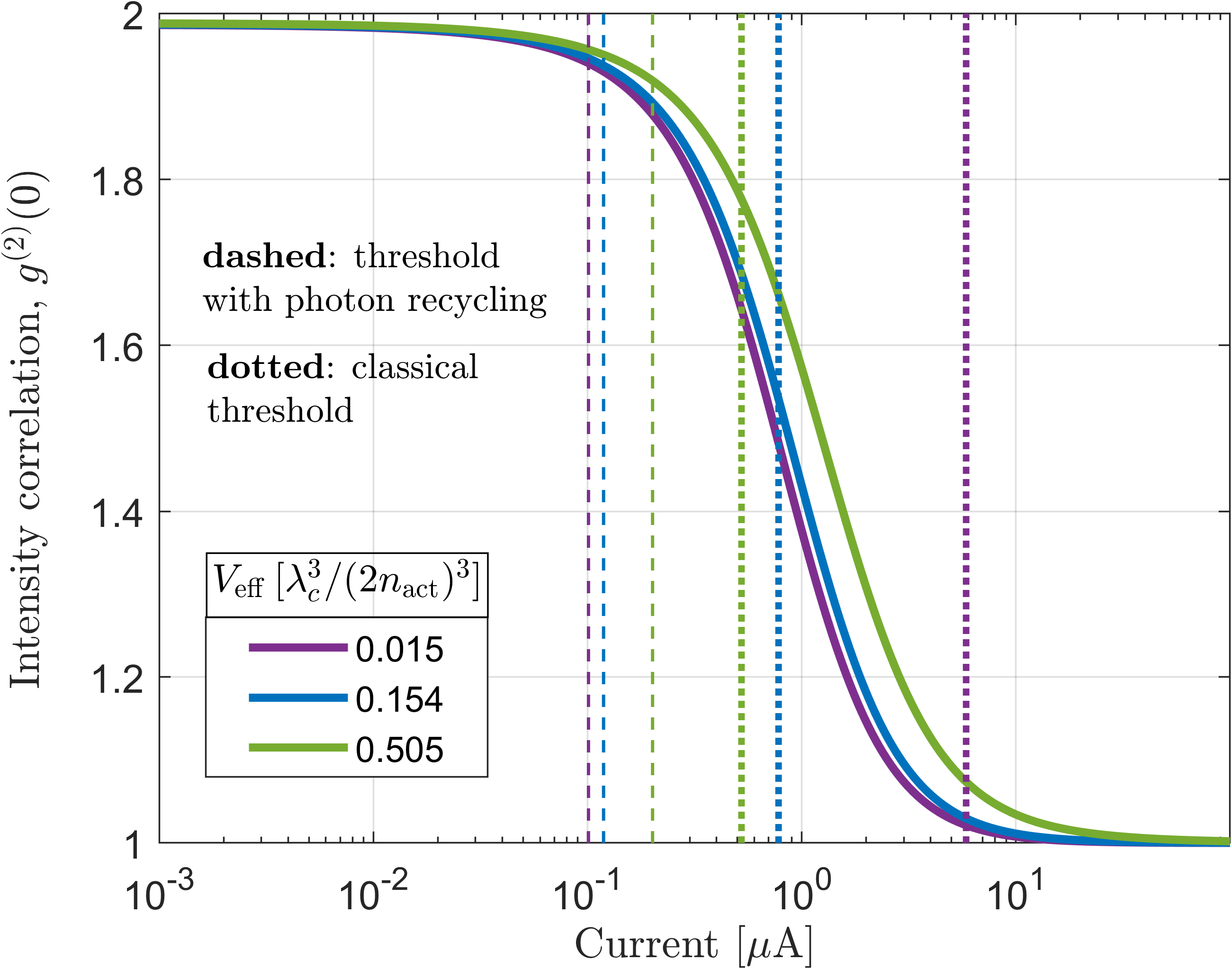}
		\caption{\label{fig:SecLangevin:g20} Intensity correlation versus current of EDC laser 1 in \tabref{tab:nano-devices} (blue) and two other EDC lasers, with a smaller (purple) and a larger (green) mode volume, as indicated in the legend. The other parameters are the same as EDC laser 1. The vertical lines indicate the threshold current with photon recycling (dashed) and the classical threshold current (dotted).}
    \end{figure}
    To further characterize the threshold with photon recycling, in \figref{fig:SecLangevin:g20} we show the intensity correlation of two other EDC lasers versus current, with a smaller (purple) and a larger (green) mode volume compared to EDC laser 1 (blue). The other parameters are unchanged. Irrespective of the mode volume, the threshold with photon recycling (dashed) consistently marks the onset of the transition toward coherent laser light. A smaller mode volume translates into better coherence at a given value of current and therefore into a lower threshold current with photon recycling. Conversely, the classical threshold current (dotted) features the opposite trend.

    We shall now reconsider \figref{fig:sec:INTRO:FIG1}, illustrating (bottom row) the threshold current versus mode volume. In all cases (EDC, PhC and macroscopic laser), the mode volume reflects a total optical confinement factor ranging from $0.3\%$ to $30\%$. In macroscopic lasers, one finds $\gamma_r\ll\gamma_\mathrm{bg}$, leading to $\beta\ll1$. As a result, the effective $\beta$-factor in \Eqref{eq:beta-tilde} reduces to $\beta/(2\xi+1)$ and the threshold with photon recycling approaches the classical threshold. At sufficiently small mode volumes, corresponding to $\xi\gg1/2$ and $\beta_{\mathrm{eff}}\approx\beta/(2\xi)$, the threshold current ultimately approaches the transparency current, \Eqref{eq:Itr-definition}. Since the carrier lifetime, $(\gamma_r+\gamma_\mathrm{bg})^{-1}$, is dominated by the background recombination, the transparency current is independent of the mode volume and the threshold current saturates. 
    
    Conversely, the two thresholds differ qualitatively as the $\beta$-factor approaches $1$, as is the case for EDC lasers. In this case, one finds $\gamma_r\gg\gamma_\mathrm{bg}$, implying that $\gamma_r$ determines the carrier lifetime. As a result, at sufficiently small mode volumes, the classical threshold current strongly increases with decreasing mode volume, reflecting the growth of the transparency current. We note that the transparency current is the lower bound to the classical threshold current, consistently with the commonly accepted requirement \cite{Coldren-book2ndEd-2012} of inverting the active medium before achieving lasing. On the other hand, the effective $\beta$-factor reduces to $\beta\approx1$ for $\beta$ approaching $1$. Correspondingly, one finds $\eta\approx1/2$ if $\xi$ is sufficiently larger than one-half. Ultimately, the threshold current with photon recycling tends to $I_\mathrm{pr}\approx q\gamma_c$, thus being only limited by the cavity decay rate. Therefore, $I_\mathrm{pr}$ saturates in EDC lasers at sufficiently small mode volumes. As already mentioned in \secref{sec:Intro}, this behavior is interpreted as an effective saturation of the carrier lifetime induced by photon recycling.    
    
    We note that, irrespective of the $\beta$-factor and similarly to the classical threshold current, $I_\mathrm{pr}$ diverges if $\xi$ tends to one-half, signifying the transition toward the LED operation regime. This is due to the inability of the maximum gain, $n_0\gamma_r$, to overcome the cavity loss. Therefore, the optimum mode volume minimizing the classical threshold current in EDC lasers is a trade-off between the larger transparency current and the higher pump injection efficiency with decreasing mode volume.

    It should be mentioned that expressions similar to $I_\mathrm{pr}$ have been reported in previous works \cite{Ma_LPR_2013,Takemura_PRA_2019,Khurgin_LPR_2021}, but solely based on the empirical consideration of the photon number second-order derivative \cite{Takemura_PRA_2019} (cf. \Eqref{eq:LIchar-2ndder-Ith2}) or as ad-hoc definitions \cite{Ma_LPR_2013,Khurgin_LPR_2021}. Here, instead, we have elucidated the physical origin of this threshold definition. Moreover, we emphasize that pump blocking, neglected in those works, is essential to include to capture phenomena such as the appearance of an inverse S-shape in the LI characteristic (see \secref{sec:beta-factor}) or the transition with decreasing $\xi$ from the laser regime to the LED regime, where no threshold exists. 

    \section{Role of the spontaneous emission factor}\label{sec:beta-factor}

   The spontaneous emission factor or $\beta$-factor quantifies the fraction of spontaneous emission funneled into the lasing mode. 
   % Generally speaking, a higher $\beta$-factor translates, below threshold, into a larger number of photons at a given current, thereby improving the energy efficiency \cite{Noda_Science_2006}. 
   %In this section, we take a closer look at the impact of the $\beta$-factor on the LI characteristic and threshold with photon recycling, clarifying some common misconceptions.    
   As long as the cavity is much larger than the wavelength, the $\beta$-factor is mainly determined by the number of longitudinal modes in the spontaneous emission bandwidth \cite{Coldren-book2ndEd-2012,Khurgin_LPR_2021}. In this case, the $\beta$-factor is much smaller than one \cite{Coldren-book2ndEd-2012} and increases with decreasing cavity length due to the larger free spectral range. As the dimensions of the cavity become comparable to the wavelength, the density of optical modes per unit frequency and unit volume departs \cite{Kleppner_PRL_1981,Yablonovitch_PRL_1987,Yamamoto_PRA_1991,Denning_PRB_2018} from its simple bulk form \cite{Coldren-book2ndEd-2012}, just as for the electronic density of states in the case of QWs and QDs. This leads to a nontrivial dependence of the spontaneous and background recombination rates on the cavity geometry via the optical density of states of the lasing mode and background emission \cite{Gregersen_APL_2012,Lorke_PRB_2013}. Furthermore, in the case of extended active media, the spontaneous and background recombination rates depend on the pumping level via the Fermi-Dirac distributions of electrons and holes, as well as on the electronic density of states, which reflects the inhomogeneous broadening (see Sec. I in the Supplementary Material). As a result, the $\beta$-factor in lasers with extended active media is also a function of the number of carriers \cite{Gregersen_APL_2012,Romeira_JQE_2018}. 
   
   In general, it should be stressed that the $\beta$-factor depends on the optical density of states, the electronic density of states, and their frequency overlap \cite{Baba_JQE_1991,Noda_NatPhot_2007,Takiguchi_APL_2013,Takiguchi_OptExpress_2016}, as well as on the pumping level \cite{Gregersen_APL_2012}. Therefore, accurate estimates of the $\beta$-factor require complex laser models \cite{Gies_PRA_2007,Gregersen_APL_2012}. Nonetheless, \Eqref{eq:beta-factor} still retains a fundamental feature, that is the increase of the $\beta$-factor with decreasing mode volume for a given value of the background recombination rate. If properly engineered, nanolasers may have a near-unity \cite{Ota_OptExp_2017,Jagsch_NatComm_2018} $\beta$-factor, unlike the case of conventional macroscopic lasers. 

    \begin{figure}[ht]
	\centering\includegraphics[width=1\linewidth]{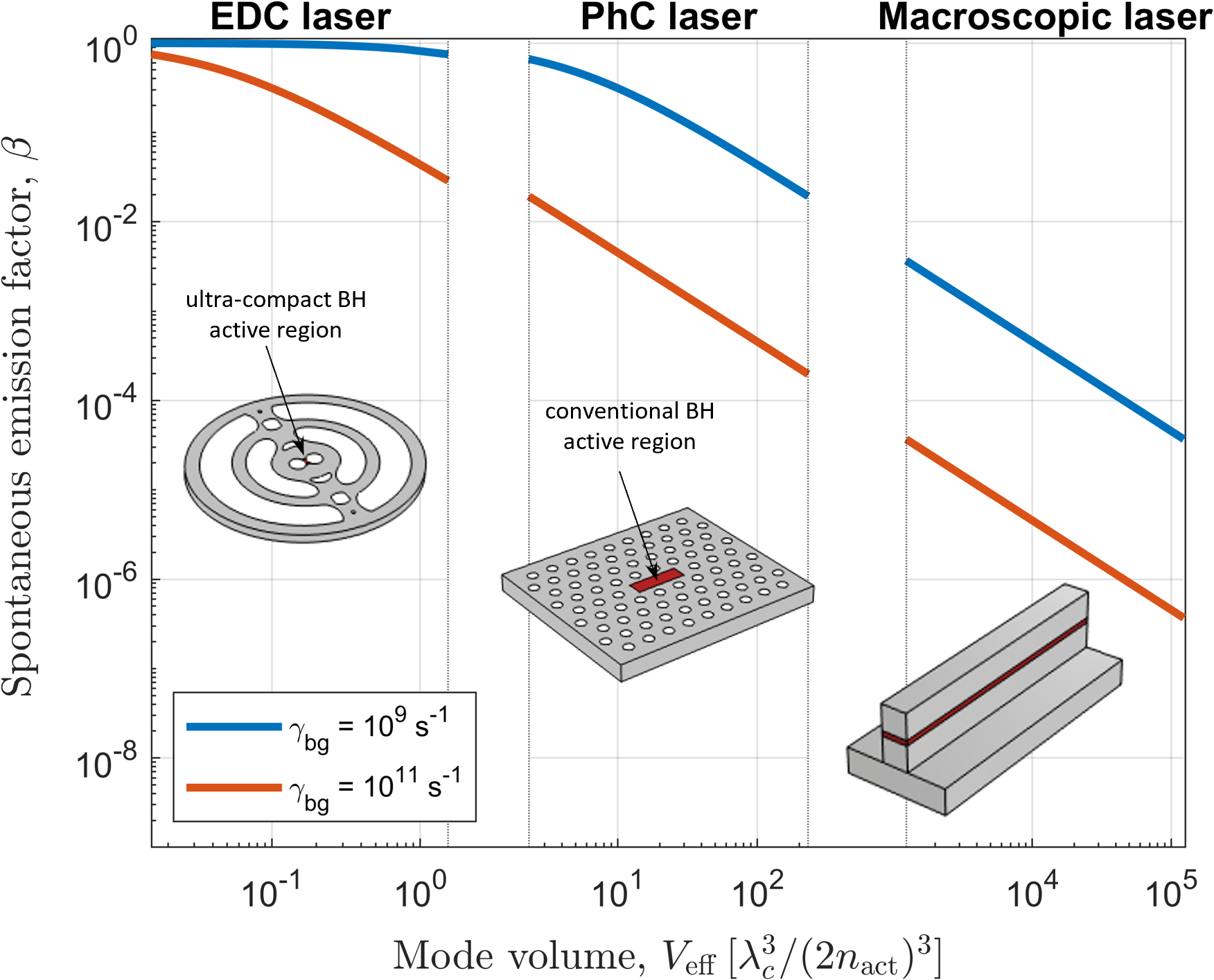}
	\caption{\label{fig:SecIthIntro:FIG1} Spontaneous emission factor versus mode volume for a nanolaser with extreme dielectric confinement (EDC laser, left), a photonic crystal laser (PhC laser, center) and a conventional macroscopic laser (right). The spontaneous emission factor is shown for a value of the background decay rate equal to $10^9\,\mathrm{s^{-1}}$ (blue) and $10^{11}\,\mathrm{s^{-1}}$ (red). The other parameters are the same as in \figref{fig:sec:INTRO:FIG1}. The number of emitters is fixed, while the mode volume varies with the optical confinement factor. 
    % The mode volume may reflect the variation of the optical confinement factor in the presence of an extended active region, as indicated in the text. In EDC lasers, the mode volume may be also interpreted as that for a single emitter. 
    %The EDC laser illustrated in the figure is inspired by recently demonstrated cavity designs \cite{Wang_APL_2018,Albrechtsen_NatComm_2022,Kountouris-OptExress-2022}.
    }
    \end{figure}
    As an example, \figref{fig:SecIthIntro:FIG1} illustrates the $\beta$-factor versus mode volume for an EDC laser (left), a PhC laser (center) and a macroscopic laser (right). The $\beta$-factor is shown for $\gamma_\mathrm{bg} = 10^9\,\mathrm{s^{-1}}$ (blue) and $\gamma_\mathrm{bg} = 10^{11}\,\mathrm{s^{-1}}$ (red). The other parameters are the same as in \figref{fig:sec:INTRO:FIG1}. As already mentioned, the mode volume reflects, in each case (EDC, PhC and macroscopic laser), a total optical confinement factor ranging from $0.3\%$ to $30\%$. PhC lasers exploiting the buried heterostructure (BH) technology already offer a small active region volume and high optical confinement factor \cite{Kuramochi_OptExpress_2018,Dimopoulos_LPR_2022}. Future cavity designs combining a BH active region and extreme dielectric confinement may further scale the active region volume while ensuring a high optical confinement factor. Such scaling may be accomplished without severely degrading the quality factor, at variance with plasmonic nanolasers \cite{Ma_LPR_2013,Gwo_RepProgrPhys_2016}. We shall now examine the role of the $\beta$-factor, $\beta$, and the number of photons at transparency, $\xi$, in shaping the LI characteristic. 

     \begin{figure}[ht]
		\centering\includegraphics[width=1\linewidth]{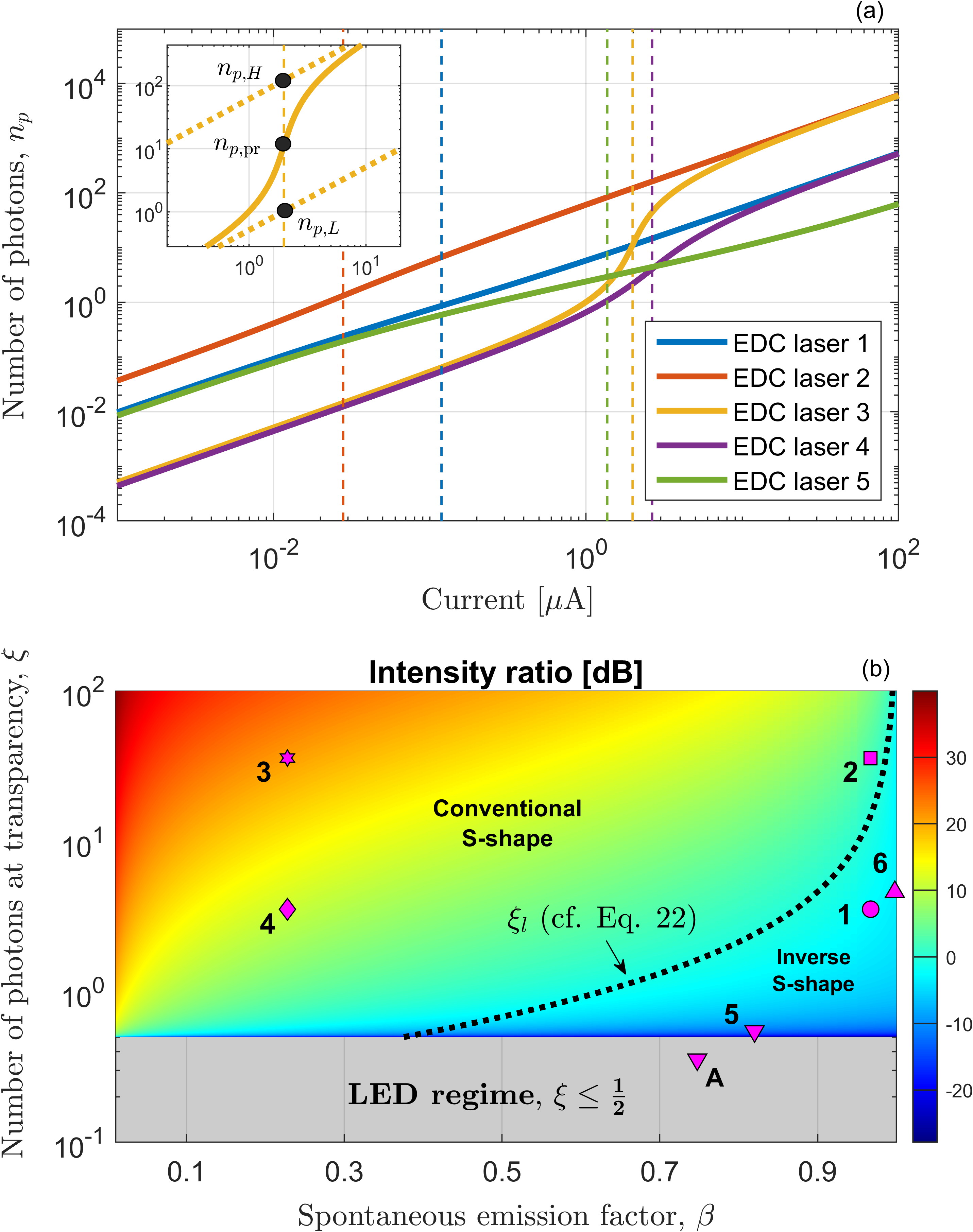}
		\caption{\label{fig:SecIth:FIG2} (a) Number of photons versus current for EDC lasers with parameters specified in \tabref{tab:nano-devices} and leading to different values $\beta$ and $\xi$. The vertical, dashed lines denote the threshold with photon recycling. The inset displays the LI characteristic (solid) of EDC laser 3 around the intensity jump and the asymptotes of the upper and lower branches (dotted). (b) Intensity ratio (cf. \Eqref{eq:intensity-jump}) in dB of the LI characteristic versus $\beta$ and $\xi$. The markers correspond to the nanoscale devices in \tabref{tab:nano-devices}. For the LI characteristic to be linear, $\beta$ and $\xi$ must follow the dotted line (cf. \Eqref{eq:linear-LI-xi-cond}).}
     \end{figure} 
    \figref{fig:SecIth:FIG2}a shows the number of photons versus current for EDC laser 1 and other EDC lasers, with parameters specified in \tabref{tab:nano-devices}. These parameters lead to different values of $\beta$ and $\xi$. The vertical, dashed lines indicate the threshold with photon recycling. 
    %Specifically, we have considered $\gamma_\mathrm{bg} = 10^{9}\,\mathrm{s^{-1}}$ (nanolasers 1 and 2) and $\gamma_\mathrm{bg} = 10^{11}\,\mathrm{s^{-1}}$ (nanolasers 3 and 4). The photon decay rate is $\gamma_c = 5\times10^{11}\,\mathrm{s^{-1}}$ (nanolasers 1 and 4) and $\gamma_c = 5\times10^{10}\,\mathrm{s^{-1}}$ (nanolasers 2 and 3). The other parameters are unchanged. 
    The $\beta$-factor of EDC lasers 1 and 2 is close to unity, and the LI characteristic is almost linear. In EDC laser 3 and 4, instead, the $\beta$-factor drops to around $0.23$ and a marked intensity jump shows up. This jump, however, is much more pronounced for EDC laser 3, which features a large value of $\xi$. Indeed, the intensity jump is determined by the pump injection efficiency, $\eta$, and \textit{effective} $\beta$-factor, $\beta_{\mathrm{eff}}$, as shown below.  
 
 The inset of \figref{fig:SecIth:FIG2}a zooms in on the LI characteristic of EDC laser 3 around threshold, displaying the asymptotes (dotted lines) and the threshold current with photon recycling (vertical line). By neglecting pump blocking and assuming $\gamma_r(2n_e-n_0)n_p\approx-\gamma_rn_0n_p$ in the carrier and photon rate equations, we find the lower asymptote, $n_p = \beta_{\mathrm{eff}}(I/q)/\gamma_c$. The upper asymptote, $n_p = \eta(I/q)/\gamma_c$, is found by neglecting the carrier recombination term compared to the net stimulated emission term in the carrier rate equation and assuming the carriers to be clamped. The upper and lower asymptotes at the threshold with photon recycling have a value of $n_{p,H} = (2\beta_{\mathrm{eff}})^{-1}$ and $n_{p,L} = 1/(2\eta)$, respectively. 
 
 To quantify the intensity jump, we introduce the intensity ratio
 \begin{equation}
  \label{eq:intensity-jump}
  \frac{n_{p,H}}{n_{p,L}} = \frac{\eta}{\beta_{\mathrm{eff}}} \quad \text{(intensity ratio)}
  \end{equation}
 Despite having the same $\beta$-factor, EDC laser 3 and 4 feature different Q-factors. Moreover, $\beta$ is close to, but not exactly equal to unity. Therefore, the larger value of $\xi$ in EDC laser 3 results in a smaller $\beta_{\mathrm{eff}}$ (cf. \Eqref{eq:beta-tilde}), as well as in a larger $\eta$ (cf. \Eqref{eq:eta}), ultimately implying a larger intensity ratio. Hence, we emphasize that the intensity jump of the LI characteristic is \textit{not} a direct measure of the $\beta$-factor, as opposed to what is commonly held (with some notable exceptions \cite{Gies_PRA_2007,Takemura_PRA_2019}).  

   Interestingly, the photon number at the threshold with photon recycling, $n_{p,\mathrm{pr}}$ (cf. \Eqref{eq:npth3}), is the geometric mean of $n_{p,H}$ and $n_{p,L}$. Furthermore, by comparing the expressions of $I_\mathrm{pr}$ and $I_\mathrm{tr}$, one readily finds that the threshold with photon recycling is attained below transparency if the following condition is fulfilled:
	\begin{equation}
		\label{eq:intensity-jump-without-inversion}
		\xi\beta>1 \quad \text{(intensity jump without inversion)}
	\end{equation} 
    This is the condition required for lasing without inversion \cite{Yamamoto_JJAP_1991}, or, more precisely, intensity jump without inversion \cite{Takemura_PRA_2019}. For example, EDC laser 1 fulfils \Eqref{eq:intensity-jump-without-inversion}, as evidenced by \figref{fig:SecIth:FIG1}, where $I_\mathrm{pr}$ is seen to be smaller than $I_\mathrm{tr}$.  
	
    \figref{fig:SecIth:FIG2}b shows, in color scale, the intensity ratio versus $\beta$ and $\xi$. The markers correspond to the nanolasers and the nanoLED in \tabref{tab:nano-devices}. For values of $\xi$ equal to or smaller than one-half, the intensity ratio is not defined, and the device operates as an LED. For larger values of $\xi$, it is seen that the intensity ratio may strongly depend on $\xi$, unless the $\beta$-factor is very close to one. The relation that $\beta$ and $\xi$ must fulfill for the LI characteristic to be linear is also shown (dotted line). This relation is found by requiring the intensity ratio to be equal to one, leading to 
    \begin{equation}
		\label{eq:linear-LI-xi-cond}
		\xi_{l} = \frac{\beta + \sqrt{1-\beta+\beta^2}}{4(1-\beta)} \quad \text{(linear LI characteristic)}
     \end{equation} 
    In contrast to what is usually stated \cite{Lohof_PhysRevApplied_2018,Kreinberg_LPR_2020,Deng_AdvOptMat_2021}, a linear LI characteristic does \textit{not} require a $\beta=1$. \textcolor{black}{We emphasize that nonradiative recombination, which is included in the $\beta$-factor definition (cf. \Eqref{eq:beta-factor}), is \textit{not} responsible for this feature. This is due, instead, to the pump blocking term. If pump blocking is neglected ($\eta=1$), the LI characteristic is only linear for $\beta=1$.} 
    
    We also note that the LI characteristic shows the conventional S-shape on a double logarithmic scale if $\xi$ is larger than $\xi_l$. For smaller values of $\xi$, instead, an \textit{inverse} S-shape appears, with an intensity ratio smaller than $0\,\mathrm{dB}$. In \figref{fig:SecIth:FIG2}a, EDC laser 1 and, more markedly, EDC laser 5 feature LI characteristics with an inverse S-shape. Furthermore, as shown in \figref{fig:SecIth:FIG2}b, the LI characteristic may evolve from the usual S-shape, to a straight line and finally to an inverse S-shape with increasing $\beta$-factor. 
    
    This behavior is consistent with the experimental results in \cite{Jagsch_NatComm_2018}, where the LI characteristic of a high-$\beta$ nanolaser was measured at decreasing values of temperature. In \cite{Jagsch_NatComm_2018}, the phenomenon was explained as a complex interplay between zero-dimensional and two-dimensional gain contributions, depending on the temperature and the excitation power. Here, we point out that the effect naturally emerges from our rate equation model as $\gamma_\mathrm{bg}$ decreases for a given value of $\xi$. Experimentally, lower temperatures may suppress nonradiative recombination, leading to lower values of $\gamma_\mathrm{bg}$ and larger values of $\beta$, with $\xi$ being fixed. In practice, the emitter dephasing rate may actually decrease with decreasing temperature, leading to larger values of $\xi$. However, the transition from a conventional to an inverse S-shape would still occur as long as $\beta$ increases sufficiently fast compared to $\xi$. It should be stressed that including pump blocking in the model is crucial to reproduce this behavior. If pump blocking is neglected, the inverse S-shape appears under no circumstances. % The intensity jump changes quantitatively, but not qualitatively. 

            \begin{figure}[ht]
		\centering\includegraphics[width=1\linewidth]{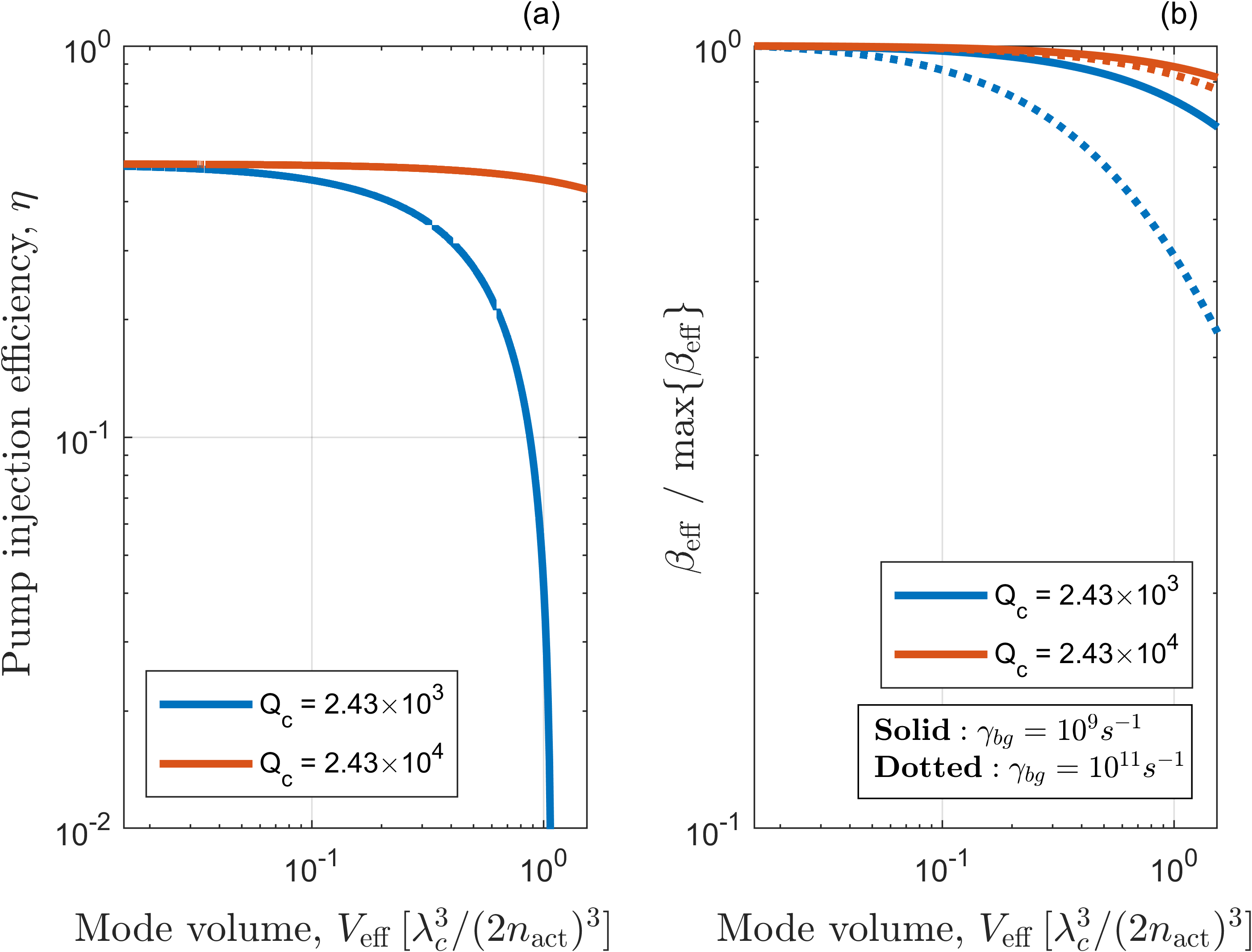}
		\caption{\label{fig:SecIthIntro:FIG2a} (a) Pump injection efficiency and (b) effective $\beta$-factor versus mode volume for the EDC laser in \figref{fig:sec:INTRO:FIG1} (blue) and for an EDC laser with larger Q-factor (red), as indicated in the legend. In (b), the background decay rate is $\gamma_\mathrm{bg}=10^9\mathrm{s^{-1}}$ (solid) and $\gamma_\mathrm{bg}=10^{11}\mathrm{s^{-1}}$ (dotted). In each case, the effective $\beta$-factor is normalized to the maximum value.}
    \end{figure}
    To conclude this section, we point out that the reduction with decreasing mode volume of the threshold current with photon recycling (cf. \figref{fig:sec:INTRO:FIG1}) is mostly due to the larger pump injection efficiency when the active region volume is fixed. As an example, \figref{fig:SecIthIntro:FIG2a} shows (a) the pump injection efficiency (blue) and (b) effective $\beta$-factor (blue, solid) versus mode volume for the EDC laser in \figref{fig:sec:INTRO:FIG1}. The effective $\beta$-factor is normalized to the maximum value. As the mode volume is reduced, the relative variation of $\beta_\mathrm{eff}$ is negligible compared to the increase in $\gamma_r$. Such an increase strongly reduces the number of carriers at the lasing threshold, thereby limiting the impact of pump blocking, enhancing the pump injection efficiency, and ultimately decreasing the threshold current. However, the contribution of the $\beta$-factor to such a strong decrease is marginal, a conclusion which agrees with \cite{Khurgin_LPR_2021}. 
    
    Indeed, a larger $\gamma_r$ not only leads to a larger $\beta$, but also to a larger $\xi$. The two effects somehow balance out, resulting in a slight variation of $\beta_\mathrm{eff}$ (cf. \Eqref{eq:beta-tilde}). Larger values of the background recombination rate (blue, dotted in \figref{fig:SecIthIntro:FIG2a}b) lead to a larger variation of $\beta$ and thereby of $\beta_{\mathrm{eff}}$, but the impact remains modest. Furthermore, larger values of the Q-factor naturally limit the variation of both $\eta$ (red, in \figref{fig:SecIthIntro:FIG2a}a) and $\beta_{\mathrm{eff}}$ (red, in \figref{fig:SecIthIntro:FIG2a}b).              
    %We note that the growth of the various threshold currents with increasing mode volume observed in \figref{fig:SecIthIntro:FIG2} is due to the higher degree of pump blocking. In practice, such an increase would not be observed in macroscopic lasers and long PhC lasers. Indeed, in the presence of multiple longitudinal modes (not taken into account in our rate equation model), mode hopping would occur, leading to lasing on a mode with a lower threshold current. On the contrary, in lasers with a single mode, such as short PhC lasers and EDC lasers, the threshold current is expected to increase with increasing mode volume.        
    %Therefore, the observation of an optimum mode volume minimizing the classical threshold current in EDC lasers is understood as a trade-off between pump blocking and inversion of the active medium.    
    %This explains why the threshold current with photon recycling in EDC lasers (cf. \figref{fig:SecIthIntro:FIG2}) saturates at sufficiently small values of mode volume, as opposed to the classical threshold current. 

\section{Other threshold definitions}\label{sec:other-threshold-definitions}

    Besides the classical threshold and the threshold with photon recycling, other threshold definitions have been proposed \cite{Bjork_JQE_1991,Bjork_PRA_1994,Takemura_PRA_2019,Carroll_PhysRevLett_2021,Lippi_ChaosSolitonsFractals_2022, Yacomotti-LPR-2022}, that we briefly review and compare in this section and the following. % In particular, we discuss the variation of the threshold current with the mode volume and further elucidate the role of the $\beta$-factor. % Furthermore, we demonstrate that some threshold definitions \cite{Bjork_PRA_1994,Takemura_PRA_2019,Lippi_ChaosSolitonsFractals_2022}, contrary to the threshold with photon recycling, cannot reliably predict the possibility of lasing. 
    
    \begin{figure}[ht]
		\centering\includegraphics[width=1\linewidth]{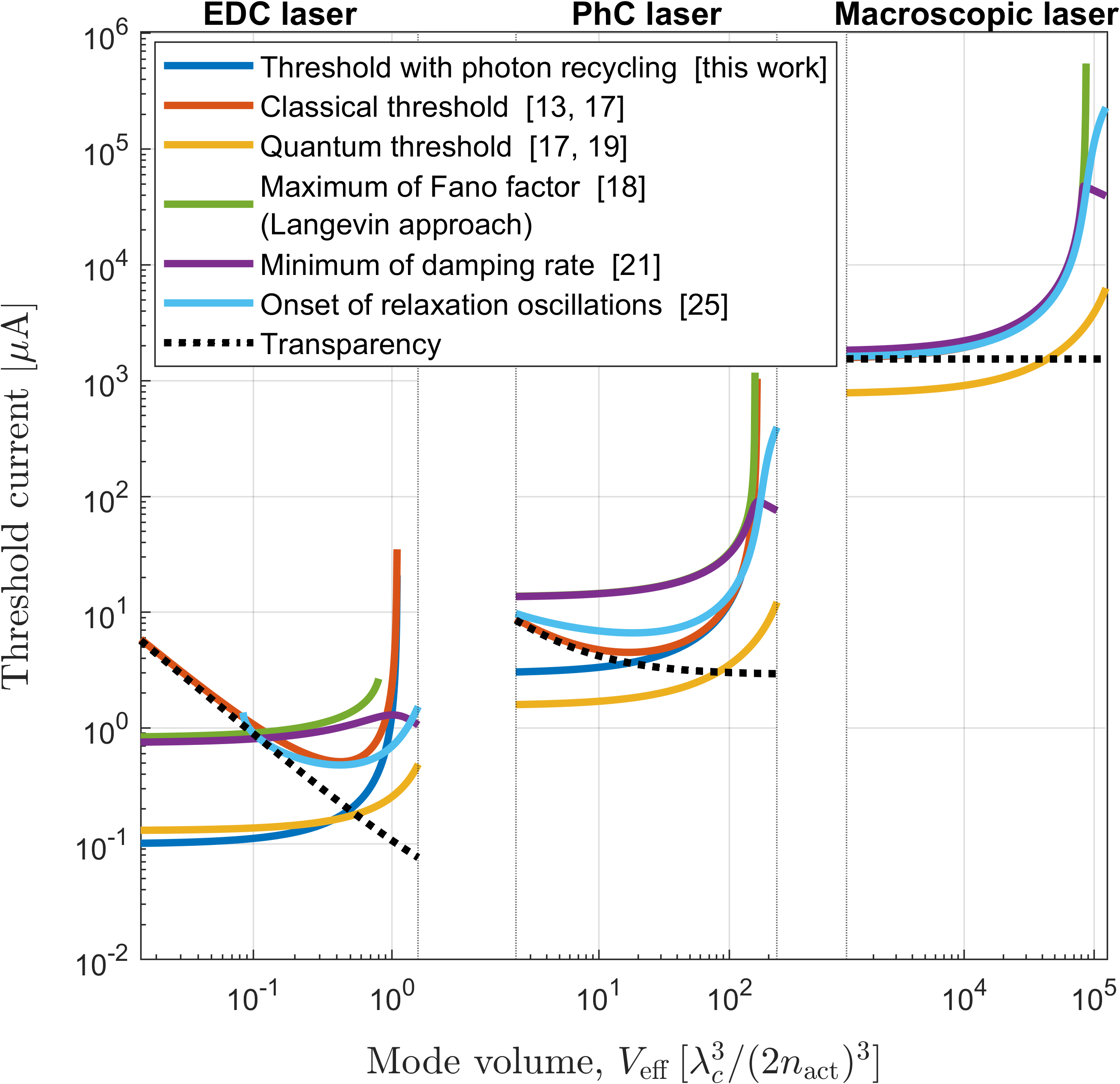}
		\caption{\label{fig:SecIthIntro:FIG2} Threshold current versus mode volume for a nanolaser with extreme dielectric confinement (EDC laser, left), a photonic crystal laser (PhC laser, center) and a conventional macroscopic laser (right). Each color corresponds to a different threshold definition, as indicated in the legend. The transparency current (dotted) is also shown. The parameters are the same as in \figref{fig:sec:INTRO:FIG1}. The number of emitters is fixed, while the mode volume varies with the optical confinement factor.}
    \end{figure}
    \figref{fig:SecIthIntro:FIG2} shows the threshold current versus mode volume for an EDC laser (left), a PhC laser (center) and a conventional macroscopic laser (right), with the same parameters as in \figref{fig:sec:INTRO:FIG1}. Each color denotes a different threshold definition, as indicated in the legend. For the sake of comparison, the transparency current (dotted) is also included. For macroscopic lasers, all threshold definitions (except the quantum threshold, see below) approximately coincide \cite{Rice-PRA-1994,Lippi_ChaosSolitonsFractals_2022}, provided that $\xi$ is larger than one-half. 
    % Perfect agreement would be observed in the so-called thermodynamic limit \cite{Rice-PRA-1994} of $\beta$ approaching zero.   
    %As long as the maximum available gain is larger than the cavity loss ($\xi>1/2$), all threshold definitions approximately coincide for macroscopic lasers, apart from the quantum threshold \cite{Rice-PRA-1994,Lippi_ChaosSolitonsFractals_2022}. 
    Conversely, quantitative and qualitative differences gradually emerge among the various threshold definitions as the $\beta$-factor approaches unity, irrespective of the value of $\xi$. This is already observed for PhC lasers and, to a larger extent, for EDC lasers. 

    For PhC and EDC lasers, as already discussed in \secref{sec:lasing-threshold-intro}, the non-monotonic trend of the classical threshold current (red) stems from the counteracting variations of the transparency current and the pump injection efficiency with decreasing mode volume. By contrast, the threshold current with photon recycling (blue) decreases and eventually saturates, due to an effective saturation of the carrier lifetime. A similar saturation is observed for other threshold definitions. We note that the classical threshold current in \Eqref{eq:Ith1} differs from another expression, $\tilde{I}_{cl} = q\gamma_c/\beta$, reported elsewhere \cite{Rice-PRA-1994,Lippi_ChaosSolitonsFractals_2022}. In those works, stimulated absorption is not considered, assuming the lower lasing level to be quickly depopulated. As a result, the net stimulated emission term in the carrier and photon rate equations reduces to $\gamma_rn_en_p$. Under this assumption, by neglecting pump blocking and applying the same procedure which leads to \Eqref{eq:Ith1}, one finds $\tilde{I}_{cl} = q\gamma_c/\beta$. However, a more realistic semiconductor laser model must include stimulated absorption \cite{Bjork_JQE_1991,Coldren-book2ndEd-2012,Takemura_PRA_2019}. Therefore, \Eqref{eq:Ith1} better describes the classical threshold current of conventional semiconductor lasers.   
    
    The so-called quantum threshold \cite{Bjork_JQE_1991,Bjork_PRA_1994} (yellow line in \figref{fig:SecIthIntro:FIG2}) is attained when the stimulated emission, $\gamma_rn_en_p$, equals the spontaneous emission into the lasing mode, $\gamma_rn_e$. Hence, the number of photons at the quantum threshold equals one. By utilizing this condition in \Eqref{eq:rate-eq-1} and \Eqref{eq:rate-eq-2}, one finds the quantum threshold current
    \begin{equation}
	\label{eq:Ith2}
	I_\mathrm{qu} = \frac{q}{\eta_{q}}\frac{1}{3}\frac{\gamma_c}{\beta}\left[1+2\beta+2\xi(1-\beta)\right] \quad \text{(quantum threshold)}
     \end{equation}
    where $\eta_{q} = \frac{2-1/(2\xi)}{3}$ and $\eta_{q} = 1$ with and without inclusion of pump blocking, respectively. For macroscopic lasers, the quantum threshold current and the classical threshold current approximately differ by a factor of two \cite{Bjork_JQE_1991}, unless the number of photons at transparency, $\xi$, tends to one-half. In this case, as discussed in \secref{sec:lasing-threshold-intro}, $I_\mathrm{cl}$ and $I_\mathrm{pr}$ diverge, since the maximum available gain cannot compensate for the cavity loss.

    \begin{figure}[ht]
		\centering\includegraphics[width=0.9\linewidth]{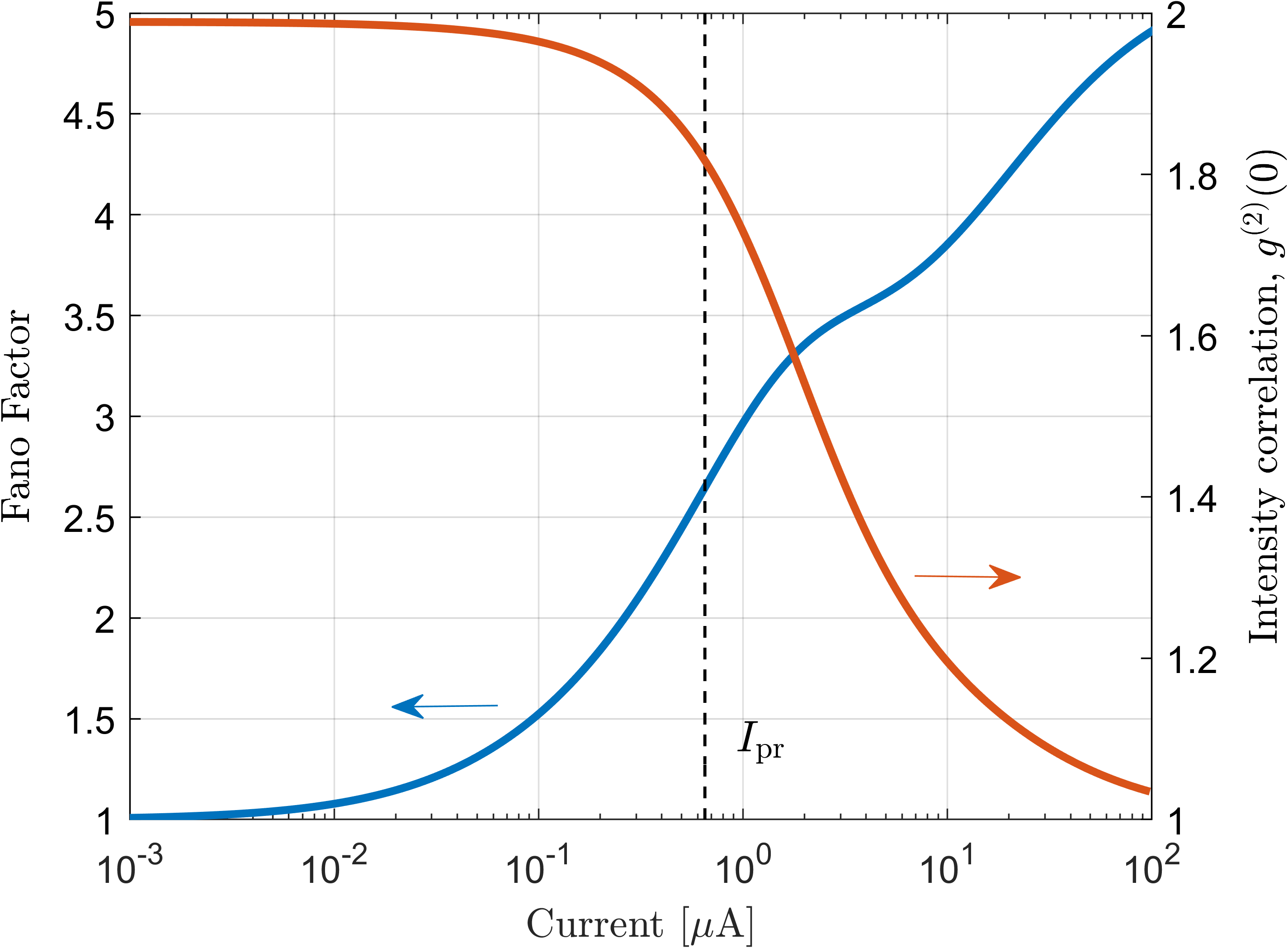}
		\caption{\label{fig:SecIthIntro:FIG2b} Fano factor (left) and intensity correlation (right) of an EDC laser with $V_{\mathrm{eff}} = 0.9\lambda_c^3/(2n_{\mathrm{act}})^3$. The other parameters are those of EDC laser 1 in \tabref{tab:nano-devices}. The vertical, dashed line marks the threshold with photon recycling.}
    \end{figure}
    The Fano threshold (green line in \figref{fig:SecIthIntro:FIG2}) corresponds to the maximum of the Fano factor (see \Eqref{eq:statistical-measures}) as obtained from the Langevin approach. The Fano threshold definition originates from the enhancement of the intensity noise typically observed close to threshold \cite{Jin_PRA_1994}. In such a case, the peak of the Fano factor roughly marks the midpoint in the variation of the intensity correlation from a value of two to a value of one (cf. \figref{fig:SecIth:FIG3}). This property often makes the Fano threshold an appealing indicator of the laser coherence. Unfortunately, the absence of a peak in the Fano factor as obtained from the Langevin approach does not necessarily imply the absence of lasing for EDC lasers. This is seen from \figref{fig:SecIthIntro:FIG2b}, showing the Fano factor (left) and the intensity correlation (right) of an EDC laser with mode volume, $V_{\mathrm{eff}}$, equal to $0.9\lambda_c^3/(2n_{\mathrm{act}})^3$. In such cases, more fundamental approaches to the quantum noise, such as stochastic simulations, may reveal a peak in the Fano factor (see \secref{Sec: Photon statistics: stochastic simulations}) and thereby the existence of the Fano threshold. However, a general conclusion on the existence and location of such a peak cannot be drawn. 

    The damping rate minimum \cite{Takemura_PRA_2019} and the onset of relaxation oscillations \cite{Lippi_ChaosSolitonsFractals_2022} have been proposed as operational lasing threshold definitions. In \figref{fig:SecIthIntro:FIG2}, the corresponding currents are shown in purple and light blue, respectively. The damping rate is easily obtained from a small-signal analysis of \Eqref{eq:rate-eq-1} and \Eqref{eq:rate-eq-2}. The eigenvalues, $\lambda_{\pm}$, of the linearized rate equation system are found by \cite{Coldren-book2ndEd-2012}
    \begin{equation}
    \label{eq:lambda-pm}
    \lambda_{\pm} = -\frac{1}{2}\gamma \pm i\sqrt{\omega_R^2-(\gamma/2)^2}
    \end{equation}
    where $\gamma$ is the damping rate and $\omega_R$ is the relaxation resonance frequency. Relaxation oscillations exist if the imaginary part of the eigenvalues is nonzero \cite{Lippi_ChaosSolitonsFractals_2022}. However, we note that the presence or absence of relaxation oscillations is not generally sufficient to confirm or rule out lasing. For instance, it is well known that lasers of class-A \cite{Takemura_PRA_2021}, characterized by $\gamma_r+\gamma_\mathrm{bg}\gg\gamma_c$, do not show relaxation oscillations. This is the reason why, for mode volumes smaller than a specific value, no threshold is found marking the onset of relaxation oscillations in  EDC lasers. We also note that the minimum of the damping rate approximately corresponds, in a wide range of mode volumes, to the maximum of the Fano factor, as already noted for PhC lasers \cite{Takemura_PRA_2019}. However, the two thresholds are not generally equivalent, especially for EDC lasers. 
    
    Besides the above observations, we emphasize that if a lasing threshold exists, the photon statistics must asymptotically approach a Poisson process as the current increases above threshold. This is a fundamental reliability test that any threshold definition should pass to be general. As we show below, neither the quantum threshold nor the definitions based on the damping rate or relaxation oscillations, necessarily pass such a test.        

   \begin{figure}[ht]
		\centering\includegraphics[width=1\linewidth]{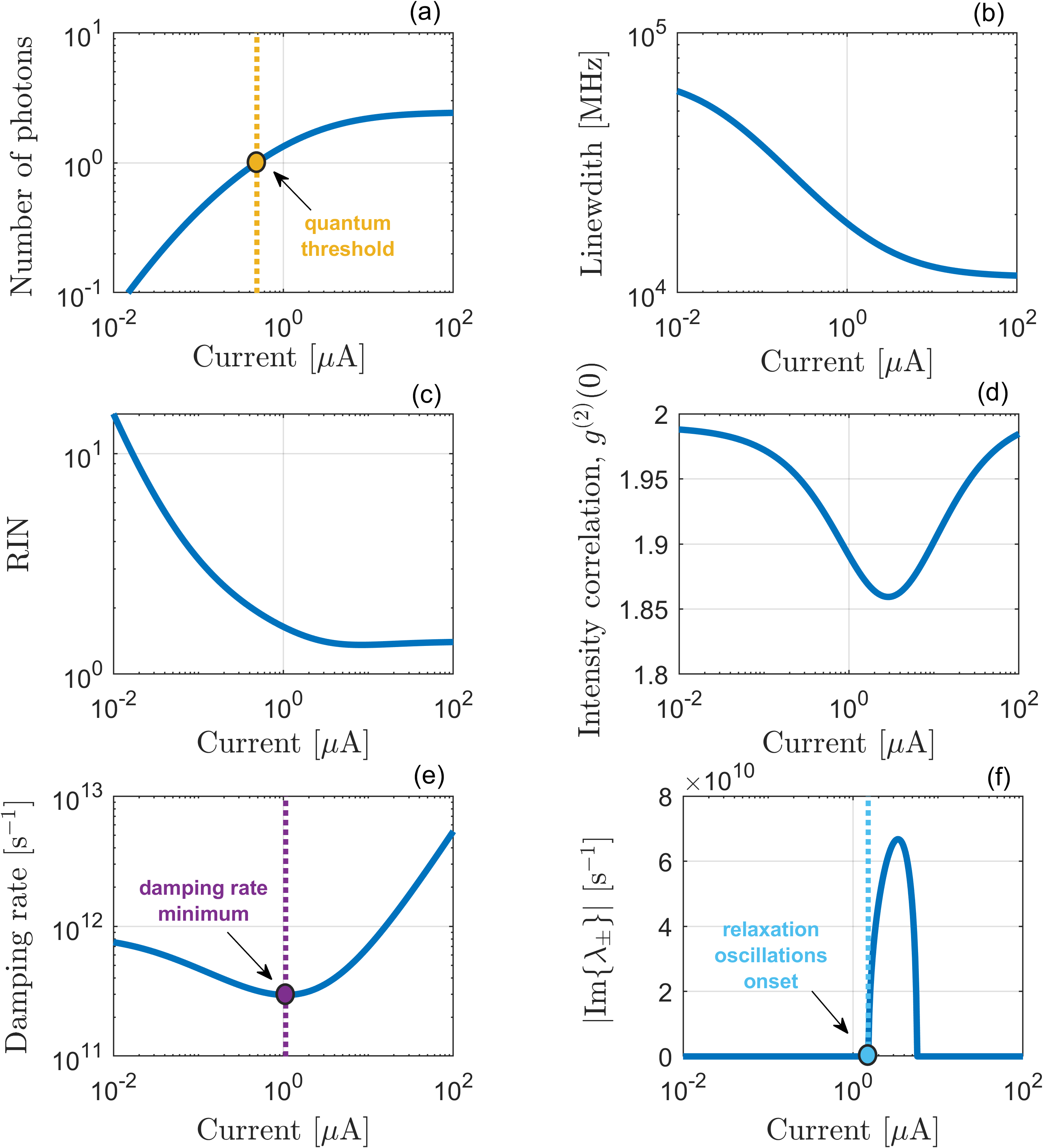}
		\caption{\label{fig:SecIth:FIG4} Current dependence of key characteristics of nanoLED A with parameters summarized in \tabref{tab:nano-devices}. (a) Photon number, (b) modified Schawlow–Townes linewidth, (c) RIN, (d) intensity correlation, (e) damping rate and (f) absolute value of the imaginary part of the rate equations eigenvalues.}
    \end{figure}
    %However, the quantum threshold cannot reliably predict the possibility of lasing \cite{Jagsch_NatComm_2018}. This shortcoming is also common to other definitions, as we demonstrate below.  
    % Indeed, if pump blocking is included, the number of photons may be larger than one and yet the laser may operate in the LED regime, as also previously noted \cite{Jagsch_NatComm_2018}. % If the $\beta$-factor, instead, is much smaller than one, the quantum threshold current is close, albeit not equal \cite{Bjork_JQE_1991}, to the classical threshold current. This non-perfect equivalence is one of the arguments raised against the quantum threshold \cite{Rice-PRA-1994}, since the concept of lasing threshold is well-defined in the thermodynamic limit of $\beta\rightarrow0$. We will suggest an additional argument later on in this article.
    In \figref{fig:SecIth:FIG4}, we consider the input-output and small-signal characteristics of nanoLED A, with parameters summarized in \tabref{tab:nano-devices}. These parameters lead to a value of $\xi$ smaller than one-half, which makes the classical threshold and the threshold with photon recycling unattainable. Conversely, the quantum threshold exists, as evident from \figref{fig:SecIth:FIG4}a, where the number of photons is seen to reach and overcome a value of one. The linewidth somewhat narrows, as displayed by \figref{fig:SecIth:FIG2}b, but the photon number saturates at current values beyond the quantum threshold. Correspondingly, the photon statistics tends to the limit of thermal light. This is seen from \figref{fig:SecIth:FIG4}c and \figref{fig:SecIth:FIG4}d, showing the RIN and intensity correlation, respectively. \textcolor{black}{Interestingly, the variation of the intensity correlation is consistent with experimental results shown in \cite{Kreinberg_LightScAppl_2017}}. This example suggests that neither the quantum threshold nor linewidth narrowing are reliable indicators of lasing, as noted in previous works \cite{Kreinberg_LightScAppl_2017,Jagsch_NatComm_2018}. \figref{fig:SecIth:FIG4}e and \figref{fig:SecIth:FIG4}f, showing, respectively, the damping rate and imaginary part of the eigenvalues, would also suggest the possibility of lasing. Nonetheless, as discussed above, coherent laser light is out of reach, irrespective of the pumping level. % Such an example does not obviously imply that these threshold definitions should be systematically discarded. Nonetheless, the example in \figref{fig:SecIth:FIG4} clearly illustrates that the quantum threshold, the minimum of the damping rate and the onset of relaxation oscillations cannot be taken, in general, as reliable indicators of lasing.  

    It should be mentioned that the onset of lasing in the thermodynamic limit is marked by the appearance of a bifurcation \cite{Lippi_ChaosSolitonsFractals_2022}. Therefore, it has been argued that the emergence of the coherent optical field from a bifurcation is the primary lasing threshold definition, a so-called bifurcation threshold \cite{Lippi_ChaosSolitonsFractals_2022}. However, rate equation models that include spontaneous emission in the lasing mode do not feature a bifurcation \cite{Lippi_ChaosSolitonsFractals_2022} and one cannot identify, therefore, a bifurcation threshold outside the thermodynamic limit. A recent work \cite{Carroll_PhysRevLett_2021}, at the center of a vivid debate \cite{Vyshnevyy_PhysRevLett_2022,Lippi_ChaosSolitonsFractals_2022,Carroll_PhysRevLett_2022}, has proposed a laser model ostensibly featuring a bifurcation all the way from the thermodynamic limit to the case of $\beta=1$. Based on this model, an expression for the pumping rate at the bifurcation threshold has been reported \cite{Carroll_PhysRevLett_2021}. Here, we briefly comment on this threshold expression. 
    
    Consistently with \secref{sec:rate-equation-model}, we assume the emitters to be in resonance with the lasing mode and consider the practically relevant case of the good-cavity limit, $\gamma_2\gg\gamma_c$. Under these assumptions, the pumping rate per emitter at threshold reported in \cite{Carroll_PhysRevLett_2021} leads to a threshold current given by
    \begin{equation}
	\label{eq:bifurcation-Ith}
	I_{b} = \frac{q}{\eta}\frac{\gamma_c}{2}(2\xi+1)\left(\frac{1-\beta}{\beta} + \frac{\gamma_2}{\gamma_c}\right)
    \end{equation}
    We note that the existence of a minimum number of emitters to achieve lasing \cite{Carroll_PhysRevLett_2021} is embedded in the pump injection efficiency, $\eta$, and stems from considering a finite maximum gain, $n_0\gamma_r$. 
    
    For macroscopic lasers ($\beta\ll1$), assuming $\xi$ larger than one-half, $I_{b}$ only coincides with the other threshold definitions if $\beta^{-1}$ is much larger than $\gamma_2/\gamma_c$, which is not necessarily the case. For EDC lasers, considering $\beta\approx1$ and $\xi$ much larger than one-half, one finds 
    \begin{equation}
    \label{eq:bifurcation-Ith-EDC}
    I_b\approx qn_0\gamma_r\frac{\gamma_2}{\gamma_c} = qN_\mathrm{tr}\Gamma \frac{2d^2}{\hbar\epsilon_0n_\mathrm{act}^2}Q_c
    \end{equation}
    In such a case, $I_b$ increases with decreasing mode volume, like the classical threshold current. Importantly, though, $I_b$ increases with increasing Q-factor, unlike, e.g., the threshold with photon recycling or the classical threshold. This feature of the bifurcation threshold appears counter-intuitive and calls for further investigations. 

\section{Photon distributions and stochastic threshold}\label{Sec: Photon statistics: stochastic simulations}

    % The conclusions of this section do not change if one models the quantum noise with a more fundamental stochastic simulation scheme \cite{Mørk_APL_2018,Andre_OptExpress_2020}. Stochastic simulations, though, offer new insights into another definition \cite{Yacomotti-LPR-2022} of lasing threshold, based on the photon statistics. This discussion is the subject of the next section.          

    In the previous sections, we used the Langevin approach to model the quantum noise. However, this approach may break down for very few emitters, in which case stochastic approaches \cite{Puccioni_OptExpress_2015,Mørk_APL_2018} are more accurate. In this section, we analyze the transition to lasing in EDC lasers by applying a stochastic approach \cite{Andre_OptExpress_2020,Bundgaard_PRL_2023} (see \secref{sec:rate-equation-model}).  
    
    Importantly, the stochastic approach gives access to the probability distribution of the photon number. Therefore, we can also explore another definition of lasing threshold, the so-called \textit{stochastic threshold}, as we name it in the following. At this threshold, the probabilities of having zero photons and one photon inside the cavity are equal \cite{Yacomotti-LPR-2022}. The fact that such a balance is fulfilled at the threshold of macroscopic lasers \cite{Scully_PhysRev_1967,Rice-PRA-1994} motivates the definition. Still, it does not ensure that the concept also applies at the nanoscale. With the aid of a density-matrix approach, the extension of the stochastic threshold to nanolasers has been argued for \cite{Yacomotti-LPR-2022}. Here, we investigate the stochastic threshold by utilizing the stochastic approach.

    \begin{figure}[ht]
	\centering\includegraphics[width=1\linewidth]{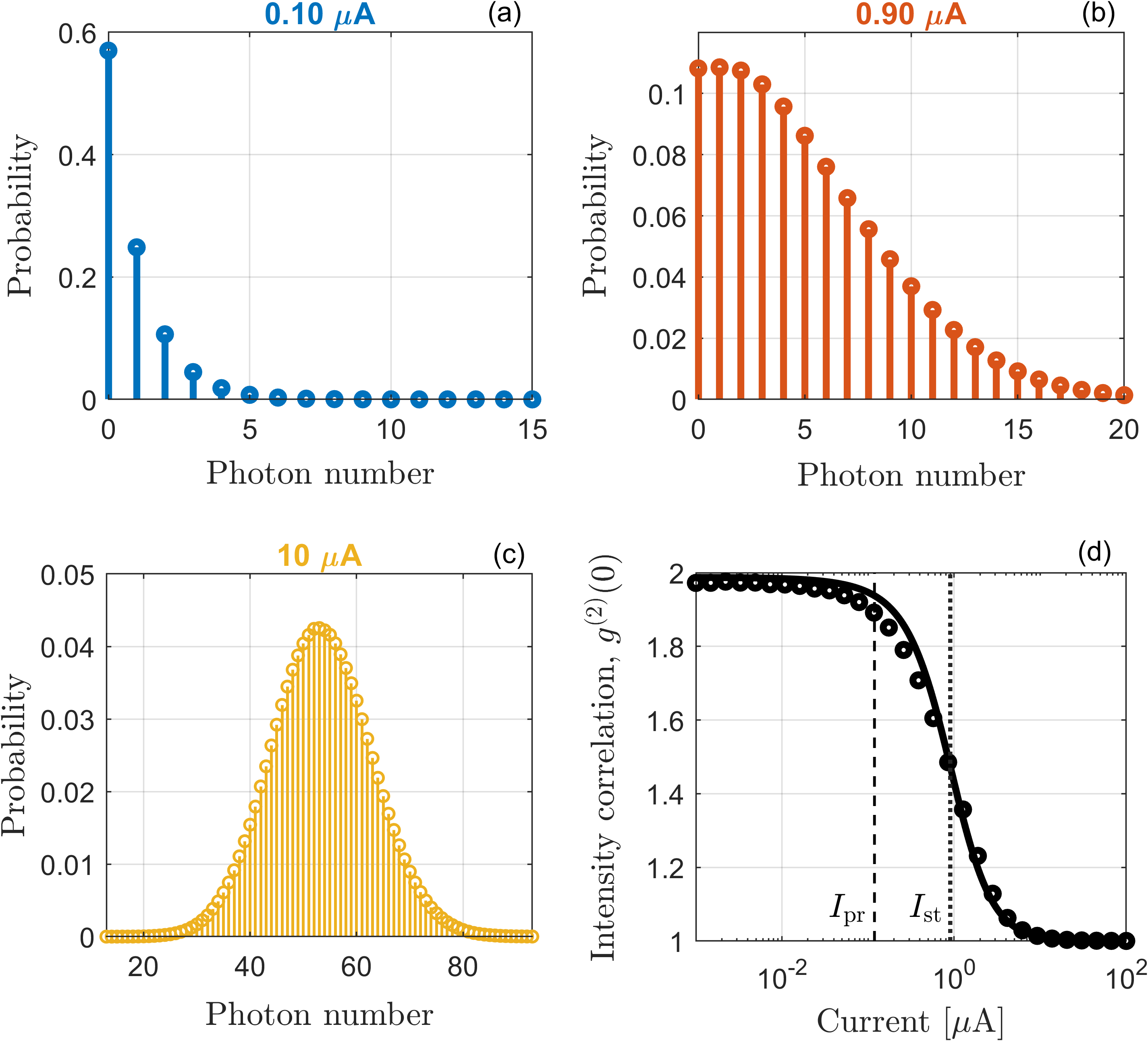}
	\caption{\label{fig:Sec7:FIG1} Probability distributions of the intracavity photons for EDC laser 1 in \tabref{tab:nano-devices} at (a) $0.10\,\mu\mathrm{A}$, (b) $0.90\,\mu\mathrm{A}$ and (c) $10\,\mu\mathrm{A}$. The current value in (b) marks the stochastic threshold. At this threshold, the probabilities of having zero photons and one photon inside the cavity are equal. (d) Intensity correlation of EDC laser 1 versus current as obtained from stochastic simulations (markers) and Langevin approach (solid line). The vertical lines mark the threshold with photon recycling, $I_\mathrm{pr}$ (dashed), and the stochastic threshold, $I_\mathrm{st}$ (dotted).}
    \end{figure} 
    We shall first reconsider the case of EDC laser 1, with parameters summarized in \tabref{tab:nano-devices}. \figref{fig:Sec7:FIG1} shows the probability distributions of the intracavity photon number at a current (a) well below, (b) corresponding to and (c) well above the stochastic threshold. The probability distribution is obtained from the time-resolved photon number as the fraction of the total simulation time spent with a given value of the photon number. Well above the stochastic threshold, a Poisson distribution is approached, which indicates lasing. 
    
    This is consistent with \figref{fig:Sec7:FIG1}d, showing the intensity correlation versus current. Results obtained from the Langevin approach (lines) and the stochastic approach (markers) match pretty well. Higher accuracy is expected from the stochastic approach, which does not rely on a small-signal approximation \cite{Bundgaard_PRL_2023}.
    % Such correlations are neglected in the Langevin approach, which inherits from the rate equations the mean-field approximation, $\langle n_pn_e\rangle=\langle n_p\rangle\langle n_e\rangle$ \cite{Mørk_APL_2018}.
    The stochastic threshold occurs in the middle of the transition regime, whereas the threshold with photon recycling indicates the onset of the transition.
    
    \begin{figure}[ht]
	\centering\includegraphics[width=1\linewidth]{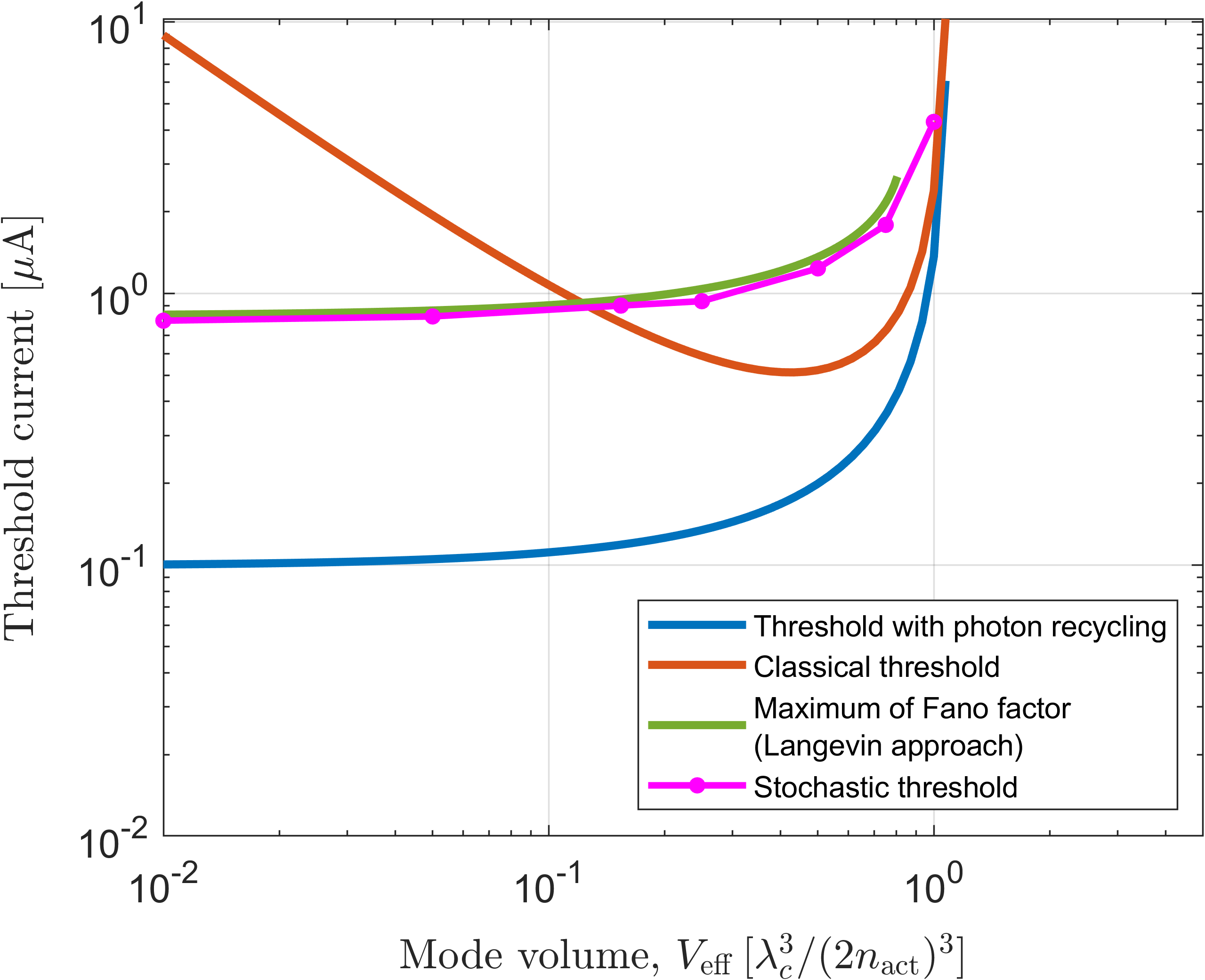}
	\caption{\label{fig:Sec7:FIG2} Threshold current versus mode volume for an EDC laser, with the same parameters as in \figref{fig:sec:INTRO:FIG1}. Each color denotes a different threshold definition, as specified in the legend. The number of emitters is fixed, while the mode volume varies with the optical confinement factor.}
    \end{figure}     
    \figref{fig:Sec7:FIG2} shows the threshold current versus mode volume, including the threshold with photon recycling (blue), the classical threshold (red), the Fano threshold from the Langevin approach (green) and the stochastic threshold (pink). The stochastic threshold current, $I_\mathrm{st}$, is computed numerically from the probability distribution of the intracavity photon number, with each marker corresponding to a given value of mode volume. The threshold current with photon recycling, $I_\mathrm{pr}$, scales similarly to $I_\mathrm{st}$, even though the absolute values generally differ. Importantly, the two thresholds tend to match as the mode volume increases and pump blocking gradually prevents the possibility of lasing. Interestingly, the stochastic threshold almost coincides with the Fano threshold. This is not, though, a general feature, as the next example shows.        

    \begin{figure}[ht]
	\centering\includegraphics[width=1\linewidth]{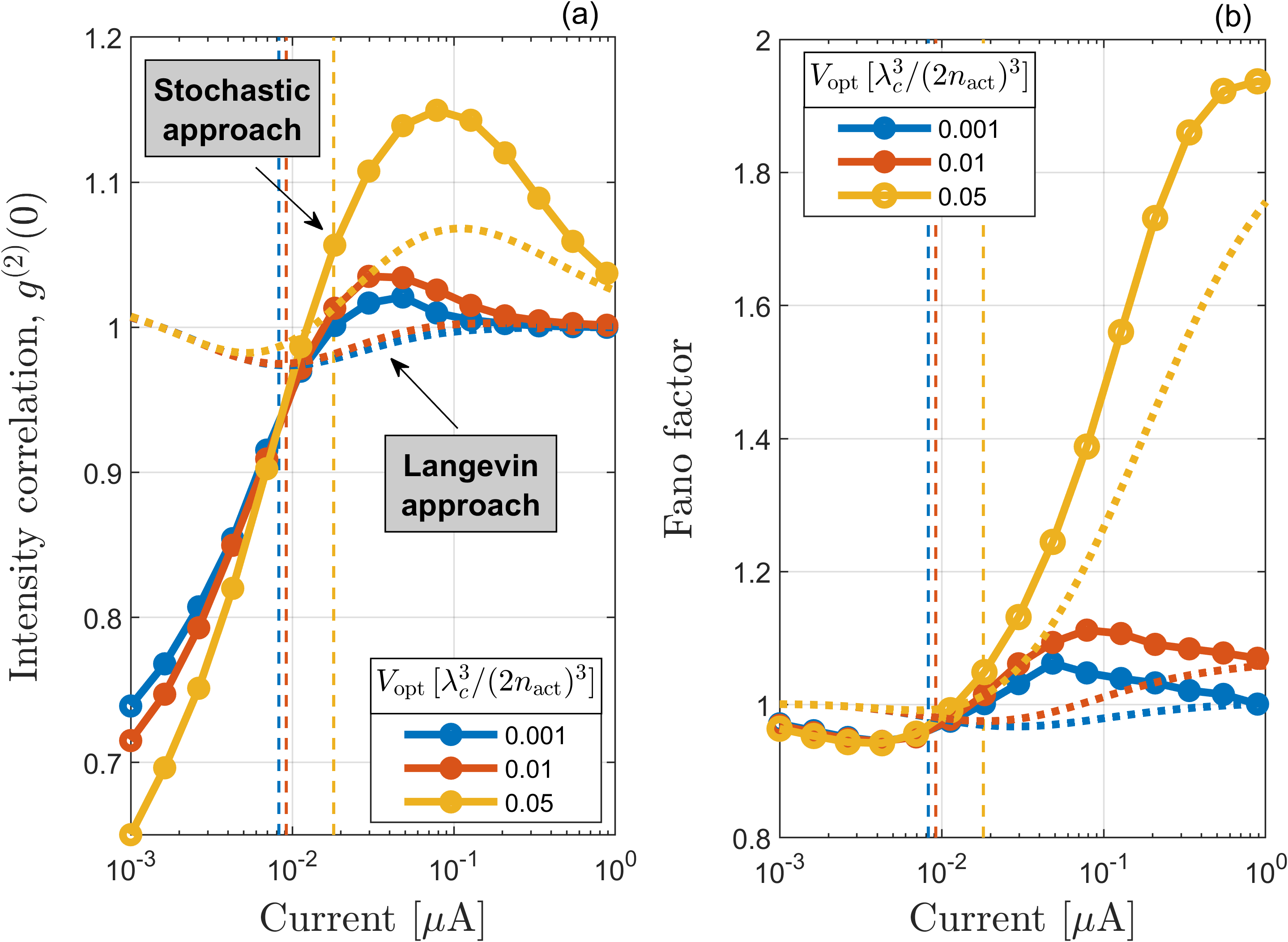}
        \caption{\label{fig:Sec7:FIG3} (a) Intensity correlation and (b) Fano factor versus current for single-emitter nanolasers: EDC laser 6 in \tabref{tab:nano-devices} (red) and two other EDC lasers, with a smaller (blue) and a larger (yellow) mode volume, as indicated in the legend. The other parameters are those of EDC laser 6. Dotted lines and markers correspond to the Langevin approach and the stochastic approach, respectively. The vertical, dashed lines indicate the threshold with photon recycling.}
    \end{figure}
    To highlight the difference between the Langevin and stochastic approaches, we consider the case of a single-emitter nanolaser. As mentioned earlier, it was shown that the stochastic approach agrees quantitatively with full quantum simulations of the intensity noise for a single-emitter laser \cite{Bundgaard_PRL_2023}. \figref{fig:Sec7:FIG3} shows (a) the intensity correlation and (b) Fano factor versus current for EDC laser 6 in \tabref{tab:nano-devices} (red). Two other EDC lasers are also considered, with a smaller (blue) and a larger (yellow) mode volume, as indicated in the legend. The larger Q-factor ($Q_c\approx2.3\times10^4$) compared to EDC laser 1 leads to threshold current values low enough that including pump broadening (see \secref{sec:rate-equation-model}) would have a negligible impact. The vertical, dashed lines denote the threshold with photon recycling. As the current increases from below to above threshold, the stochastic approach (markers) successfully captures the change in the photon statistics from anti-bunched ($g^{(2)}(0)<1$) to bunched ($g^{(2)}(0)>1$) and finally coherent light ($g^{(2)}(0)\approx1$), as expected for a single-emitter nanolaser \cite{Nomura_OptExpress_2009,Bundgaard_PRL_2023}. On the other hand, results from the Langevin approach (dotted lines) are qualitatively wrong. At larger current values (not shown in the figures), both approaches would reveal a further transition to thermal light if pump broadening were included \cite{Bundgaard_PRL_2023}.

    Interestingly, the threshold with photon recycling marks, with good approximation, the transition from the non-classical regime of anti-bunched light to the classical photon bunching regime which precedes lasing, as confirmed by full quantum simulations \cite{Bundgaard_PRL_2023}. In this sense, the threshold with photon recycling does mark the onset of the change in the photon statistics toward coherent laser light, all the way from the macro-scale (see \secref{sec:lasing-threshold-intro}) to the extreme case of a single emitter.   

    \begin{figure}[ht]
	\centering\includegraphics[width=1\linewidth]{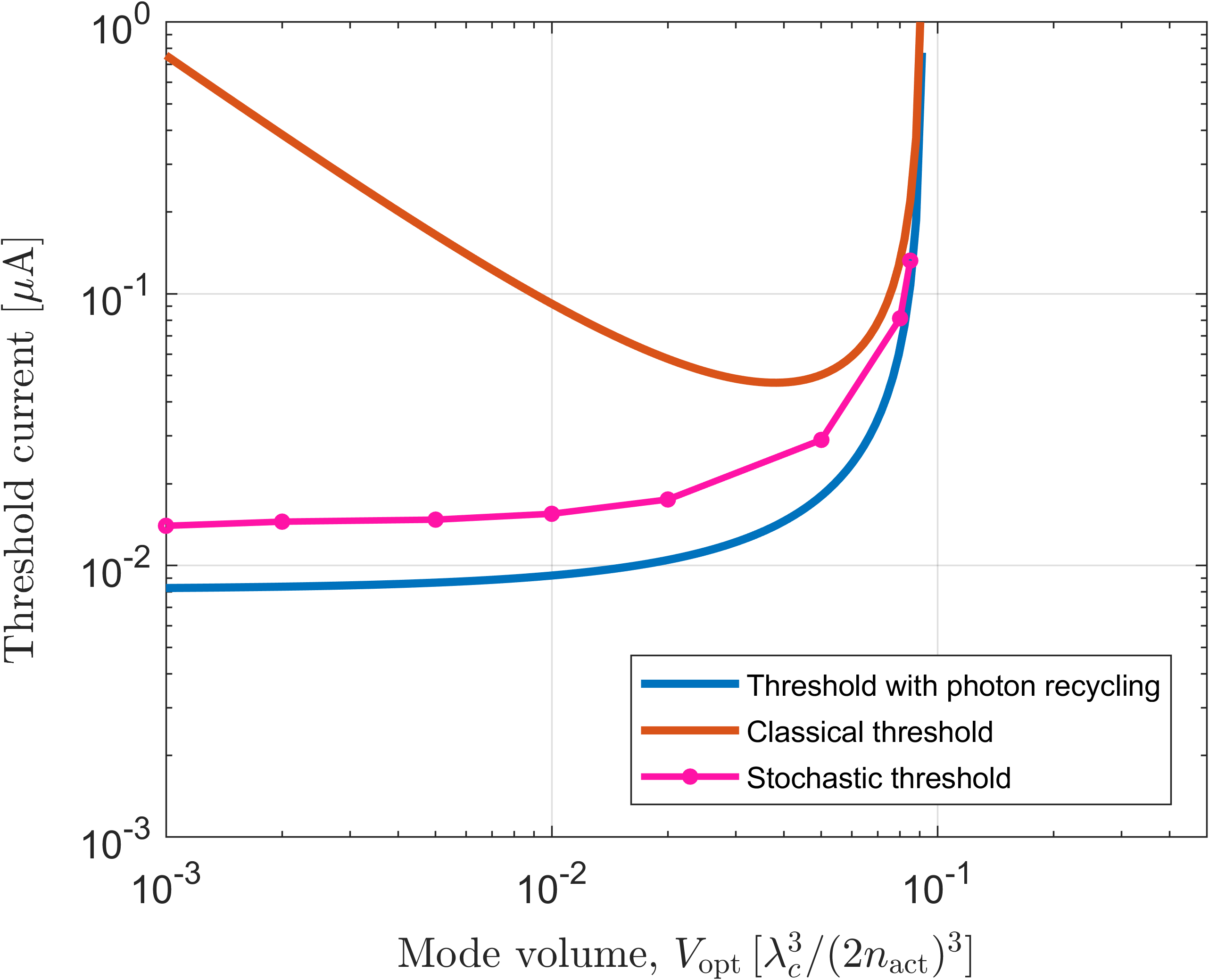}
        \caption{\label{fig:Sec7:FIG4} Threshold current versus mode volume for a single-emitter EDC laser. Other than the mode volume, the parameters are those of EDC laser 6 in \tabref{tab:nano-devices}. Each color denotes a different threshold definition, as specified in the legend.}
    \end{figure}
   The variation with mode volume of the threshold current is illustrated in \figref{fig:Sec7:FIG4}, showing the threshold with photon recycling (blue), the classical threshold (red) and the stochastic threshold (pink). The stochastic threshold and the threshold with photon recycling are again seen to scale similarly and to gradually match as pump blocking becomes important. Note that the Fano factor computed by the Langevin approach features no peak (cf. \figref{fig:Sec7:FIG3}b). Hence, \figref{fig:Sec7:FIG4} does not include the Fano threshold. On the other hand, depending on the mode volume, the stochastic approach may bring out a peak in the Fano factor. In such cases (blue and red markers in \figref{fig:Sec7:FIG3}b), the stochastic approach indicates, unlike the Langevin approach, that the Fano threshold does exist. 

   A systematic investigation of the Fano threshold as obtained from stochastic simulations is beyond the scope of this article. Nonetheless, here we emphasize that the Fano and stochastic threshold are not generally equivalent, despite what \figref{fig:Sec7:FIG2} might suggest. In fact, the Fano threshold current emerging from the stochastic approach (cf. \figref{fig:Sec7:FIG3}b) is found to be quite different from the stochastic threshold current (cf. \figref{fig:Sec7:FIG4}).

\section{Conclusions}\label{sec:discussion}

    \textcolor{black}{Identifying the onset of lasing is both a question of fundamental nature and technological importance, which recent advances in nanotechnology and semiconductor nanolasers \cite{Hill_NatPhot_2014,Ning-SPIE-2019,Ma_NatNanotech_2019,Deng_AdvOptMat_2021,Dimopoulos_Optica_2023} are making as relevant as ever.} In particular, emerging dielectric cavities with deep sub-wavelength optical confinement (so-called extreme dielectric confinement, EDC) \cite{Hu_ACSPhot_2016,Choi_PRL_2017,Wang_APL_2018,Albrechtsen_NatComm_2022,Kountouris-OptExress-2022} are an attractive platform for ultra-compact and energy-efficient nanolasers with a near-unity $\beta$-factor, $\beta$. 
    %EDC cavities may strongly enhance the spontaneous emission rate per emitter, $\gamma_r$, thus featuring a near-unity $\beta$-factor, $\beta$. Cavity designs combining a compact active region \cite{Kuramochi_OptExpress_2018,Dimopoulos_LPR_2022} or single-emitter \cite{Nomura_OptExpress_2009,Bundgaard_PRL_2023} with extreme dielectric confinement may culminate in EDC lasers featuring an ultra-low threshold current. 
   
    % In this complicated arena, one may take the stand that no threshold exists outside the thermodynamic limit of $\beta\rightarrow0$ \cite{Rice-PRA-1994}. On these grounds, it has been argued that a laser with $\beta=1$ is a truly thresholdless device, in the sense that the concept itself of lasing threshold is just meaningless \cite{Rice-PRA-1994}. The relevance of this argument, however, does not address the equally relevant problem of a definition at least functional to applications. The flourishing of publications on the matter proves, at the very least, that the issue is far from being resolved and not entirely semantic. 
    % The large $\gamma_r$ and near-unity $\beta$-factor, offered by EDC lasers, call into question the classical approach \cite{Coldren-book2ndEd-2012} to computing the threshold current. 
    \textcolor{black}{Rivers of ink have been poured on the lasing threshold of high-$\beta$ micro- and nanolasers \cite{Bjork_JQE_1991,Jin_PRA_1994,Bjork_PRA_1994,Rice-PRA-1994,Ning_JSTQE_2013,Ma_LPR_2013,Kreinberg_LightScAppl_2017,Lohof_PhysRevApplied_2018,Takemura_PRA_2019,Kreinberg_LPR_2020,Carroll_PhysRevLett_2021,Khurgin_LPR_2021,Lippi_ChaosSolitonsFractals_2022,Yacomotti-LPR-2022,Carroll_PRA_2023}. Yet, a unified and clear understanding of the onset of lasing is still elusive. Consistently with previous works \cite{Bjork_PRA_1994,Kreinberg_LightScAppl_2017,Lohof_PhysRevApplied_2018}, here we argue that the transition to lasing begins when a given average photon number is built up inside the cavity. However, a simple and unique criterion was missing for the photon number at threshold. The central and original contribution of this paper is determining this photon number by including the effect of photon recycling \cite{Yamamoto_JJAP_1991} in the classical balance between gain and loss. Photon recycling is already accounted for by standard rate equations \cite{Bjork_JQE_1991}, but the effect was not considered, so far, when defining the lasing threshold.} 
    % The classical approach \cite{Coldren-book2ndEd-2012} to compute the threshold current shows its limits in EDC lasers, where most of the spontaneously emitted photons are funneled into the lasing mode and are more likely to be re-absorbed than decaying. 
    
    %In this article, we have proposed, derived, and thoroughly discussed a new threshold definition, the so-called \textit{threshold with photon recycling}. 

    \textcolor{black}{When $\beta$ is close to unity and the average photon number at transparency, $\xi$ (cf. \Eqref{eq:LED-regime}), is sufficiently large, the classical approach to computing the threshold current - neglecting the \textit{net} stimulated emission below threshold \cite{Coldren-book2ndEd-2012} - is invalid. Conversely, the number of carriers below threshold is accurately estimated by neglecting the stimulated emission, but retaining the stimulated absorption (cf. \figref{fig:Sec4:FIGphot-res}). By requiring this number of carriers to fulfill the balance between gain and loss, we arrive at a new threshold definition - the \textit{threshold with photon recycling}.}    
    This threshold current, \Eqref{eq:Ith3}, reduces to the classical threshold current, \Eqref{eq:Ith1}, in the limit of $\beta\ll1$, whereas quantitative, as well as qualitative differences emerge for $\beta\rightarrow1$ (cf. \figref{fig:sec:INTRO:FIG1}). \textcolor{black}{The photon number at the threshold with photon recycling is given by \Eqref{eq:npth3}}. 
    
    We make the following general observations regarding the threshold with photon recycling:
    \begin{itemize}
        \item It marks the onset of the transition in the second-order intensity correlation, $g^{(2)}(0)$, toward coherent laser light, irrespective of the laser size and down to the case of a single emitter.
        \item It identifies the onset of the transition based on physical, rather than empirical arguments.
        \item It reliably distinguishes between lasers and LEDs.
    \end{itemize}
    \textcolor{black}{The caveat is that coherent effects such as Rabi oscillations \cite{Nomura_NatPhys_2010,Gies_PRA_2017} and superradiance \cite{Leymann_PRA_2015,Jahnke_NatCom_2016} are not accounted for, which limits the conclusions to the particular, yet practically relevant case of room-temperature operation.}

    Our investigations provide a systematic overview of the many different threshold definitions for micro- and nanolasers (cf. \figref{fig:SecIthIntro:FIG2}) that have emerged during a long and vivid debate. Importantly, we have clarified that some threshold currents, such as the quantum threshold \cite{Bjork_PRA_1994}, are unreliable indicators of lasing (cf. \figref{fig:SecIth:FIG4}), and others \cite{Carroll_PhysRevLett_2021} may scale, for EDC lasers, in a counter-intuitive manner. 
    %Stochastic simulations reveal that the stochastic threshold, recently proposed \cite{Yacomotti-LPR-2022}, scales with the mode volume in a similar fashion as the threshold with photon recycling. 
    Furthermore, we have shown that lasers with a linear light-current characteristic do \textit{not} require $\beta=1$ if one takes into account the finite number of electronic states that may contribute to lasing (cf. \figref{fig:SecIth:FIG2}). This is important to consider in single-mode nanolasers, such as EDC lasers, even in the presence of extended active media. This observation contrasts with common belief, but appears to agree with experimental results \cite{Jagsch_NatComm_2018}. 
    
    % Modeling the quantum noise with a stochastic approach \cite{Andre_OptExpress_2020} also corroborates the concept of stochastic threshold, as we have named it. This threshold marks the pumping level at which the probabilities of having zero photons and one photon inside the cavity balance out \cite{Scully_PhysRev_1967,Rice-PRA-1994}. Its extension to the nanoscale has been recently proposed \cite{Yacomotti-LPR-2022}, but no analytical expression is available, unless approximations are made \cite{Yacomotti-LPR-2022}. Interestingly, the stochastic threshold scales with the mode volume in a similar fashion as the threshold with photon recycling.

    Besides shedding light on the fundamental question of the onset of lasing, these findings may guide future nanolaser designs. 

\section*{Acknowledgements} \label{sec:acknowledgements}

Authors gratefully acknowledge funding by the European Research Council (ERC) under the European Union’s Horizon 2020 Research and Innovation Programme (Grant No. 834410 FANO), and the Danish National Research Foundation (Grant No. DNRF147 NanoPhoton). Y.Y. acknowledges the support from Villum Fonden via the Young Investigator Programme (Grant No. 42026). J. Mørk acknowledges helpful discussions with Gian Luca Lippi.

%\begin{thebibliography}{4}
%\bibitem{Griffiths}
%D. J. Griffiths,
%\textit{Introduction to Electrodynamics}
%(Cambridge University Press, Cambridge, 2017).
%\end{thebibliography}

% Bibliography
\bibliography{main.bib}

%merlin.mbs apsrev4-1.bst 2010-07-25 4.21a (PWD, AO, DPC) hacked
%Control: key (0)
%Control: author (8) initials jnrlst
%Control: editor formatted (1) identically to author
%Control: production of article title (-1) disabled
%Control: page (0) single
%Control: year (1) truncated
%Control: production of eprint (0) enabled
\begin{thebibliography}{107}%
\makeatletter
\providecommand \@ifxundefined [1]{%
 \@ifx{#1\undefined}
}%
\providecommand \@ifnum [1]{%
 \ifnum #1\expandafter \@firstoftwo
 \else \expandafter \@secondoftwo
 \fi
}%
\providecommand \@ifx [1]{%
 \ifx #1\expandafter \@firstoftwo
 \else \expandafter \@secondoftwo
 \fi
}%
\providecommand \natexlab [1]{#1}%
\providecommand \enquote  [1]{``#1''}%
\providecommand \bibnamefont  [1]{#1}%
\providecommand \bibfnamefont [1]{#1}%
\providecommand \citenamefont [1]{#1}%
\providecommand \href@noop [0]{\@secondoftwo}%
\providecommand \href [0]{\begingroup \@sanitize@url \@href}%
\providecommand \@href[1]{\@@startlink{#1}\@@href}%
\providecommand \@@href[1]{\endgroup#1\@@endlink}%
\providecommand \@sanitize@url [0]{\catcode `\\12\catcode `\$12\catcode
  `\&12\catcode `\#12\catcode `\^12\catcode `\_12\catcode `\%12\relax}%
\providecommand \@@startlink[1]{}%
\providecommand \@@endlink[0]{}%
\providecommand \url  [0]{\begingroup\@sanitize@url \@url }%
\providecommand \@url [1]{\endgroup\@href {#1}{\urlprefix }}%
\providecommand \urlprefix  [0]{URL }%
\providecommand \Eprint [0]{\href }%
\providecommand \doibase [0]{http://dx.doi.org/}%
\providecommand \selectlanguage [0]{\@gobble}%
\providecommand \bibinfo  [0]{\@secondoftwo}%
\providecommand \bibfield  [0]{\@secondoftwo}%
\providecommand \translation [1]{[#1]}%
\providecommand \BibitemOpen [0]{}%
\providecommand \bibitemStop [0]{}%
\providecommand \bibitemNoStop [0]{.\EOS\space}%
\providecommand \EOS [0]{\spacefactor3000\relax}%
\providecommand \BibitemShut  [1]{\csname bibitem#1\endcsname}%
\let\auto@bib@innerbib\@empty
%</preamble>
\bibitem [{\citenamefont {Hecht}(2010)}]{Hecht_ApplOpt_2010}%
  \BibitemOpen
  \bibfield  {author} {\bibinfo {author} {\bibfnamefont {J.}~\bibnamefont
  {Hecht}},\ }\href {\doibase 10.1364/AO.49.000F99} {\bibfield  {journal}
  {\bibinfo  {journal} {Appl. Opt.}\ }\textbf {\bibinfo {volume} {49}},\
  \bibinfo {pages} {F99} (\bibinfo {year} {2010})}\BibitemShut {NoStop}%
\bibitem [{\citenamefont {Hill}\ and\ \citenamefont
  {Gather}(2014)}]{Hill_NatPhot_2014}%
  \BibitemOpen
  \bibfield  {author} {\bibinfo {author} {\bibfnamefont {M.~T.}\ \bibnamefont
  {Hill}}\ and\ \bibinfo {author} {\bibfnamefont {M.~C.}\ \bibnamefont
  {Gather}},\ }\href {\doibase 10.1038/nphoton.2014.239} {\bibfield  {journal}
  {\bibinfo  {journal} {Nature Photonics}\ }\textbf {\bibinfo {volume} {8}},\
  \bibinfo {pages} {908} (\bibinfo {year} {2014})}\BibitemShut {NoStop}%
\bibitem [{\citenamefont {Ning}(2019)}]{Ning-SPIE-2019}%
  \BibitemOpen
  \bibfield  {author} {\bibinfo {author} {\bibfnamefont {C.-Z.}\ \bibnamefont
  {Ning}},\ }\href {\doibase 10.1117/1.AP.1.1.014002} {\bibfield  {journal}
  {\bibinfo  {journal} {Advanced Photonics}\ }\textbf {\bibinfo {volume} {1}},\
  \bibinfo {pages} {014002} (\bibinfo {year} {2019})}\BibitemShut {NoStop}%
\bibitem [{\citenamefont {Ma}\ and\ \citenamefont
  {Oulton}(2019)}]{Ma_NatNanotech_2019}%
  \BibitemOpen
  \bibfield  {author} {\bibinfo {author} {\bibfnamefont {R.-M.}\ \bibnamefont
  {Ma}}\ and\ \bibinfo {author} {\bibfnamefont {R.~F.}\ \bibnamefont
  {Oulton}},\ }\href@noop {} {\bibfield  {journal} {\bibinfo  {journal} {Nature
  Nanotechnology}\ }\textbf {\bibinfo {volume} {14}},\ \bibinfo {pages} {12}
  (\bibinfo {year} {2019})}\BibitemShut {NoStop}%
\bibitem [{\citenamefont {Deng}\ \emph {et~al.}(2021)\citenamefont {Deng},
  \citenamefont {Lippi}, \citenamefont {Mørk}, \citenamefont {Wiersig},\ and\
  \citenamefont {Reitzenstein}}]{Deng_AdvOptMat_2021}%
  \BibitemOpen
  \bibfield  {author} {\bibinfo {author} {\bibfnamefont {H.}~\bibnamefont
  {Deng}}, \bibinfo {author} {\bibfnamefont {G.~L.}\ \bibnamefont {Lippi}},
  \bibinfo {author} {\bibfnamefont {J.}~\bibnamefont {Mørk}}, \bibinfo
  {author} {\bibfnamefont {J.}~\bibnamefont {Wiersig}}, \ and\ \bibinfo
  {author} {\bibfnamefont {S.}~\bibnamefont {Reitzenstein}},\ }\href {\doibase
  https://doi.org/10.1002/adom.202100415} {\bibfield  {journal} {\bibinfo
  {journal} {Advanced Optical Materials}\ }\textbf {\bibinfo {volume} {9}},\
  \bibinfo {pages} {2100415} (\bibinfo {year} {2021})}\BibitemShut {NoStop}%
\bibitem [{\citenamefont {Nozaki}\ \emph {et~al.}(2019)\citenamefont {Nozaki},
  \citenamefont {Matsuo}, \citenamefont {Fujii}, \citenamefont {Takeda},
  \citenamefont {Shinya}, \citenamefont {Kuramochi},\ and\ \citenamefont
  {Notomi}}]{Nozaki_NatPhot_2019}%
  \BibitemOpen
  \bibfield  {author} {\bibinfo {author} {\bibfnamefont {K.}~\bibnamefont
  {Nozaki}}, \bibinfo {author} {\bibfnamefont {S.}~\bibnamefont {Matsuo}},
  \bibinfo {author} {\bibfnamefont {T.}~\bibnamefont {Fujii}}, \bibinfo
  {author} {\bibfnamefont {K.}~\bibnamefont {Takeda}}, \bibinfo {author}
  {\bibfnamefont {A.}~\bibnamefont {Shinya}}, \bibinfo {author} {\bibfnamefont
  {E.}~\bibnamefont {Kuramochi}}, \ and\ \bibinfo {author} {\bibfnamefont
  {M.}~\bibnamefont {Notomi}},\ }\href {\doibase 10.1038/s41566-019-0397-3}
  {\bibfield  {journal} {\bibinfo  {journal} {Nature Photonics}\ }\textbf
  {\bibinfo {volume} {13}},\ \bibinfo {pages} {454} (\bibinfo {year}
  {2019})}\BibitemShut {NoStop}%
\bibitem [{\citenamefont {Takeda}\ \emph {et~al.}(2021)\citenamefont {Takeda},
  \citenamefont {Tsurugaya}, \citenamefont {Fujii}, \citenamefont {Shinya},
  \citenamefont {Maeda}, \citenamefont {Tsuchizawa}, \citenamefont {Nishi},
  \citenamefont {Notomi}, \citenamefont {Kakitsuka},\ and\ \citenamefont
  {Matsuo}}]{Takeda_OptExpress_2021}%
  \BibitemOpen
  \bibfield  {author} {\bibinfo {author} {\bibfnamefont {K.}~\bibnamefont
  {Takeda}}, \bibinfo {author} {\bibfnamefont {T.}~\bibnamefont {Tsurugaya}},
  \bibinfo {author} {\bibfnamefont {T.}~\bibnamefont {Fujii}}, \bibinfo
  {author} {\bibfnamefont {A.}~\bibnamefont {Shinya}}, \bibinfo {author}
  {\bibfnamefont {Y.}~\bibnamefont {Maeda}}, \bibinfo {author} {\bibfnamefont
  {T.}~\bibnamefont {Tsuchizawa}}, \bibinfo {author} {\bibfnamefont
  {H.}~\bibnamefont {Nishi}}, \bibinfo {author} {\bibfnamefont
  {M.}~\bibnamefont {Notomi}}, \bibinfo {author} {\bibfnamefont
  {T.}~\bibnamefont {Kakitsuka}}, \ and\ \bibinfo {author} {\bibfnamefont
  {S.}~\bibnamefont {Matsuo}},\ }\href {\doibase 10.1364/OE.427843} {\bibfield
  {journal} {\bibinfo  {journal} {Opt. Express}\ }\textbf {\bibinfo {volume}
  {29}},\ \bibinfo {pages} {26082} (\bibinfo {year} {2021})}\BibitemShut
  {NoStop}%
\bibitem [{\citenamefont {Dimopoulos}\ \emph {et~al.}(2023)\citenamefont
  {Dimopoulos}, \citenamefont {Xiong}, \citenamefont {Sakanas}, \citenamefont
  {Marchevsky}, \citenamefont {Dong}, \citenamefont {Yu}, \citenamefont
  {Semenova}, \citenamefont {M{\o}rk},\ and\ \citenamefont
  {Yvind}}]{Dimopoulos_Optica_2023}%
  \BibitemOpen
  \bibfield  {author} {\bibinfo {author} {\bibfnamefont {E.}~\bibnamefont
  {Dimopoulos}}, \bibinfo {author} {\bibfnamefont {M.}~\bibnamefont {Xiong}},
  \bibinfo {author} {\bibfnamefont {A.}~\bibnamefont {Sakanas}}, \bibinfo
  {author} {\bibfnamefont {A.}~\bibnamefont {Marchevsky}}, \bibinfo {author}
  {\bibfnamefont {G.}~\bibnamefont {Dong}}, \bibinfo {author} {\bibfnamefont
  {Y.}~\bibnamefont {Yu}}, \bibinfo {author} {\bibfnamefont {E.}~\bibnamefont
  {Semenova}}, \bibinfo {author} {\bibfnamefont {J.}~\bibnamefont {M{\o}rk}}, \
  and\ \bibinfo {author} {\bibfnamefont {K.}~\bibnamefont {Yvind}},\ }\href
  {\doibase 10.1364/OPTICA.488604} {\bibfield  {journal} {\bibinfo  {journal}
  {Optica}\ }\textbf {\bibinfo {volume} {10}},\ \bibinfo {pages} {973}
  (\bibinfo {year} {2023})}\BibitemShut {NoStop}%
\bibitem [{\citenamefont {Zhang}\ \emph {et~al.}(2021)\citenamefont {Zhang},
  \citenamefont {Nest}, \citenamefont {Wang}, \citenamefont {Wang},\ and\
  \citenamefont {Ma}}]{Zhang_PhotRes_2021}%
  \BibitemOpen
  \bibfield  {author} {\bibinfo {author} {\bibfnamefont {Z.}~\bibnamefont
  {Zhang}}, \bibinfo {author} {\bibfnamefont {L.}~\bibnamefont {Nest}},
  \bibinfo {author} {\bibfnamefont {S.}~\bibnamefont {Wang}}, \bibinfo {author}
  {\bibfnamefont {S.-Y.}\ \bibnamefont {Wang}}, \ and\ \bibinfo {author}
  {\bibfnamefont {R.-M.}\ \bibnamefont {Ma}},\ }\href {\doibase
  10.1364/PRJ.431612} {\bibfield  {journal} {\bibinfo  {journal} {Photon.
  Res.}\ }\textbf {\bibinfo {volume} {9}},\ \bibinfo {pages} {1699} (\bibinfo
  {year} {2021})}\BibitemShut {NoStop}%
\bibitem [{\citenamefont {Kreinberg}\ \emph {et~al.}(2018)\citenamefont
  {Kreinberg}, \citenamefont {Grbešić}, \citenamefont {Strauß},
  \citenamefont {Carmele}, \citenamefont {Emmerling}, \citenamefont
  {Schneider}, \citenamefont {Höfling}, \citenamefont {Porte},\ and\
  \citenamefont {Reitzenstein}}]{Kreinberg_LightScAppl_2018}%
  \BibitemOpen
  \bibfield  {author} {\bibinfo {author} {\bibfnamefont {S.}~\bibnamefont
  {Kreinberg}}, \bibinfo {author} {\bibfnamefont {T.}~\bibnamefont
  {Grbešić}}, \bibinfo {author} {\bibfnamefont {M.}~\bibnamefont {Strauß}},
  \bibinfo {author} {\bibfnamefont {A.}~\bibnamefont {Carmele}}, \bibinfo
  {author} {\bibfnamefont {M.}~\bibnamefont {Emmerling}}, \bibinfo {author}
  {\bibfnamefont {C.}~\bibnamefont {Schneider}}, \bibinfo {author}
  {\bibfnamefont {S.}~\bibnamefont {Höfling}}, \bibinfo {author}
  {\bibfnamefont {X.}~\bibnamefont {Porte}}, \ and\ \bibinfo {author}
  {\bibfnamefont {S.}~\bibnamefont {Reitzenstein}},\ }\href {\doibase
  10.1038/s41377-018-0045-6} {\bibfield  {journal} {\bibinfo  {journal} {Light:
  Science \& Applications}\ }\textbf {\bibinfo {volume} {7}},\ \bibinfo {pages}
  {41} (\bibinfo {year} {2018})}\BibitemShut {NoStop}%
\bibitem [{\citenamefont {Zhang}\ \emph {et~al.}(2022)\citenamefont {Zhang},
  \citenamefont {Zhao}, \citenamefont {Wu}, \citenamefont {Wu}, \citenamefont
  {Qiao}, \citenamefont {Gao}, \citenamefont {Agarwal}, \citenamefont {Longhi},
  \citenamefont {Litchinitser}, \citenamefont {Ge},\ and\ \citenamefont
  {Feng}}]{Zhang_Nature_2022}%
  \BibitemOpen
  \bibfield  {author} {\bibinfo {author} {\bibfnamefont {Z.}~\bibnamefont
  {Zhang}}, \bibinfo {author} {\bibfnamefont {H.}~\bibnamefont {Zhao}},
  \bibinfo {author} {\bibfnamefont {S.}~\bibnamefont {Wu}}, \bibinfo {author}
  {\bibfnamefont {T.}~\bibnamefont {Wu}}, \bibinfo {author} {\bibfnamefont
  {X.}~\bibnamefont {Qiao}}, \bibinfo {author} {\bibfnamefont {Z.}~\bibnamefont
  {Gao}}, \bibinfo {author} {\bibfnamefont {R.}~\bibnamefont {Agarwal}},
  \bibinfo {author} {\bibfnamefont {S.}~\bibnamefont {Longhi}}, \bibinfo
  {author} {\bibfnamefont {N.~M.}\ \bibnamefont {Litchinitser}}, \bibinfo
  {author} {\bibfnamefont {L.}~\bibnamefont {Ge}}, \ and\ \bibinfo {author}
  {\bibfnamefont {L.}~\bibnamefont {Feng}},\ }\href {\doibase
  10.1038/s41586-022-05339-z} {\bibfield  {journal} {\bibinfo  {journal}
  {Nature}\ }\textbf {\bibinfo {volume} {612}},\ \bibinfo {pages} {246}
  (\bibinfo {year} {2022})}\BibitemShut {NoStop}%
\bibitem [{\citenamefont {Shastri}\ \emph {et~al.}(2021)\citenamefont
  {Shastri}, \citenamefont {Tait}, \citenamefont {Ferreira~de Lima},
  \citenamefont {Pernice}, \citenamefont {Bhaskaran}, \citenamefont {Wright},\
  and\ \citenamefont {Prucnal}}]{Shastri_NatPhot_2021}%
  \BibitemOpen
  \bibfield  {author} {\bibinfo {author} {\bibfnamefont {B.~J.}\ \bibnamefont
  {Shastri}}, \bibinfo {author} {\bibfnamefont {A.~N.}\ \bibnamefont {Tait}},
  \bibinfo {author} {\bibfnamefont {T.}~\bibnamefont {Ferreira~de Lima}},
  \bibinfo {author} {\bibfnamefont {W.~H.~P.}\ \bibnamefont {Pernice}},
  \bibinfo {author} {\bibfnamefont {H.}~\bibnamefont {Bhaskaran}}, \bibinfo
  {author} {\bibfnamefont {C.~D.}\ \bibnamefont {Wright}}, \ and\ \bibinfo
  {author} {\bibfnamefont {P.~R.}\ \bibnamefont {Prucnal}},\ }\href {\doibase
  10.1038/s41566-020-00754-y} {\bibfield  {journal} {\bibinfo  {journal}
  {Nature Photonics}\ }\textbf {\bibinfo {volume} {15}},\ \bibinfo {pages}
  {102} (\bibinfo {year} {2021})}\BibitemShut {NoStop}%
\bibitem [{\citenamefont {Coldren}\ \emph {et~al.}(2012)\citenamefont
  {Coldren}, \citenamefont {Corzine},\ and\ \citenamefont
  {Ma\v{s}anovi\'{c}}}]{Coldren-book2ndEd-2012}%
  \BibitemOpen
  \bibfield  {author} {\bibinfo {author} {\bibfnamefont {L.~A.}\ \bibnamefont
  {Coldren}}, \bibinfo {author} {\bibfnamefont {S.~W.}\ \bibnamefont
  {Corzine}}, \ and\ \bibinfo {author} {\bibfnamefont {M.}~\bibnamefont
  {Ma\v{s}anovi\'{c}}},\ }\href@noop {} {\emph {\bibinfo {title} {Diode Lasers
  and Photonic Integrated Circuits}}},\ \bibinfo {edition} {2nd}\ ed.\
  (\bibinfo  {publisher} {John Wiley \& Sons, Inc.},\ \bibinfo {year}
  {2012})\BibitemShut {NoStop}%
\bibitem [{\citenamefont {Rice}\ and\ \citenamefont
  {Carmichael}(1994)}]{Rice-PRA-1994}%
  \BibitemOpen
  \bibfield  {author} {\bibinfo {author} {\bibfnamefont {P.~R.}\ \bibnamefont
  {Rice}}\ and\ \bibinfo {author} {\bibfnamefont {H.~J.}\ \bibnamefont
  {Carmichael}},\ }\href {\doibase 10.1103/PhysRevA.50.4318} {\bibfield
  {journal} {\bibinfo  {journal} {Phys. Rev. A}\ }\textbf {\bibinfo {volume}
  {50}},\ \bibinfo {pages} {4318} (\bibinfo {year} {1994})}\BibitemShut
  {NoStop}%
\bibitem [{\citenamefont {Ning}(2013)}]{Ning_JSTQE_2013}%
  \BibitemOpen
  \bibfield  {author} {\bibinfo {author} {\bibfnamefont {C.~Z.}\ \bibnamefont
  {Ning}},\ }\href {\doibase 10.1109/JSTQE.2013.2259222} {\bibfield  {journal}
  {\bibinfo  {journal} {IEEE Journal of Selected Topics in Quantum
  Electronics}\ }\textbf {\bibinfo {volume} {19}},\ \bibinfo {pages} {1503604}
  (\bibinfo {year} {2013})}\BibitemShut {NoStop}%
\bibitem [{\citenamefont {Kreinberg}\ \emph {et~al.}(2017)\citenamefont
  {Kreinberg}, \citenamefont {Chow}, \citenamefont {Wolters}, \citenamefont
  {Schneider}, \citenamefont {Gies}, \citenamefont {Jahnke}, \citenamefont
  {Höfling}, \citenamefont {Kamp},\ and\ \citenamefont
  {Reitzenstein}}]{Kreinberg_LightScAppl_2017}%
  \BibitemOpen
  \bibfield  {author} {\bibinfo {author} {\bibfnamefont {S.}~\bibnamefont
  {Kreinberg}}, \bibinfo {author} {\bibfnamefont {W.~W.}\ \bibnamefont {Chow}},
  \bibinfo {author} {\bibfnamefont {J.}~\bibnamefont {Wolters}}, \bibinfo
  {author} {\bibfnamefont {C.}~\bibnamefont {Schneider}}, \bibinfo {author}
  {\bibfnamefont {C.}~\bibnamefont {Gies}}, \bibinfo {author} {\bibfnamefont
  {F.}~\bibnamefont {Jahnke}}, \bibinfo {author} {\bibfnamefont
  {S.}~\bibnamefont {Höfling}}, \bibinfo {author} {\bibfnamefont
  {M.}~\bibnamefont {Kamp}}, \ and\ \bibinfo {author} {\bibfnamefont
  {S.}~\bibnamefont {Reitzenstein}},\ }\href {\doibase 10.1038/lsa.2017.30}
  {\bibfield  {journal} {\bibinfo  {journal} {Light: Science \& Applications}\
  }\textbf {\bibinfo {volume} {6}},\ \bibinfo {pages} {e17030} (\bibinfo {year}
  {2017})}\BibitemShut {NoStop}%
\bibitem [{\citenamefont {Bjork}\ and\ \citenamefont
  {Yamamoto}(1991)}]{Bjork_JQE_1991}%
  \BibitemOpen
  \bibfield  {author} {\bibinfo {author} {\bibfnamefont {G.}~\bibnamefont
  {Bjork}}\ and\ \bibinfo {author} {\bibfnamefont {Y.}~\bibnamefont
  {Yamamoto}},\ }\href {\doibase 10.1109/3.100877} {\bibfield  {journal}
  {\bibinfo  {journal} {IEEE Journal of Quantum Electronics}\ }\textbf
  {\bibinfo {volume} {27}},\ \bibinfo {pages} {2386} (\bibinfo {year}
  {1991})}\BibitemShut {NoStop}%
\bibitem [{\citenamefont {Jin}\ \emph {et~al.}(1994)\citenamefont {Jin},
  \citenamefont {Boggavarapu}, \citenamefont {Sargent}, \citenamefont
  {Meystre}, \citenamefont {Gibbs},\ and\ \citenamefont
  {Khitrova}}]{Jin_PRA_1994}%
  \BibitemOpen
  \bibfield  {author} {\bibinfo {author} {\bibfnamefont {R.}~\bibnamefont
  {Jin}}, \bibinfo {author} {\bibfnamefont {D.}~\bibnamefont {Boggavarapu}},
  \bibinfo {author} {\bibfnamefont {M.}~\bibnamefont {Sargent}}, \bibinfo
  {author} {\bibfnamefont {P.}~\bibnamefont {Meystre}}, \bibinfo {author}
  {\bibfnamefont {H.~M.}\ \bibnamefont {Gibbs}}, \ and\ \bibinfo {author}
  {\bibfnamefont {G.}~\bibnamefont {Khitrova}},\ }\href {\doibase
  10.1103/PhysRevA.49.4038} {\bibfield  {journal} {\bibinfo  {journal} {Phys.
  Rev. A}\ }\textbf {\bibinfo {volume} {49}},\ \bibinfo {pages} {4038}
  (\bibinfo {year} {1994})}\BibitemShut {NoStop}%
\bibitem [{\citenamefont {Bj\"ork}\ \emph {et~al.}(1994)\citenamefont
  {Bj\"ork}, \citenamefont {Karlsson},\ and\ \citenamefont
  {Yamamoto}}]{Bjork_PRA_1994}%
  \BibitemOpen
  \bibfield  {author} {\bibinfo {author} {\bibfnamefont {G.}~\bibnamefont
  {Bj\"ork}}, \bibinfo {author} {\bibfnamefont {A.}~\bibnamefont {Karlsson}}, \
  and\ \bibinfo {author} {\bibfnamefont {Y.}~\bibnamefont {Yamamoto}},\ }\href
  {\doibase 10.1103/PhysRevA.50.1675} {\bibfield  {journal} {\bibinfo
  {journal} {Phys. Rev. A}\ }\textbf {\bibinfo {volume} {50}},\ \bibinfo
  {pages} {1675} (\bibinfo {year} {1994})}\BibitemShut {NoStop}%
\bibitem [{\citenamefont {Lohof}\ \emph {et~al.}(2018)\citenamefont {Lohof},
  \citenamefont {Barzel}, \citenamefont {Gartner},\ and\ \citenamefont
  {Gies}}]{Lohof_PhysRevApplied_2018}%
  \BibitemOpen
  \bibfield  {author} {\bibinfo {author} {\bibfnamefont {F.}~\bibnamefont
  {Lohof}}, \bibinfo {author} {\bibfnamefont {R.}~\bibnamefont {Barzel}},
  \bibinfo {author} {\bibfnamefont {P.}~\bibnamefont {Gartner}}, \ and\
  \bibinfo {author} {\bibfnamefont {C.}~\bibnamefont {Gies}},\ }\href {\doibase
  10.1103/PhysRevApplied.10.054055} {\bibfield  {journal} {\bibinfo  {journal}
  {Phys. Rev. Appl.}\ }\textbf {\bibinfo {volume} {10}},\ \bibinfo {pages}
  {054055} (\bibinfo {year} {2018})}\BibitemShut {NoStop}%
\bibitem [{\citenamefont {Takemura}\ \emph {et~al.}(2019)\citenamefont
  {Takemura}, \citenamefont {Takiguchi}, \citenamefont {Kuramochi},
  \citenamefont {Shinya}, \citenamefont {Sato}, \citenamefont {Takeda},
  \citenamefont {Matsuo},\ and\ \citenamefont {Notomi}}]{Takemura_PRA_2019}%
  \BibitemOpen
  \bibfield  {author} {\bibinfo {author} {\bibfnamefont {N.}~\bibnamefont
  {Takemura}}, \bibinfo {author} {\bibfnamefont {M.}~\bibnamefont {Takiguchi}},
  \bibinfo {author} {\bibfnamefont {E.}~\bibnamefont {Kuramochi}}, \bibinfo
  {author} {\bibfnamefont {A.}~\bibnamefont {Shinya}}, \bibinfo {author}
  {\bibfnamefont {T.}~\bibnamefont {Sato}}, \bibinfo {author} {\bibfnamefont
  {K.}~\bibnamefont {Takeda}}, \bibinfo {author} {\bibfnamefont
  {S.}~\bibnamefont {Matsuo}}, \ and\ \bibinfo {author} {\bibfnamefont
  {M.}~\bibnamefont {Notomi}},\ }\href {\doibase 10.1103/PhysRevA.99.053820}
  {\bibfield  {journal} {\bibinfo  {journal} {Phys. Rev. A}\ }\textbf {\bibinfo
  {volume} {99}},\ \bibinfo {pages} {053820} (\bibinfo {year}
  {2019})}\BibitemShut {NoStop}%
\bibitem [{\citenamefont {Kreinberg}\ \emph {et~al.}(2020)\citenamefont
  {Kreinberg}, \citenamefont {Laiho}, \citenamefont {Lohof}, \citenamefont
  {Hayenga}, \citenamefont {Holewa}, \citenamefont {Gies}, \citenamefont
  {Khajavikhan},\ and\ \citenamefont {Reitzenstein}}]{Kreinberg_LPR_2020}%
  \BibitemOpen
  \bibfield  {author} {\bibinfo {author} {\bibfnamefont {S.}~\bibnamefont
  {Kreinberg}}, \bibinfo {author} {\bibfnamefont {K.}~\bibnamefont {Laiho}},
  \bibinfo {author} {\bibfnamefont {F.}~\bibnamefont {Lohof}}, \bibinfo
  {author} {\bibfnamefont {W.~E.}\ \bibnamefont {Hayenga}}, \bibinfo {author}
  {\bibfnamefont {P.}~\bibnamefont {Holewa}}, \bibinfo {author} {\bibfnamefont
  {C.}~\bibnamefont {Gies}}, \bibinfo {author} {\bibfnamefont {M.}~\bibnamefont
  {Khajavikhan}}, \ and\ \bibinfo {author} {\bibfnamefont {S.}~\bibnamefont
  {Reitzenstein}},\ }\href {\doibase https://doi.org/10.1002/lpor.202000065}
  {\bibfield  {journal} {\bibinfo  {journal} {Laser \& Photonics Reviews}\
  }\textbf {\bibinfo {volume} {14}},\ \bibinfo {pages} {2000065} (\bibinfo
  {year} {2020})}\BibitemShut {NoStop}%
\bibitem [{\citenamefont {Carroll}\ \emph {et~al.}(2021)\citenamefont
  {Carroll}, \citenamefont {D'Alessandro}, \citenamefont {Lippi}, \citenamefont
  {Oppo},\ and\ \citenamefont {Papoff}}]{Carroll_PhysRevLett_2021}%
  \BibitemOpen
  \bibfield  {author} {\bibinfo {author} {\bibfnamefont {M.~A.}\ \bibnamefont
  {Carroll}}, \bibinfo {author} {\bibfnamefont {G.}~\bibnamefont
  {D'Alessandro}}, \bibinfo {author} {\bibfnamefont {G.~L.}\ \bibnamefont
  {Lippi}}, \bibinfo {author} {\bibfnamefont {G.-L.}\ \bibnamefont {Oppo}}, \
  and\ \bibinfo {author} {\bibfnamefont {F.}~\bibnamefont {Papoff}},\ }\href
  {\doibase 10.1103/PhysRevLett.126.063902} {\bibfield  {journal} {\bibinfo
  {journal} {Phys. Rev. Lett.}\ }\textbf {\bibinfo {volume} {126}},\ \bibinfo
  {pages} {063902} (\bibinfo {year} {2021})}\BibitemShut {NoStop}%
\bibitem [{\citenamefont {Khurgin}\ and\ \citenamefont
  {Noginov}(2021)}]{Khurgin_LPR_2021}%
  \BibitemOpen
  \bibfield  {author} {\bibinfo {author} {\bibfnamefont {J.~B.}\ \bibnamefont
  {Khurgin}}\ and\ \bibinfo {author} {\bibfnamefont {M.~A.}\ \bibnamefont
  {Noginov}},\ }\href {\doibase https://doi.org/10.1002/lpor.202000250}
  {\bibfield  {journal} {\bibinfo  {journal} {Laser \& Photonics Reviews}\
  }\textbf {\bibinfo {volume} {15}},\ \bibinfo {pages} {2000250} (\bibinfo
  {year} {2021})}\BibitemShut {NoStop}%
\bibitem [{\citenamefont {Lippi}\ \emph {et~al.}(2022)\citenamefont {Lippi},
  \citenamefont {Wang},\ and\ \citenamefont
  {Puccioni}}]{Lippi_ChaosSolitonsFractals_2022}%
  \BibitemOpen
  \bibfield  {author} {\bibinfo {author} {\bibfnamefont {G.}~\bibnamefont
  {Lippi}}, \bibinfo {author} {\bibfnamefont {T.}~\bibnamefont {Wang}}, \ and\
  \bibinfo {author} {\bibfnamefont {G.}~\bibnamefont {Puccioni}},\ }\href
  {\doibase https://doi.org/10.1016/j.chaos.2022.111850} {\bibfield  {journal}
  {\bibinfo  {journal} {Chaos, Solitons \& Fractals}\ }\textbf {\bibinfo
  {volume} {157}},\ \bibinfo {pages} {111850} (\bibinfo {year}
  {2022})}\BibitemShut {NoStop}%
\bibitem [{\citenamefont {Yacomotti}\ \emph {et~al.}(2022)\citenamefont
  {Yacomotti}, \citenamefont {Denis}, \citenamefont {Biella},\ and\
  \citenamefont {Ciuti}}]{Yacomotti-LPR-2022}%
  \BibitemOpen
  \bibfield  {author} {\bibinfo {author} {\bibfnamefont {A.~M.}\ \bibnamefont
  {Yacomotti}}, \bibinfo {author} {\bibfnamefont {Z.}~\bibnamefont {Denis}},
  \bibinfo {author} {\bibfnamefont {A.}~\bibnamefont {Biella}}, \ and\ \bibinfo
  {author} {\bibfnamefont {C.}~\bibnamefont {Ciuti}},\ }\href {\doibase
  https://doi.org/10.1002/lpor.202200377} {\bibfield  {journal} {\bibinfo
  {journal} {Laser \& Photonics Reviews}\ }\textbf {\bibinfo {volume} {17}},\
  \bibinfo {pages} {2200377} (\bibinfo {year} {2022})}\BibitemShut {NoStop}%
\bibitem [{\citenamefont {Carroll}\ \emph {et~al.}(2023)\citenamefont
  {Carroll}, \citenamefont {D'Alessandro}, \citenamefont {Lippi}, \citenamefont
  {Oppo},\ and\ \citenamefont {Papoff}}]{Carroll_PRA_2023}%
  \BibitemOpen
  \bibfield  {author} {\bibinfo {author} {\bibfnamefont {M.~A.}\ \bibnamefont
  {Carroll}}, \bibinfo {author} {\bibfnamefont {G.}~\bibnamefont
  {D'Alessandro}}, \bibinfo {author} {\bibfnamefont {G.~L.}\ \bibnamefont
  {Lippi}}, \bibinfo {author} {\bibfnamefont {G.-L.}\ \bibnamefont {Oppo}}, \
  and\ \bibinfo {author} {\bibfnamefont {F.}~\bibnamefont {Papoff}},\ }\href
  {\doibase 10.1103/PhysRevA.107.063710} {\bibfield  {journal} {\bibinfo
  {journal} {Phys. Rev. A}\ }\textbf {\bibinfo {volume} {107}},\ \bibinfo
  {pages} {063710} (\bibinfo {year} {2023})}\BibitemShut {NoStop}%
\bibitem [{\citenamefont {Miller}(2017)}]{Miller_JLWT_2017}%
  \BibitemOpen
  \bibfield  {author} {\bibinfo {author} {\bibfnamefont {D.~A.~B.}\
  \bibnamefont {Miller}},\ }\href@noop {} {\bibfield  {journal} {\bibinfo
  {journal} {J. Lightwave Technol.}\ }\textbf {\bibinfo {volume} {35}},\
  \bibinfo {pages} {346} (\bibinfo {year} {2017})}\BibitemShut {NoStop}%
\bibitem [{\citenamefont {De~Martini}\ and\ \citenamefont
  {Jacobovitz}(1988)}]{DeMartini_PRL_1988}%
  \BibitemOpen
  \bibfield  {author} {\bibinfo {author} {\bibfnamefont {F.}~\bibnamefont
  {De~Martini}}\ and\ \bibinfo {author} {\bibfnamefont {G.~R.}\ \bibnamefont
  {Jacobovitz}},\ }\href {\doibase 10.1103/PhysRevLett.60.1711} {\bibfield
  {journal} {\bibinfo  {journal} {Phys. Rev. Lett.}\ }\textbf {\bibinfo
  {volume} {60}},\ \bibinfo {pages} {1711} (\bibinfo {year}
  {1988})}\BibitemShut {NoStop}%
\bibitem [{\citenamefont {Yamamoto}\ and\ \citenamefont
  {Björk}(1991)}]{Yamamoto_JJAP_1991}%
  \BibitemOpen
  \bibfield  {author} {\bibinfo {author} {\bibfnamefont {Y.}~\bibnamefont
  {Yamamoto}}\ and\ \bibinfo {author} {\bibfnamefont {G.}~\bibnamefont
  {Björk}},\ }\href {\doibase 10.1143/jjap.30.l2039} {\bibfield  {journal}
  {\bibinfo  {journal} {Japanese Journal of Applied Physics}\ }\textbf
  {\bibinfo {volume} {30}},\ \bibinfo {pages} {L2039} (\bibinfo {year}
  {1991})}\BibitemShut {NoStop}%
\bibitem [{\citenamefont {Gies}\ \emph {et~al.}(2007)\citenamefont {Gies},
  \citenamefont {Wiersig}, \citenamefont {Lorke},\ and\ \citenamefont
  {Jahnke}}]{Gies_PRA_2007}%
  \BibitemOpen
  \bibfield  {author} {\bibinfo {author} {\bibfnamefont {C.}~\bibnamefont
  {Gies}}, \bibinfo {author} {\bibfnamefont {J.}~\bibnamefont {Wiersig}},
  \bibinfo {author} {\bibfnamefont {M.}~\bibnamefont {Lorke}}, \ and\ \bibinfo
  {author} {\bibfnamefont {F.}~\bibnamefont {Jahnke}},\ }\href {\doibase
  10.1103/PhysRevA.75.013803} {\bibfield  {journal} {\bibinfo  {journal} {Phys.
  Rev. A}\ }\textbf {\bibinfo {volume} {75}},\ \bibinfo {pages} {013803}
  (\bibinfo {year} {2007})}\BibitemShut {NoStop}%
\bibitem [{\citenamefont {Wiersig}\ \emph {et~al.}(2009)\citenamefont
  {Wiersig}, \citenamefont {Gies}, \citenamefont {Jahnke}, \citenamefont
  {Aßmann}, \citenamefont {Berstermann}, \citenamefont {Bayer}, \citenamefont
  {Kistner}, \citenamefont {Reitzenstein}, \citenamefont {Schneider},
  \citenamefont {Höfling}, \citenamefont {Forchel}, \citenamefont {Kruse},
  \citenamefont {Kalden},\ and\ \citenamefont {Hommel}}]{Wiersig_Nature_2009}%
  \BibitemOpen
  \bibfield  {author} {\bibinfo {author} {\bibfnamefont {J.}~\bibnamefont
  {Wiersig}}, \bibinfo {author} {\bibfnamefont {C.}~\bibnamefont {Gies}},
  \bibinfo {author} {\bibfnamefont {F.}~\bibnamefont {Jahnke}}, \bibinfo
  {author} {\bibfnamefont {M.}~\bibnamefont {Aßmann}}, \bibinfo {author}
  {\bibfnamefont {T.}~\bibnamefont {Berstermann}}, \bibinfo {author}
  {\bibfnamefont {M.}~\bibnamefont {Bayer}}, \bibinfo {author} {\bibfnamefont
  {C.}~\bibnamefont {Kistner}}, \bibinfo {author} {\bibfnamefont
  {S.}~\bibnamefont {Reitzenstein}}, \bibinfo {author} {\bibfnamefont
  {C.}~\bibnamefont {Schneider}}, \bibinfo {author} {\bibfnamefont
  {S.}~\bibnamefont {Höfling}}, \bibinfo {author} {\bibfnamefont
  {A.}~\bibnamefont {Forchel}}, \bibinfo {author} {\bibfnamefont
  {C.}~\bibnamefont {Kruse}}, \bibinfo {author} {\bibfnamefont
  {J.}~\bibnamefont {Kalden}}, \ and\ \bibinfo {author} {\bibfnamefont
  {D.}~\bibnamefont {Hommel}},\ }\href {\doibase 10.1038/nature08126}
  {\bibfield  {journal} {\bibinfo  {journal} {Nature}\ }\textbf {\bibinfo
  {volume} {460}},\ \bibinfo {pages} {5081} (\bibinfo {year}
  {2009})}\BibitemShut {NoStop}%
\bibitem [{\citenamefont {Nomura}\ \emph {et~al.}(2010)\citenamefont {Nomura},
  \citenamefont {Kumagai}, \citenamefont {Iwamoto}, \citenamefont {Ota},\ and\
  \citenamefont {Arakawa}}]{Nomura_NatPhys_2010}%
  \BibitemOpen
  \bibfield  {author} {\bibinfo {author} {\bibfnamefont {M.}~\bibnamefont
  {Nomura}}, \bibinfo {author} {\bibfnamefont {N.}~\bibnamefont {Kumagai}},
  \bibinfo {author} {\bibfnamefont {S.}~\bibnamefont {Iwamoto}}, \bibinfo
  {author} {\bibfnamefont {Y.}~\bibnamefont {Ota}}, \ and\ \bibinfo {author}
  {\bibfnamefont {Y.}~\bibnamefont {Arakawa}},\ }\href {\doibase
  10.1038/nphys1518} {\bibfield  {journal} {\bibinfo  {journal} {Nature
  Physics}\ }\textbf {\bibinfo {volume} {6}},\ \bibinfo {pages} {279} (\bibinfo
  {year} {2010})}\BibitemShut {NoStop}%
\bibitem [{\citenamefont {Gies}\ \emph {et~al.}(2017)\citenamefont {Gies},
  \citenamefont {Gericke}, \citenamefont {Gartner}, \citenamefont {Holzinger},
  \citenamefont {Hopfmann}, \citenamefont {Heindel}, \citenamefont {Wolters},
  \citenamefont {Schneider}, \citenamefont {Florian}, \citenamefont {Jahnke},
  \citenamefont {H\"ofling}, \citenamefont {Kamp},\ and\ \citenamefont
  {Reitzenstein}}]{Gies_PRA_2017}%
  \BibitemOpen
  \bibfield  {author} {\bibinfo {author} {\bibfnamefont {C.}~\bibnamefont
  {Gies}}, \bibinfo {author} {\bibfnamefont {F.}~\bibnamefont {Gericke}},
  \bibinfo {author} {\bibfnamefont {P.}~\bibnamefont {Gartner}}, \bibinfo
  {author} {\bibfnamefont {S.}~\bibnamefont {Holzinger}}, \bibinfo {author}
  {\bibfnamefont {C.}~\bibnamefont {Hopfmann}}, \bibinfo {author}
  {\bibfnamefont {T.}~\bibnamefont {Heindel}}, \bibinfo {author} {\bibfnamefont
  {J.}~\bibnamefont {Wolters}}, \bibinfo {author} {\bibfnamefont
  {C.}~\bibnamefont {Schneider}}, \bibinfo {author} {\bibfnamefont
  {M.}~\bibnamefont {Florian}}, \bibinfo {author} {\bibfnamefont
  {F.}~\bibnamefont {Jahnke}}, \bibinfo {author} {\bibfnamefont
  {S.}~\bibnamefont {H\"ofling}}, \bibinfo {author} {\bibfnamefont
  {M.}~\bibnamefont {Kamp}}, \ and\ \bibinfo {author} {\bibfnamefont
  {S.}~\bibnamefont {Reitzenstein}},\ }\href {\doibase
  10.1103/PhysRevA.96.023806} {\bibfield  {journal} {\bibinfo  {journal} {Phys.
  Rev. A}\ }\textbf {\bibinfo {volume} {96}},\ \bibinfo {pages} {023806}
  (\bibinfo {year} {2017})}\BibitemShut {NoStop}%
\bibitem [{\citenamefont {Leymann}\ \emph {et~al.}(2015)\citenamefont
  {Leymann}, \citenamefont {Foerster}, \citenamefont {Jahnke}, \citenamefont
  {Wiersig},\ and\ \citenamefont {Gies}}]{Leymann_PRA_2015}%
  \BibitemOpen
  \bibfield  {author} {\bibinfo {author} {\bibfnamefont {H.~A.~M.}\
  \bibnamefont {Leymann}}, \bibinfo {author} {\bibfnamefont {A.}~\bibnamefont
  {Foerster}}, \bibinfo {author} {\bibfnamefont {F.}~\bibnamefont {Jahnke}},
  \bibinfo {author} {\bibfnamefont {J.}~\bibnamefont {Wiersig}}, \ and\
  \bibinfo {author} {\bibfnamefont {C.}~\bibnamefont {Gies}},\ }\href {\doibase
  10.1103/PhysRevApplied.4.044018} {\bibfield  {journal} {\bibinfo  {journal}
  {Phys. Rev. Appl.}\ }\textbf {\bibinfo {volume} {4}},\ \bibinfo {pages}
  {044018} (\bibinfo {year} {2015})}\BibitemShut {NoStop}%
\bibitem [{\citenamefont {Jahnke}\ \emph {et~al.}(2016)\citenamefont {Jahnke},
  \citenamefont {Gies}, \citenamefont {Aßmann}, \citenamefont {Bayer},
  \citenamefont {Leymann}, \citenamefont {Foerster}, \citenamefont {Wiersig},
  \citenamefont {Schneider}, \citenamefont {Kamp},\ and\ \citenamefont
  {Höfling}}]{Jahnke_NatCom_2016}%
  \BibitemOpen
  \bibfield  {author} {\bibinfo {author} {\bibfnamefont {F.}~\bibnamefont
  {Jahnke}}, \bibinfo {author} {\bibfnamefont {C.}~\bibnamefont {Gies}},
  \bibinfo {author} {\bibfnamefont {M.}~\bibnamefont {Aßmann}}, \bibinfo
  {author} {\bibfnamefont {M.}~\bibnamefont {Bayer}}, \bibinfo {author}
  {\bibfnamefont {H.~A.~M.}\ \bibnamefont {Leymann}}, \bibinfo {author}
  {\bibfnamefont {A.}~\bibnamefont {Foerster}}, \bibinfo {author}
  {\bibfnamefont {J.}~\bibnamefont {Wiersig}}, \bibinfo {author} {\bibfnamefont
  {C.}~\bibnamefont {Schneider}}, \bibinfo {author} {\bibfnamefont
  {M.}~\bibnamefont {Kamp}}, \ and\ \bibinfo {author} {\bibfnamefont
  {S.}~\bibnamefont {Höfling}},\ }\href {\doibase 10.1038/ncomms11540}
  {\bibfield  {journal} {\bibinfo  {journal} {Nature Communications}\ }\textbf
  {\bibinfo {volume} {7}},\ \bibinfo {pages} {11540} (\bibinfo {year}
  {2016})}\BibitemShut {NoStop}%
\bibitem [{\citenamefont {Auffèves}\ \emph {et~al.}(2011)\citenamefont
  {Auffèves}, \citenamefont {Gerace}, \citenamefont {Portolan}, \citenamefont
  {Drezet},\ and\ \citenamefont {Santos}}]{Auffèves_NewJournPhys_2011}%
  \BibitemOpen
  \bibfield  {author} {\bibinfo {author} {\bibfnamefont {A.}~\bibnamefont
  {Auffèves}}, \bibinfo {author} {\bibfnamefont {D.}~\bibnamefont {Gerace}},
  \bibinfo {author} {\bibfnamefont {S.}~\bibnamefont {Portolan}}, \bibinfo
  {author} {\bibfnamefont {A.}~\bibnamefont {Drezet}}, \ and\ \bibinfo {author}
  {\bibfnamefont {M.~F.}\ \bibnamefont {Santos}},\ }\href {\doibase
  10.1088/1367-2630/13/9/093020} {\bibfield  {journal} {\bibinfo  {journal}
  {New Journal of Physics}\ }\textbf {\bibinfo {volume} {13}},\ \bibinfo
  {pages} {093020} (\bibinfo {year} {2011})}\BibitemShut {NoStop}%
\bibitem [{\citenamefont {Mørk}\ and\ \citenamefont
  {Lippi}(2018)}]{Mørk_APL_2018}%
  \BibitemOpen
  \bibfield  {author} {\bibinfo {author} {\bibfnamefont {J.}~\bibnamefont
  {Mørk}}\ and\ \bibinfo {author} {\bibfnamefont {G.~L.}\ \bibnamefont
  {Lippi}},\ }\href {\doibase 10.1063/1.5022958} {\bibfield  {journal}
  {\bibinfo  {journal} {Applied Physics Letters}\ }\textbf {\bibinfo {volume}
  {112}},\ \bibinfo {pages} {141103} (\bibinfo {year} {2018})}\BibitemShut
  {NoStop}%
\bibitem [{\citenamefont {Bundgaard-Nielsen}\ \emph {et~al.}(2023)\citenamefont
  {Bundgaard-Nielsen}, \citenamefont {Denning}, \citenamefont {Saldutti},\ and\
  \citenamefont {M\o{}rk}}]{Bundgaard_PRL_2023}%
  \BibitemOpen
  \bibfield  {author} {\bibinfo {author} {\bibfnamefont {M.}~\bibnamefont
  {Bundgaard-Nielsen}}, \bibinfo {author} {\bibfnamefont {E.~V.}\ \bibnamefont
  {Denning}}, \bibinfo {author} {\bibfnamefont {M.}~\bibnamefont {Saldutti}}, \
  and\ \bibinfo {author} {\bibfnamefont {J.}~\bibnamefont {M\o{}rk}},\ }\href
  {\doibase 10.1103/PhysRevLett.130.253801} {\bibfield  {journal} {\bibinfo
  {journal} {Phys. Rev. Lett.}\ }\textbf {\bibinfo {volume} {130}},\ \bibinfo
  {pages} {253801} (\bibinfo {year} {2023})}\BibitemShut {NoStop}%
\bibitem [{\citenamefont {Hu}\ and\ \citenamefont
  {Weiss}(2016)}]{Hu_ACSPhot_2016}%
  \BibitemOpen
  \bibfield  {author} {\bibinfo {author} {\bibfnamefont {S.}~\bibnamefont
  {Hu}}\ and\ \bibinfo {author} {\bibfnamefont {S.~M.}\ \bibnamefont {Weiss}},\
  }\href {\doibase 10.1021/acsphotonics.6b00219} {\bibfield  {journal}
  {\bibinfo  {journal} {ACS Photonics}\ }\textbf {\bibinfo {volume} {3}},\
  \bibinfo {pages} {1647} (\bibinfo {year} {2016})}\BibitemShut {NoStop}%
\bibitem [{\citenamefont {Choi}\ \emph {et~al.}(2017)\citenamefont {Choi},
  \citenamefont {Heuck},\ and\ \citenamefont {Englund}}]{Choi_PRL_2017}%
  \BibitemOpen
  \bibfield  {author} {\bibinfo {author} {\bibfnamefont {H.}~\bibnamefont
  {Choi}}, \bibinfo {author} {\bibfnamefont {M.}~\bibnamefont {Heuck}}, \ and\
  \bibinfo {author} {\bibfnamefont {D.}~\bibnamefont {Englund}},\ }\href
  {\doibase 10.1103/PhysRevLett.118.223605} {\bibfield  {journal} {\bibinfo
  {journal} {Phys. Rev. Lett.}\ }\textbf {\bibinfo {volume} {118}},\ \bibinfo
  {pages} {223605} (\bibinfo {year} {2017})}\BibitemShut {NoStop}%
\bibitem [{\citenamefont {Wang}\ \emph {et~al.}(2018)\citenamefont {Wang},
  \citenamefont {Christiansen}, \citenamefont {Yu}, \citenamefont {Mørk},\
  and\ \citenamefont {Sigmund}}]{Wang_APL_2018}%
  \BibitemOpen
  \bibfield  {author} {\bibinfo {author} {\bibfnamefont {F.}~\bibnamefont
  {Wang}}, \bibinfo {author} {\bibfnamefont {R.~E.}\ \bibnamefont
  {Christiansen}}, \bibinfo {author} {\bibfnamefont {Y.}~\bibnamefont {Yu}},
  \bibinfo {author} {\bibfnamefont {J.}~\bibnamefont {Mørk}}, \ and\ \bibinfo
  {author} {\bibfnamefont {O.}~\bibnamefont {Sigmund}},\ }\href {\doibase
  10.1063/1.5064468} {\bibfield  {journal} {\bibinfo  {journal} {Applied
  Physics Letters}\ }\textbf {\bibinfo {volume} {113}},\ \bibinfo {pages}
  {241101} (\bibinfo {year} {2018})}\BibitemShut {NoStop}%
\bibitem [{\citenamefont {Albrechtsen}\ \emph
  {et~al.}(2022{\natexlab{a}})\citenamefont {Albrechtsen}, \citenamefont
  {Vosoughi~Lahijani}, \citenamefont {Christiansen}, \citenamefont {Nguyen},
  \citenamefont {Casses}, \citenamefont {Hansen}, \citenamefont {Stenger},
  \citenamefont {Sigmund}, \citenamefont {Jansen}, \citenamefont {Mørk},\ and\
  \citenamefont {Stobbe}}]{Albrechtsen_NatComm_2022}%
  \BibitemOpen
  \bibfield  {author} {\bibinfo {author} {\bibfnamefont {M.}~\bibnamefont
  {Albrechtsen}}, \bibinfo {author} {\bibfnamefont {B.}~\bibnamefont
  {Vosoughi~Lahijani}}, \bibinfo {author} {\bibfnamefont {R.~E.}\ \bibnamefont
  {Christiansen}}, \bibinfo {author} {\bibfnamefont {V.~T.~H.}\ \bibnamefont
  {Nguyen}}, \bibinfo {author} {\bibfnamefont {L.~N.}\ \bibnamefont {Casses}},
  \bibinfo {author} {\bibfnamefont {S.~E.}\ \bibnamefont {Hansen}}, \bibinfo
  {author} {\bibfnamefont {N.}~\bibnamefont {Stenger}}, \bibinfo {author}
  {\bibfnamefont {O.}~\bibnamefont {Sigmund}}, \bibinfo {author} {\bibfnamefont
  {H.}~\bibnamefont {Jansen}}, \bibinfo {author} {\bibfnamefont
  {J.}~\bibnamefont {Mørk}}, \ and\ \bibinfo {author} {\bibfnamefont
  {S.}~\bibnamefont {Stobbe}},\ }\href {\doibase 10.1038/s41467-022-33874-w}
  {\bibfield  {journal} {\bibinfo  {journal} {Nature Communications}\ }\textbf
  {\bibinfo {volume} {13}},\ \bibinfo {pages} {6281} (\bibinfo {year}
  {2022}{\natexlab{a}})}\BibitemShut {NoStop}%
\bibitem [{\citenamefont {Albrechtsen}\ \emph
  {et~al.}(2022{\natexlab{b}})\citenamefont {Albrechtsen}, \citenamefont
  {Lahijani},\ and\ \citenamefont {Stobbe}}]{Albrechtsen_OptExpress_2022}%
  \BibitemOpen
  \bibfield  {author} {\bibinfo {author} {\bibfnamefont {M.}~\bibnamefont
  {Albrechtsen}}, \bibinfo {author} {\bibfnamefont {B.~V.}\ \bibnamefont
  {Lahijani}}, \ and\ \bibinfo {author} {\bibfnamefont {S.}~\bibnamefont
  {Stobbe}},\ }\href {\doibase 10.1364/OE.448929} {\bibfield  {journal}
  {\bibinfo  {journal} {Opt. Express}\ }\textbf {\bibinfo {volume} {30}},\
  \bibinfo {pages} {15458} (\bibinfo {year} {2022}{\natexlab{b}})}\BibitemShut
  {NoStop}%
\bibitem [{\citenamefont {Kountouris}\ \emph {et~al.}(2022)\citenamefont
  {Kountouris}, \citenamefont {M{\o}rk}, \citenamefont {Denning},\ and\
  \citenamefont {Kristensen}}]{Kountouris-OptExress-2022}%
  \BibitemOpen
  \bibfield  {author} {\bibinfo {author} {\bibfnamefont {G.}~\bibnamefont
  {Kountouris}}, \bibinfo {author} {\bibfnamefont {J.}~\bibnamefont {M{\o}rk}},
  \bibinfo {author} {\bibfnamefont {E.~V.}\ \bibnamefont {Denning}}, \ and\
  \bibinfo {author} {\bibfnamefont {P.~T.}\ \bibnamefont {Kristensen}},\ }\href
  {\doibase 10.1364/OE.472793} {\bibfield  {journal} {\bibinfo  {journal} {Opt.
  Express}\ }\textbf {\bibinfo {volume} {30}},\ \bibinfo {pages} {40367}
  (\bibinfo {year} {2022})}\BibitemShut {NoStop}%
\bibitem [{\citenamefont {Kristensen}\ \emph {et~al.}(2012)\citenamefont
  {Kristensen}, \citenamefont {Vlack},\ and\ \citenamefont
  {Hughes}}]{Kristensen_OptLett_2012}%
  \BibitemOpen
  \bibfield  {author} {\bibinfo {author} {\bibfnamefont {P.~T.}\ \bibnamefont
  {Kristensen}}, \bibinfo {author} {\bibfnamefont {C.~V.}\ \bibnamefont
  {Vlack}}, \ and\ \bibinfo {author} {\bibfnamefont {S.}~\bibnamefont
  {Hughes}},\ }\href {\doibase 10.1364/OL.37.001649} {\bibfield  {journal}
  {\bibinfo  {journal} {Opt. Lett.}\ }\textbf {\bibinfo {volume} {37}},\
  \bibinfo {pages} {1649} (\bibinfo {year} {2012})}\BibitemShut {NoStop}%
\bibitem [{\citenamefont {Sauvan}\ \emph {et~al.}(2013)\citenamefont {Sauvan},
  \citenamefont {Hugonin}, \citenamefont {Maksymov},\ and\ \citenamefont
  {Lalanne}}]{Sauvan_PRL_2013}%
  \BibitemOpen
  \bibfield  {author} {\bibinfo {author} {\bibfnamefont {C.}~\bibnamefont
  {Sauvan}}, \bibinfo {author} {\bibfnamefont {J.~P.}\ \bibnamefont {Hugonin}},
  \bibinfo {author} {\bibfnamefont {I.~S.}\ \bibnamefont {Maksymov}}, \ and\
  \bibinfo {author} {\bibfnamefont {P.}~\bibnamefont {Lalanne}},\ }\href
  {\doibase 10.1103/PhysRevLett.110.237401} {\bibfield  {journal} {\bibinfo
  {journal} {Phys. Rev. Lett.}\ }\textbf {\bibinfo {volume} {110}},\ \bibinfo
  {pages} {237401} (\bibinfo {year} {2013})}\BibitemShut {NoStop}%
\bibitem [{\citenamefont {Coccioli}\ \emph {et~al.}(1998)\citenamefont
  {Coccioli}, \citenamefont {Boroditsky}, \citenamefont {Kim}, \citenamefont
  {Rahmat-Samii},\ and\ \citenamefont {Yablonovitch}}]{Coccioli_1998}%
  \BibitemOpen
  \bibfield  {author} {\bibinfo {author} {\bibfnamefont {R.}~\bibnamefont
  {Coccioli}}, \bibinfo {author} {\bibfnamefont {M.}~\bibnamefont
  {Boroditsky}}, \bibinfo {author} {\bibfnamefont {K.}~\bibnamefont {Kim}},
  \bibinfo {author} {\bibfnamefont {Y.}~\bibnamefont {Rahmat-Samii}}, \ and\
  \bibinfo {author} {\bibfnamefont {E.}~\bibnamefont {Yablonovitch}},\ }\href
  {\doibase 10.1049/ip-opt:19982468} {\bibfield  {journal} {\bibinfo  {journal}
  {IEE Proceedings - Optoelectronics}\ }\textbf {\bibinfo {volume} {145}},\
  \bibinfo {pages} {391} (\bibinfo {year} {1998})}\BibitemShut {NoStop}%
\bibitem [{\citenamefont {Khurgin}(2015)}]{Khurgin_NatNanotech_2015}%
  \BibitemOpen
  \bibfield  {author} {\bibinfo {author} {\bibfnamefont {J.~B.}\ \bibnamefont
  {Khurgin}},\ }\href {\doibase 10.1038/nnano.2014.310} {\bibfield  {journal}
  {\bibinfo  {journal} {Nature Nanotechnology}\ }\textbf {\bibinfo {volume}
  {10}},\ \bibinfo {pages} {2} (\bibinfo {year} {2015})}\BibitemShut {NoStop}%
\bibitem [{\citenamefont {Bozhevolnyi}\ and\ \citenamefont
  {Khurgin}(2016)}]{Bozhevolnyi_Optica_2016}%
  \BibitemOpen
  \bibfield  {author} {\bibinfo {author} {\bibfnamefont {S.~I.}\ \bibnamefont
  {Bozhevolnyi}}\ and\ \bibinfo {author} {\bibfnamefont {J.~B.}\ \bibnamefont
  {Khurgin}},\ }\href {\doibase 10.1364/OPTICA.3.001418} {\bibfield  {journal}
  {\bibinfo  {journal} {Optica}\ }\textbf {\bibinfo {volume} {3}},\ \bibinfo
  {pages} {1418} (\bibinfo {year} {2016})}\BibitemShut {NoStop}%
\bibitem [{\citenamefont {Andr\'{e}}\ \emph {et~al.}(2020)\citenamefont
  {Andr\'{e}}, \citenamefont {M{\o}rk},\ and\ \citenamefont
  {Wubs}}]{Andre_OptExpress_2020}%
  \BibitemOpen
  \bibfield  {author} {\bibinfo {author} {\bibfnamefont {E.~C.}\ \bibnamefont
  {Andr\'{e}}}, \bibinfo {author} {\bibfnamefont {J.}~\bibnamefont {M{\o}rk}},
  \ and\ \bibinfo {author} {\bibfnamefont {M.}~\bibnamefont {Wubs}},\ }\href
  {\doibase 10.1364/OE.405979} {\bibfield  {journal} {\bibinfo  {journal} {Opt.
  Express}\ }\textbf {\bibinfo {volume} {28}},\ \bibinfo {pages} {32632}
  (\bibinfo {year} {2020})}\BibitemShut {NoStop}%
\bibitem [{\citenamefont {Notomi}(2010)}]{Notomi_RepProgrPhys_2010}%
  \BibitemOpen
  \bibfield  {author} {\bibinfo {author} {\bibfnamefont {M.}~\bibnamefont
  {Notomi}},\ }\href {\doibase 10.1088/0034-4885/73/9/096501} {\bibfield
  {journal} {\bibinfo  {journal} {Reports on Progress in Physics}\ }\textbf
  {\bibinfo {volume} {73}},\ \bibinfo {pages} {096501} (\bibinfo {year}
  {2010})}\BibitemShut {NoStop}%
\bibitem [{\citenamefont {Maier}(2007)}]{Maier_book_2007}%
  \BibitemOpen
  \bibfield  {author} {\bibinfo {author} {\bibfnamefont {S.~A.}\ \bibnamefont
  {Maier}},\ }\href@noop {} {\emph {\bibinfo {title} {Plasmonics: Fundamentals
  and Applications}}}\ (\bibinfo  {publisher} {Springer New York, NY},\
  \bibinfo {year} {2007})\BibitemShut {NoStop}%
\bibitem [{\citenamefont {Wang}\ and\ \citenamefont
  {Shen}(2006)}]{Wang_PRL_2006}%
  \BibitemOpen
  \bibfield  {author} {\bibinfo {author} {\bibfnamefont {F.}~\bibnamefont
  {Wang}}\ and\ \bibinfo {author} {\bibfnamefont {Y.~R.}\ \bibnamefont
  {Shen}},\ }\href {\doibase 10.1103/PhysRevLett.97.206806} {\bibfield
  {journal} {\bibinfo  {journal} {Phys. Rev. Lett.}\ }\textbf {\bibinfo
  {volume} {97}},\ \bibinfo {pages} {206806} (\bibinfo {year}
  {2006})}\BibitemShut {NoStop}%
\bibitem [{\citenamefont {Saldutti}\ \emph {et~al.}(2021)\citenamefont
  {Saldutti}, \citenamefont {Xiong}, \citenamefont {Dimopoulos}, \citenamefont
  {Yu}, \citenamefont {Gioannini},\ and\ \citenamefont
  {Mørk}}]{Saldutti_Nanomat_2021}%
  \BibitemOpen
  \bibfield  {author} {\bibinfo {author} {\bibfnamefont {M.}~\bibnamefont
  {Saldutti}}, \bibinfo {author} {\bibfnamefont {M.}~\bibnamefont {Xiong}},
  \bibinfo {author} {\bibfnamefont {E.}~\bibnamefont {Dimopoulos}}, \bibinfo
  {author} {\bibfnamefont {Y.}~\bibnamefont {Yu}}, \bibinfo {author}
  {\bibfnamefont {M.}~\bibnamefont {Gioannini}}, \ and\ \bibinfo {author}
  {\bibfnamefont {J.}~\bibnamefont {Mørk}},\ }\href {\doibase
  10.3390/nano11113030} {\bibfield  {journal} {\bibinfo  {journal}
  {Nanomaterials}\ }\textbf {\bibinfo {volume} {11}} (\bibinfo {year} {2021}),\
  10.3390/nano11113030}\BibitemShut {NoStop}%
\bibitem [{\citenamefont {Asano}\ \emph {et~al.}(2017)\citenamefont {Asano},
  \citenamefont {Ochi}, \citenamefont {Takahashi}, \citenamefont {Kishimoto},\
  and\ \citenamefont {Noda}}]{Asano_OptExpress_2017}%
  \BibitemOpen
  \bibfield  {author} {\bibinfo {author} {\bibfnamefont {T.}~\bibnamefont
  {Asano}}, \bibinfo {author} {\bibfnamefont {Y.}~\bibnamefont {Ochi}},
  \bibinfo {author} {\bibfnamefont {Y.}~\bibnamefont {Takahashi}}, \bibinfo
  {author} {\bibfnamefont {K.}~\bibnamefont {Kishimoto}}, \ and\ \bibinfo
  {author} {\bibfnamefont {S.}~\bibnamefont {Noda}},\ }\href {\doibase
  10.1364/OE.25.001769} {\bibfield  {journal} {\bibinfo  {journal} {Opt.
  Express}\ }\textbf {\bibinfo {volume} {25}},\ \bibinfo {pages} {1769}
  (\bibinfo {year} {2017})}\BibitemShut {NoStop}%
\bibitem [{\citenamefont {Zhang}\ and\ \citenamefont
  {Qiu}(2004)}]{Zhang_OE_2004}%
  \BibitemOpen
  \bibfield  {author} {\bibinfo {author} {\bibfnamefont {Z.}~\bibnamefont
  {Zhang}}\ and\ \bibinfo {author} {\bibfnamefont {M.}~\bibnamefont {Qiu}},\
  }\href {\doibase 10.1364/OPEX.12.003988} {\bibfield  {journal} {\bibinfo
  {journal} {Opt. Express}\ }\textbf {\bibinfo {volume} {12}},\ \bibinfo
  {pages} {3988} (\bibinfo {year} {2004})}\BibitemShut {NoStop}%
\bibitem [{\citenamefont {Saldutti}\ \emph {et~al.}(2022)\citenamefont
  {Saldutti}, \citenamefont {Yu}, \citenamefont {Kristensen}, \citenamefont
  {Kountouris},\ and\ \citenamefont {Mørk}}]{Saldutti_IEEE_2022}%
  \BibitemOpen
  \bibfield  {author} {\bibinfo {author} {\bibfnamefont {M.}~\bibnamefont
  {Saldutti}}, \bibinfo {author} {\bibfnamefont {Y.}~\bibnamefont {Yu}},
  \bibinfo {author} {\bibfnamefont {P.~T.}\ \bibnamefont {Kristensen}},
  \bibinfo {author} {\bibfnamefont {G.}~\bibnamefont {Kountouris}}, \ and\
  \bibinfo {author} {\bibfnamefont {J.}~\bibnamefont {Mørk}},\ }in\ \href
  {\doibase 10.1109/IPC53466.2022.9975589} {\emph {\bibinfo {booktitle} {2022
  IEEE Photonics Conference (IPC)}}}\ (\bibinfo {year} {2022})\ pp.\ \bibinfo
  {pages} {1--2}\BibitemShut {NoStop}%
\bibitem [{\citenamefont {Mork}\ and\ \citenamefont
  {Yvind}(2020)}]{Mørk_Optica_2020}%
  \BibitemOpen
  \bibfield  {author} {\bibinfo {author} {\bibfnamefont {J.}~\bibnamefont
  {Mork}}\ and\ \bibinfo {author} {\bibfnamefont {K.}~\bibnamefont {Yvind}},\
  }\href {\doibase 10.1364/OPTICA.402190} {\bibfield  {journal} {\bibinfo
  {journal} {Optica}\ }\textbf {\bibinfo {volume} {7}},\ \bibinfo {pages}
  {1641} (\bibinfo {year} {2020})}\BibitemShut {NoStop}%
\bibitem [{\citenamefont {Jackson}(1999)}]{Jackson-book3rdEd-1999}%
  \BibitemOpen
  \bibfield  {author} {\bibinfo {author} {\bibfnamefont {J.~D.}\ \bibnamefont
  {Jackson}},\ }\href@noop {} {\emph {\bibinfo {title} {Classical
  electrodynamics}}},\ \bibinfo {edition} {3rd}\ ed.\ (\bibinfo  {publisher}
  {John Wiley \& Sons, Inc.},\ \bibinfo {year} {1999})\BibitemShut {NoStop}%
\bibitem [{\citenamefont {Romeira}\ and\ \citenamefont
  {Fiore}(2018)}]{Romeira_JQE_2018}%
  \BibitemOpen
  \bibfield  {author} {\bibinfo {author} {\bibfnamefont {B.}~\bibnamefont
  {Romeira}}\ and\ \bibinfo {author} {\bibfnamefont {A.}~\bibnamefont
  {Fiore}},\ }\href {\doibase 10.1109/JQE.2018.2802464} {\bibfield  {journal}
  {\bibinfo  {journal} {IEEE Journal of Quantum Electronics}\ }\textbf
  {\bibinfo {volume} {54}},\ \bibinfo {pages} {1} (\bibinfo {year}
  {2018})}\BibitemShut {NoStop}%
\bibitem [{\citenamefont {Nomura}\ \emph {et~al.}(2009)\citenamefont {Nomura},
  \citenamefont {Kumagai}, \citenamefont {Iwamoto}, \citenamefont {Ota},\ and\
  \citenamefont {Arakawa}}]{Nomura_OptExpress_2009}%
  \BibitemOpen
  \bibfield  {author} {\bibinfo {author} {\bibfnamefont {M.}~\bibnamefont
  {Nomura}}, \bibinfo {author} {\bibfnamefont {N.}~\bibnamefont {Kumagai}},
  \bibinfo {author} {\bibfnamefont {S.}~\bibnamefont {Iwamoto}}, \bibinfo
  {author} {\bibfnamefont {Y.}~\bibnamefont {Ota}}, \ and\ \bibinfo {author}
  {\bibfnamefont {Y.}~\bibnamefont {Arakawa}},\ }\href {\doibase
  10.1364/OE.17.015975} {\bibfield  {journal} {\bibinfo  {journal} {Opt.
  Express}\ }\textbf {\bibinfo {volume} {17}},\ \bibinfo {pages} {15975}
  (\bibinfo {year} {2009})}\BibitemShut {NoStop}%
\bibitem [{\citenamefont {G{\'e}rard}(2003)}]{Gérard_Springer_2003}%
  \BibitemOpen
  \bibfield  {author} {\bibinfo {author} {\bibfnamefont {J.-M.}\ \bibnamefont
  {G{\'e}rard}},\ }\enquote {\bibinfo {title} {Solid-state cavity-quantum
  electrodynamics with self-assembled quantum dots},}\ in\ \href {\doibase
  10.1007/978-3-540-39180-7_7} {\emph {\bibinfo {booktitle} {Single Quantum
  Dots: Fundamentals, Applications, and New Concepts}}}\ (\bibinfo  {publisher}
  {Springer Berlin Heidelberg},\ \bibinfo {address} {Berlin, Heidelberg},\
  \bibinfo {year} {2003})\ pp.\ \bibinfo {pages} {269--314}\BibitemShut
  {NoStop}%
\bibitem [{\citenamefont {Lalanne}\ \emph {et~al.}(2018)\citenamefont
  {Lalanne}, \citenamefont {Yan}, \citenamefont {Vynck}, \citenamefont
  {Sauvan},\ and\ \citenamefont {Hugonin}}]{Lalanne_LPR_2018}%
  \BibitemOpen
  \bibfield  {author} {\bibinfo {author} {\bibfnamefont {P.}~\bibnamefont
  {Lalanne}}, \bibinfo {author} {\bibfnamefont {W.}~\bibnamefont {Yan}},
  \bibinfo {author} {\bibfnamefont {K.}~\bibnamefont {Vynck}}, \bibinfo
  {author} {\bibfnamefont {C.}~\bibnamefont {Sauvan}}, \ and\ \bibinfo {author}
  {\bibfnamefont {J.-P.}\ \bibnamefont {Hugonin}},\ }\href {\doibase
  https://doi.org/10.1002/lpor.201700113} {\bibfield  {journal} {\bibinfo
  {journal} {Laser \& Photonics Reviews}\ }\textbf {\bibinfo {volume} {12}},\
  \bibinfo {pages} {1700113} (\bibinfo {year} {2018})}\BibitemShut {NoStop}%
\bibitem [{\citenamefont {Kristensen}\ \emph {et~al.}(2020)\citenamefont
  {Kristensen}, \citenamefont {Herrmann}, \citenamefont {Intravaia},\ and\
  \citenamefont {Busch}}]{Kristensen_AdvOptPhot_2020}%
  \BibitemOpen
  \bibfield  {author} {\bibinfo {author} {\bibfnamefont {P.~T.}\ \bibnamefont
  {Kristensen}}, \bibinfo {author} {\bibfnamefont {K.}~\bibnamefont
  {Herrmann}}, \bibinfo {author} {\bibfnamefont {F.}~\bibnamefont {Intravaia}},
  \ and\ \bibinfo {author} {\bibfnamefont {K.}~\bibnamefont {Busch}},\ }\href
  {\doibase 10.1364/AOP.377940} {\bibfield  {journal} {\bibinfo  {journal}
  {Adv. Opt. Photon.}\ }\textbf {\bibinfo {volume} {12}},\ \bibinfo {pages}
  {612} (\bibinfo {year} {2020})}\BibitemShut {NoStop}%
\bibitem [{\citenamefont {Jensen}\ and\ \citenamefont
  {Sigmund}(2011)}]{Jensen-Sigmnud_LPR_2011}%
  \BibitemOpen
  \bibfield  {author} {\bibinfo {author} {\bibfnamefont {J.}~\bibnamefont
  {Jensen}}\ and\ \bibinfo {author} {\bibfnamefont {O.}~\bibnamefont
  {Sigmund}},\ }\href {\doibase https://doi.org/10.1002/lpor.201000014}
  {\bibfield  {journal} {\bibinfo  {journal} {Laser \& Photonics Reviews}\
  }\textbf {\bibinfo {volume} {5}},\ \bibinfo {pages} {308} (\bibinfo {year}
  {2011})}\BibitemShut {NoStop}%
\bibitem [{\citenamefont {Kuramochi}\ \emph {et~al.}(2018)\citenamefont
  {Kuramochi}, \citenamefont {Duprez}, \citenamefont {Kim}, \citenamefont
  {Takiguchi}, \citenamefont {Takeda}, \citenamefont {Fujii}, \citenamefont
  {Nozaki}, \citenamefont {Shinya}, \citenamefont {Sumikura}, \citenamefont
  {Taniyama}, \citenamefont {Matsuo},\ and\ \citenamefont
  {Notomi}}]{Kuramochi_OptExpress_2018}%
  \BibitemOpen
  \bibfield  {author} {\bibinfo {author} {\bibfnamefont {E.}~\bibnamefont
  {Kuramochi}}, \bibinfo {author} {\bibfnamefont {H.}~\bibnamefont {Duprez}},
  \bibinfo {author} {\bibfnamefont {J.}~\bibnamefont {Kim}}, \bibinfo {author}
  {\bibfnamefont {M.}~\bibnamefont {Takiguchi}}, \bibinfo {author}
  {\bibfnamefont {K.}~\bibnamefont {Takeda}}, \bibinfo {author} {\bibfnamefont
  {T.}~\bibnamefont {Fujii}}, \bibinfo {author} {\bibfnamefont
  {K.}~\bibnamefont {Nozaki}}, \bibinfo {author} {\bibfnamefont
  {A.}~\bibnamefont {Shinya}}, \bibinfo {author} {\bibfnamefont
  {H.}~\bibnamefont {Sumikura}}, \bibinfo {author} {\bibfnamefont
  {H.}~\bibnamefont {Taniyama}}, \bibinfo {author} {\bibfnamefont
  {S.}~\bibnamefont {Matsuo}}, \ and\ \bibinfo {author} {\bibfnamefont
  {M.}~\bibnamefont {Notomi}},\ }\href {\doibase 10.1364/OE.26.026598}
  {\bibfield  {journal} {\bibinfo  {journal} {Opt. Express}\ }\textbf {\bibinfo
  {volume} {26}},\ \bibinfo {pages} {26598} (\bibinfo {year}
  {2018})}\BibitemShut {NoStop}%
\bibitem [{\citenamefont {Dimopoulos}\ \emph {et~al.}()\citenamefont
  {Dimopoulos}, \citenamefont {Sakanas}, \citenamefont {Marchevsky},
  \citenamefont {Xiong}, \citenamefont {Yu}, \citenamefont {Semenova},
  \citenamefont {Mørk},\ and\ \citenamefont {Yvind}}]{Dimopoulos_LPR_2022}%
  \BibitemOpen
  \bibfield  {author} {\bibinfo {author} {\bibfnamefont {E.}~\bibnamefont
  {Dimopoulos}}, \bibinfo {author} {\bibfnamefont {A.}~\bibnamefont {Sakanas}},
  \bibinfo {author} {\bibfnamefont {A.}~\bibnamefont {Marchevsky}}, \bibinfo
  {author} {\bibfnamefont {M.}~\bibnamefont {Xiong}}, \bibinfo {author}
  {\bibfnamefont {Y.}~\bibnamefont {Yu}}, \bibinfo {author} {\bibfnamefont
  {E.}~\bibnamefont {Semenova}}, \bibinfo {author} {\bibfnamefont
  {J.}~\bibnamefont {Mørk}}, \ and\ \bibinfo {author} {\bibfnamefont
  {K.}~\bibnamefont {Yvind}},\ }\href {\doibase
  https://doi.org/10.1002/lpor.202200109} {\bibfield  {journal} {\bibinfo
  {journal} {Laser \& Photonics Reviews}\ }\textbf {\bibinfo {volume} {n/a}},\
  \bibinfo {pages} {2200109}}\BibitemShut {NoStop}%
\bibitem [{\citenamefont {Lorke}\ \emph {et~al.}(2013)\citenamefont {Lorke},
  \citenamefont {Suhr}, \citenamefont {Gregersen},\ and\ \citenamefont
  {M\o{}rk}}]{Lorke_PRB_2013}%
  \BibitemOpen
  \bibfield  {author} {\bibinfo {author} {\bibfnamefont {M.}~\bibnamefont
  {Lorke}}, \bibinfo {author} {\bibfnamefont {T.}~\bibnamefont {Suhr}},
  \bibinfo {author} {\bibfnamefont {N.}~\bibnamefont {Gregersen}}, \ and\
  \bibinfo {author} {\bibfnamefont {J.}~\bibnamefont {M\o{}rk}},\ }\href
  {\doibase 10.1103/PhysRevB.87.205310} {\bibfield  {journal} {\bibinfo
  {journal} {Phys. Rev. B}\ }\textbf {\bibinfo {volume} {87}},\ \bibinfo
  {pages} {205310} (\bibinfo {year} {2013})}\BibitemShut {NoStop}%
\bibitem [{\citenamefont {Moelbjerg}\ \emph {et~al.}(2013)\citenamefont
  {Moelbjerg}, \citenamefont {Kaer}, \citenamefont {Lorke}, \citenamefont
  {Tromborg},\ and\ \citenamefont {Mørk}}]{Moelbjerg_JQE_2013}%
  \BibitemOpen
  \bibfield  {author} {\bibinfo {author} {\bibfnamefont {A.}~\bibnamefont
  {Moelbjerg}}, \bibinfo {author} {\bibfnamefont {P.}~\bibnamefont {Kaer}},
  \bibinfo {author} {\bibfnamefont {M.}~\bibnamefont {Lorke}}, \bibinfo
  {author} {\bibfnamefont {B.}~\bibnamefont {Tromborg}}, \ and\ \bibinfo
  {author} {\bibfnamefont {J.}~\bibnamefont {Mørk}},\ }\href {\doibase
  10.1109/JQE.2013.2282464} {\bibfield  {journal} {\bibinfo  {journal} {IEEE
  Journal of Quantum Electronics}\ }\textbf {\bibinfo {volume} {49}},\ \bibinfo
  {pages} {945} (\bibinfo {year} {2013})}\BibitemShut {NoStop}%
\bibitem [{\citenamefont {Gregersen}\ \emph {et~al.}(2012)\citenamefont
  {Gregersen}, \citenamefont {Suhr}, \citenamefont {Lorke},\ and\ \citenamefont
  {Mørk}}]{Gregersen_APL_2012}%
  \BibitemOpen
  \bibfield  {author} {\bibinfo {author} {\bibfnamefont {N.}~\bibnamefont
  {Gregersen}}, \bibinfo {author} {\bibfnamefont {T.}~\bibnamefont {Suhr}},
  \bibinfo {author} {\bibfnamefont {M.}~\bibnamefont {Lorke}}, \ and\ \bibinfo
  {author} {\bibfnamefont {J.}~\bibnamefont {Mørk}},\ }\href {\doibase
  10.1063/1.3697702} {\bibfield  {journal} {\bibinfo  {journal} {Applied
  Physics Letters}\ }\textbf {\bibinfo {volume} {100}},\ \bibinfo {pages}
  {131107} (\bibinfo {year} {2012})}\BibitemShut {NoStop}%
\bibitem [{\citenamefont {Ota}\ \emph {et~al.}(2017)\citenamefont {Ota},
  \citenamefont {Kakuda}, \citenamefont {Watanabe}, \citenamefont {Iwamoto},\
  and\ \citenamefont {Arakawa}}]{Ota_OptExp_2017}%
  \BibitemOpen
  \bibfield  {author} {\bibinfo {author} {\bibfnamefont {Y.}~\bibnamefont
  {Ota}}, \bibinfo {author} {\bibfnamefont {M.}~\bibnamefont {Kakuda}},
  \bibinfo {author} {\bibfnamefont {K.}~\bibnamefont {Watanabe}}, \bibinfo
  {author} {\bibfnamefont {S.}~\bibnamefont {Iwamoto}}, \ and\ \bibinfo
  {author} {\bibfnamefont {Y.}~\bibnamefont {Arakawa}},\ }\href {\doibase
  10.1364/OE.25.019981} {\bibfield  {journal} {\bibinfo  {journal} {Opt.
  Express}\ }\textbf {\bibinfo {volume} {25}},\ \bibinfo {pages} {19981}
  (\bibinfo {year} {2017})}\BibitemShut {NoStop}%
\bibitem [{\citenamefont {Kaganskiy}\ \emph {et~al.}(2019)\citenamefont
  {Kaganskiy}, \citenamefont {Kreinberg}, \citenamefont {Porte},\ and\
  \citenamefont {Reitzenstein}}]{Kaganskiy_Optica_2019}%
  \BibitemOpen
  \bibfield  {author} {\bibinfo {author} {\bibfnamefont {A.}~\bibnamefont
  {Kaganskiy}}, \bibinfo {author} {\bibfnamefont {S.}~\bibnamefont
  {Kreinberg}}, \bibinfo {author} {\bibfnamefont {X.}~\bibnamefont {Porte}}, \
  and\ \bibinfo {author} {\bibfnamefont {S.}~\bibnamefont {Reitzenstein}},\
  }\href {\doibase 10.1364/OPTICA.6.000404} {\bibfield  {journal} {\bibinfo
  {journal} {Optica}\ }\textbf {\bibinfo {volume} {6}},\ \bibinfo {pages} {404}
  (\bibinfo {year} {2019})}\BibitemShut {NoStop}%
\bibitem [{\citenamefont {Pan}\ \emph {et~al.}(2016)\citenamefont {Pan},
  \citenamefont {Gu}, \citenamefont {Amili}, \citenamefont {Vallini},\ and\
  \citenamefont {Fainman}}]{Pan_Optica_2016}%
  \BibitemOpen
  \bibfield  {author} {\bibinfo {author} {\bibfnamefont {S.~H.}\ \bibnamefont
  {Pan}}, \bibinfo {author} {\bibfnamefont {Q.}~\bibnamefont {Gu}}, \bibinfo
  {author} {\bibfnamefont {A.~E.}\ \bibnamefont {Amili}}, \bibinfo {author}
  {\bibfnamefont {F.}~\bibnamefont {Vallini}}, \ and\ \bibinfo {author}
  {\bibfnamefont {Y.}~\bibnamefont {Fainman}},\ }\href {\doibase
  10.1364/OPTICA.3.001260} {\bibfield  {journal} {\bibinfo  {journal} {Optica}\
  }\textbf {\bibinfo {volume} {3}},\ \bibinfo {pages} {1260} (\bibinfo {year}
  {2016})}\BibitemShut {NoStop}%
\bibitem [{\citenamefont {Jagsch}\ \emph {et~al.}(2018)\citenamefont {Jagsch},
  \citenamefont {Triviño}, \citenamefont {Lohof}, \citenamefont {Callsen},
  \citenamefont {Kalinowski}, \citenamefont {Rousseau}, \citenamefont {Barzel},
  \citenamefont {Carlin}, \citenamefont {Jahnke}, \citenamefont {Butté},
  \citenamefont {Gies}, \citenamefont {Hoffmann}, \citenamefont {Grandjean},\
  and\ \citenamefont {Reitzenstein}}]{Jagsch_NatComm_2018}%
  \BibitemOpen
  \bibfield  {author} {\bibinfo {author} {\bibfnamefont {S.~T.}\ \bibnamefont
  {Jagsch}}, \bibinfo {author} {\bibfnamefont {N.~V.}\ \bibnamefont
  {Triviño}}, \bibinfo {author} {\bibfnamefont {F.}~\bibnamefont {Lohof}},
  \bibinfo {author} {\bibfnamefont {G.}~\bibnamefont {Callsen}}, \bibinfo
  {author} {\bibfnamefont {S.}~\bibnamefont {Kalinowski}}, \bibinfo {author}
  {\bibfnamefont {I.~M.}\ \bibnamefont {Rousseau}}, \bibinfo {author}
  {\bibfnamefont {R.}~\bibnamefont {Barzel}}, \bibinfo {author} {\bibfnamefont
  {J.-F.}\ \bibnamefont {Carlin}}, \bibinfo {author} {\bibfnamefont
  {F.}~\bibnamefont {Jahnke}}, \bibinfo {author} {\bibfnamefont
  {R.}~\bibnamefont {Butté}}, \bibinfo {author} {\bibfnamefont
  {C.}~\bibnamefont {Gies}}, \bibinfo {author} {\bibfnamefont {A.}~\bibnamefont
  {Hoffmann}}, \bibinfo {author} {\bibfnamefont {N.}~\bibnamefont {Grandjean}},
  \ and\ \bibinfo {author} {\bibfnamefont {S.}~\bibnamefont {Reitzenstein}},\
  }\href {\doibase 10.1038/s41467-018-02999-2} {\bibfield  {journal} {\bibinfo
  {journal} {Nature Communications}\ }\textbf {\bibinfo {volume} {9}},\
  \bibinfo {pages} {564} (\bibinfo {year} {2018})}\BibitemShut {NoStop}%
\bibitem [{\citenamefont {Xu}\ \emph {et~al.}(2000)\citenamefont {Xu},
  \citenamefont {Lee},\ and\ \citenamefont {Yariv}}]{Xu-Yariv_PRA_2000}%
  \BibitemOpen
  \bibfield  {author} {\bibinfo {author} {\bibfnamefont {Y.}~\bibnamefont
  {Xu}}, \bibinfo {author} {\bibfnamefont {R.~K.}\ \bibnamefont {Lee}}, \ and\
  \bibinfo {author} {\bibfnamefont {A.}~\bibnamefont {Yariv}},\ }\href
  {\doibase 10.1103/PhysRevA.61.033807} {\bibfield  {journal} {\bibinfo
  {journal} {Phys. Rev. A}\ }\textbf {\bibinfo {volume} {61}},\ \bibinfo
  {pages} {033807} (\bibinfo {year} {2000})}\BibitemShut {NoStop}%
\bibitem [{\citenamefont {Mark}\ and\ \citenamefont
  {Mørk}(1992)}]{Mark_APL_1992}%
  \BibitemOpen
  \bibfield  {author} {\bibinfo {author} {\bibfnamefont {J.}~\bibnamefont
  {Mark}}\ and\ \bibinfo {author} {\bibfnamefont {J.}~\bibnamefont {Mørk}},\
  }\href {\doibase 10.1063/1.108265} {\bibfield  {journal} {\bibinfo  {journal}
  {Applied Physics Letters}\ }\textbf {\bibinfo {volume} {61}},\ \bibinfo
  {pages} {2281} (\bibinfo {year} {1992})}\BibitemShut {NoStop}%
\bibitem [{\citenamefont {Fox}(2006)}]{Fox-book-2006}%
  \BibitemOpen
  \bibfield  {author} {\bibinfo {author} {\bibfnamefont {M.}~\bibnamefont
  {Fox}},\ }\href@noop {} {\emph {\bibinfo {title} {Quantum Optics: An
  Introduction}}}\ (\bibinfo  {publisher} {Oxford University Press},\ \bibinfo
  {year} {2006})\BibitemShut {NoStop}%
\bibitem [{\citenamefont {Yamamoto}\ \emph {et~al.}(1986)\citenamefont
  {Yamamoto}, \citenamefont {Machida},\ and\ \citenamefont
  {Nilsson}}]{Yamamoto_PRA_1986}%
  \BibitemOpen
  \bibfield  {author} {\bibinfo {author} {\bibfnamefont {Y.}~\bibnamefont
  {Yamamoto}}, \bibinfo {author} {\bibfnamefont {S.}~\bibnamefont {Machida}}, \
  and\ \bibinfo {author} {\bibfnamefont {O.}~\bibnamefont {Nilsson}},\ }\href
  {\doibase 10.1103/PhysRevA.34.4025} {\bibfield  {journal} {\bibinfo
  {journal} {Phys. Rev. A}\ }\textbf {\bibinfo {volume} {34}},\ \bibinfo
  {pages} {4025} (\bibinfo {year} {1986})}\BibitemShut {NoStop}%
\bibitem [{\citenamefont {Roy-Choudhury}\ and\ \citenamefont
  {Levi}(2010)}]{Roy-Choudhury_PRA_2010}%
  \BibitemOpen
  \bibfield  {author} {\bibinfo {author} {\bibfnamefont {K.}~\bibnamefont
  {Roy-Choudhury}}\ and\ \bibinfo {author} {\bibfnamefont {A.~F.~J.}\
  \bibnamefont {Levi}},\ }\href {\doibase 10.1103/PhysRevA.81.013827}
  {\bibfield  {journal} {\bibinfo  {journal} {Phys. Rev. A}\ }\textbf {\bibinfo
  {volume} {81}},\ \bibinfo {pages} {013827} (\bibinfo {year}
  {2010})}\BibitemShut {NoStop}%
\bibitem [{\citenamefont {Gillespie}(1976)}]{Gillespie_JCP_1976}%
  \BibitemOpen
  \bibfield  {author} {\bibinfo {author} {\bibfnamefont {D.~T.}\ \bibnamefont
  {Gillespie}},\ }\href {\doibase https://doi.org/10.1016/0021-9991(76)90041-3}
  {\bibfield  {journal} {\bibinfo  {journal} {Journal of Computational
  Physics}\ }\textbf {\bibinfo {volume} {22}},\ \bibinfo {pages} {403}
  (\bibinfo {year} {1976})}\BibitemShut {NoStop}%
\bibitem [{\citenamefont {Puccioni}\ and\ \citenamefont
  {Lippi}(2015)}]{Puccioni_OptExpress_2015}%
  \BibitemOpen
  \bibfield  {author} {\bibinfo {author} {\bibfnamefont {G.~P.}\ \bibnamefont
  {Puccioni}}\ and\ \bibinfo {author} {\bibfnamefont {G.~L.}\ \bibnamefont
  {Lippi}},\ }\href {\doibase 10.1364/OE.23.002369} {\bibfield  {journal}
  {\bibinfo  {journal} {Opt. Express}\ }\textbf {\bibinfo {volume} {23}},\
  \bibinfo {pages} {2369} (\bibinfo {year} {2015})}\BibitemShut {NoStop}%
\bibitem [{\citenamefont {Gartner}(2011)}]{Gartner_PhysRevA_2011}%
  \BibitemOpen
  \bibfield  {author} {\bibinfo {author} {\bibfnamefont {P.}~\bibnamefont
  {Gartner}},\ }\href {\doibase 10.1103/PhysRevA.84.053804} {\bibfield
  {journal} {\bibinfo  {journal} {Phys. Rev. A}\ }\textbf {\bibinfo {volume}
  {84}},\ \bibinfo {pages} {053804} (\bibinfo {year} {2011})}\BibitemShut
  {NoStop}%
\bibitem [{\citenamefont {Yamada}(2014)}]{Yamada_book_2014}%
  \BibitemOpen
  \bibfield  {author} {\bibinfo {author} {\bibfnamefont {M.}~\bibnamefont
  {Yamada}},\ }\href {\doibase 10.1007/978-4-431-54889-8_5} {\emph {\bibinfo
  {title} {Theory of Semiconductor Lasers: From Basis of Quantum Electronics to
  Analyses of the Mode Competition Phenomena and Noise}}}\ (\bibinfo
  {publisher} {Springer Japan},\ \bibinfo {address} {Tokyo},\ \bibinfo {year}
  {2014})\BibitemShut {NoStop}%
\bibitem [{\citenamefont {Borri}\ \emph {et~al.}(2000)\citenamefont {Borri},
  \citenamefont {Langbein}, \citenamefont {Hvam}, \citenamefont
  {Heinrichsdorff}, \citenamefont {Mao},\ and\ \citenamefont
  {Bimberg}}]{Borri_APL_2000}%
  \BibitemOpen
  \bibfield  {author} {\bibinfo {author} {\bibfnamefont {P.}~\bibnamefont
  {Borri}}, \bibinfo {author} {\bibfnamefont {W.}~\bibnamefont {Langbein}},
  \bibinfo {author} {\bibfnamefont {J.~M.}\ \bibnamefont {Hvam}}, \bibinfo
  {author} {\bibfnamefont {F.}~\bibnamefont {Heinrichsdorff}}, \bibinfo
  {author} {\bibfnamefont {M.-H.}\ \bibnamefont {Mao}}, \ and\ \bibinfo
  {author} {\bibfnamefont {D.}~\bibnamefont {Bimberg}},\ }\href {\doibase
  10.1063/1.126038} {\bibfield  {journal} {\bibinfo  {journal} {Applied Physics
  Letters}\ }\textbf {\bibinfo {volume} {76}},\ \bibinfo {pages} {1380}
  (\bibinfo {year} {2000})}\BibitemShut {NoStop}%
\bibitem [{\citenamefont {Yokoyama}\ and\ \citenamefont
  {Brorson}(1989)}]{Yokoyama_JAP_1989}%
  \BibitemOpen
  \bibfield  {author} {\bibinfo {author} {\bibfnamefont {H.}~\bibnamefont
  {Yokoyama}}\ and\ \bibinfo {author} {\bibfnamefont {S.~D.}\ \bibnamefont
  {Brorson}},\ }\href {\doibase 10.1063/1.343793} {\bibfield  {journal}
  {\bibinfo  {journal} {Journal of Applied Physics}\ }\textbf {\bibinfo
  {volume} {66}},\ \bibinfo {pages} {4801} (\bibinfo {year}
  {1989})}\BibitemShut {NoStop}%
\bibitem [{\citenamefont {Yokoyama}(1992)}]{Yokohama_Science_1992}%
  \BibitemOpen
  \bibfield  {author} {\bibinfo {author} {\bibfnamefont {H.}~\bibnamefont
  {Yokoyama}},\ }\href {\doibase 10.1126/science.256.5053.66} {\bibfield
  {journal} {\bibinfo  {journal} {Science}\ }\textbf {\bibinfo {volume}
  {256}},\ \bibinfo {pages} {66} (\bibinfo {year} {1992})}\BibitemShut
  {NoStop}%
\bibitem [{\citenamefont {Henry}(1986)}]{Henry_JLWT_1986}%
  \BibitemOpen
  \bibfield  {author} {\bibinfo {author} {\bibfnamefont {C.}~\bibnamefont
  {Henry}},\ }\href {\doibase 10.1109/JLT.1986.1074721} {\bibfield  {journal}
  {\bibinfo  {journal} {Journal of Lightwave Technology}\ }\textbf {\bibinfo
  {volume} {4}},\ \bibinfo {pages} {298} (\bibinfo {year} {1986})}\BibitemShut
  {NoStop}%
\bibitem [{\citenamefont {Schawlow}\ and\ \citenamefont
  {Townes}(1958)}]{Schawlow_PR_1958}%
  \BibitemOpen
  \bibfield  {author} {\bibinfo {author} {\bibfnamefont {A.~L.}\ \bibnamefont
  {Schawlow}}\ and\ \bibinfo {author} {\bibfnamefont {C.~H.}\ \bibnamefont
  {Townes}},\ }\href {\doibase 10.1103/PhysRev.112.1940} {\bibfield  {journal}
  {\bibinfo  {journal} {Phys. Rev.}\ }\textbf {\bibinfo {volume} {112}},\
  \bibinfo {pages} {1940} (\bibinfo {year} {1958})}\BibitemShut {NoStop}%
\bibitem [{\citenamefont {Henry}(1982)}]{Henry_JQE_1982}%
  \BibitemOpen
  \bibfield  {author} {\bibinfo {author} {\bibfnamefont {C.}~\bibnamefont
  {Henry}},\ }\href {\doibase 10.1109/JQE.1982.1071522} {\bibfield  {journal}
  {\bibinfo  {journal} {IEEE Journal of Quantum Electronics}\ }\textbf
  {\bibinfo {volume} {18}},\ \bibinfo {pages} {259} (\bibinfo {year}
  {1982})}\BibitemShut {NoStop}%
\bibitem [{\citenamefont {Pick}\ \emph {et~al.}(2015)\citenamefont {Pick},
  \citenamefont {Cerjan}, \citenamefont {Liu}, \citenamefont {Rodriguez},
  \citenamefont {Stone}, \citenamefont {Chong},\ and\ \citenamefont
  {Johnson}}]{Pick_PRA_2015}%
  \BibitemOpen
  \bibfield  {author} {\bibinfo {author} {\bibfnamefont {A.}~\bibnamefont
  {Pick}}, \bibinfo {author} {\bibfnamefont {A.}~\bibnamefont {Cerjan}},
  \bibinfo {author} {\bibfnamefont {D.}~\bibnamefont {Liu}}, \bibinfo {author}
  {\bibfnamefont {A.~W.}\ \bibnamefont {Rodriguez}}, \bibinfo {author}
  {\bibfnamefont {A.~D.}\ \bibnamefont {Stone}}, \bibinfo {author}
  {\bibfnamefont {Y.~D.}\ \bibnamefont {Chong}}, \ and\ \bibinfo {author}
  {\bibfnamefont {S.~G.}\ \bibnamefont {Johnson}},\ }\href {\doibase
  10.1103/PhysRevA.91.063806} {\bibfield  {journal} {\bibinfo  {journal} {Phys.
  Rev. A}\ }\textbf {\bibinfo {volume} {91}},\ \bibinfo {pages} {063806}
  (\bibinfo {year} {2015})}\BibitemShut {NoStop}%
\bibitem [{\citenamefont {Cerjan}\ \emph {et~al.}(2015)\citenamefont {Cerjan},
  \citenamefont {Pick}, \citenamefont {Chong}, \citenamefont {Johnson},\ and\
  \citenamefont {Stone}}]{Cerjan_OptExpress_2015}%
  \BibitemOpen
  \bibfield  {author} {\bibinfo {author} {\bibfnamefont {A.}~\bibnamefont
  {Cerjan}}, \bibinfo {author} {\bibfnamefont {A.}~\bibnamefont {Pick}},
  \bibinfo {author} {\bibfnamefont {Y.~D.}\ \bibnamefont {Chong}}, \bibinfo
  {author} {\bibfnamefont {S.~G.}\ \bibnamefont {Johnson}}, \ and\ \bibinfo
  {author} {\bibfnamefont {A.~D.}\ \bibnamefont {Stone}},\ }\href {\doibase
  10.1364/OE.23.028316} {\bibfield  {journal} {\bibinfo  {journal} {Opt.
  Express}\ }\textbf {\bibinfo {volume} {23}},\ \bibinfo {pages} {28316}
  (\bibinfo {year} {2015})}\BibitemShut {NoStop}%
\bibitem [{\citenamefont {Wang}\ \emph {et~al.}(2021)\citenamefont {Wang},
  \citenamefont {Zou}, \citenamefont {Puccioni}, \citenamefont {Zhao},
  \citenamefont {Lin}, \citenamefont {Chen}, \citenamefont {Wang},\ and\
  \citenamefont {Lippi}}]{Wang_OptExpress_2021}%
  \BibitemOpen
  \bibfield  {author} {\bibinfo {author} {\bibfnamefont {T.}~\bibnamefont
  {Wang}}, \bibinfo {author} {\bibfnamefont {J.}~\bibnamefont {Zou}}, \bibinfo
  {author} {\bibfnamefont {G.~P.}\ \bibnamefont {Puccioni}}, \bibinfo {author}
  {\bibfnamefont {W.}~\bibnamefont {Zhao}}, \bibinfo {author} {\bibfnamefont
  {X.}~\bibnamefont {Lin}}, \bibinfo {author} {\bibfnamefont {H.}~\bibnamefont
  {Chen}}, \bibinfo {author} {\bibfnamefont {G.}~\bibnamefont {Wang}}, \ and\
  \bibinfo {author} {\bibfnamefont {G.~L.}\ \bibnamefont {Lippi}},\ }\href
  {\doibase 10.1364/OE.416934} {\bibfield  {journal} {\bibinfo  {journal} {Opt.
  Express}\ }\textbf {\bibinfo {volume} {29}},\ \bibinfo {pages} {5081}
  (\bibinfo {year} {2021})}\BibitemShut {NoStop}%
\bibitem [{\citenamefont {Ma}\ \emph {et~al.}(2013)\citenamefont {Ma},
  \citenamefont {Oulton}, \citenamefont {Sorger},\ and\ \citenamefont
  {Zhang}}]{Ma_LPR_2013}%
  \BibitemOpen
  \bibfield  {author} {\bibinfo {author} {\bibfnamefont {R.-M.}\ \bibnamefont
  {Ma}}, \bibinfo {author} {\bibfnamefont {R.~F.}\ \bibnamefont {Oulton}},
  \bibinfo {author} {\bibfnamefont {V.~J.}\ \bibnamefont {Sorger}}, \ and\
  \bibinfo {author} {\bibfnamefont {X.}~\bibnamefont {Zhang}},\ }\href
  {\doibase https://doi.org/10.1002/lpor.201100040} {\bibfield  {journal}
  {\bibinfo  {journal} {Laser \& Photonics Reviews}\ }\textbf {\bibinfo
  {volume} {7}},\ \bibinfo {pages} {1} (\bibinfo {year} {2013})}\BibitemShut
  {NoStop}%
\bibitem [{\citenamefont {Kleppner}(1981)}]{Kleppner_PRL_1981}%
  \BibitemOpen
  \bibfield  {author} {\bibinfo {author} {\bibfnamefont {D.}~\bibnamefont
  {Kleppner}},\ }\href {\doibase 10.1103/PhysRevLett.47.233} {\bibfield
  {journal} {\bibinfo  {journal} {Phys. Rev. Lett.}\ }\textbf {\bibinfo
  {volume} {47}},\ \bibinfo {pages} {233} (\bibinfo {year} {1981})}\BibitemShut
  {NoStop}%
\bibitem [{\citenamefont {Yablonovitch}(1987)}]{Yablonovitch_PRL_1987}%
  \BibitemOpen
  \bibfield  {author} {\bibinfo {author} {\bibfnamefont {E.}~\bibnamefont
  {Yablonovitch}},\ }\href {\doibase 10.1103/PhysRevLett.58.2059} {\bibfield
  {journal} {\bibinfo  {journal} {Phys. Rev. Lett.}\ }\textbf {\bibinfo
  {volume} {58}},\ \bibinfo {pages} {2059} (\bibinfo {year}
  {1987})}\BibitemShut {NoStop}%
\bibitem [{\citenamefont {Yamamoto}\ \emph {et~al.}(1991)\citenamefont
  {Yamamoto}, \citenamefont {Machida},\ and\ \citenamefont
  {Bj\"ork}}]{Yamamoto_PRA_1991}%
  \BibitemOpen
  \bibfield  {author} {\bibinfo {author} {\bibfnamefont {Y.}~\bibnamefont
  {Yamamoto}}, \bibinfo {author} {\bibfnamefont {S.}~\bibnamefont {Machida}}, \
  and\ \bibinfo {author} {\bibfnamefont {G.}~\bibnamefont {Bj\"ork}},\ }\href
  {\doibase 10.1103/PhysRevA.44.657} {\bibfield  {journal} {\bibinfo  {journal}
  {Phys. Rev. A}\ }\textbf {\bibinfo {volume} {44}},\ \bibinfo {pages} {657}
  (\bibinfo {year} {1991})}\BibitemShut {NoStop}%
\bibitem [{\citenamefont {Denning}\ \emph {et~al.}(2018)\citenamefont
  {Denning}, \citenamefont {Iles-Smith}, \citenamefont {Osterkryger},
  \citenamefont {Gregersen},\ and\ \citenamefont {Mørk}}]{Denning_PRB_2018}%
  \BibitemOpen
  \bibfield  {author} {\bibinfo {author} {\bibfnamefont {E.~V.}\ \bibnamefont
  {Denning}}, \bibinfo {author} {\bibfnamefont {J.}~\bibnamefont {Iles-Smith}},
  \bibinfo {author} {\bibfnamefont {A.~D.}\ \bibnamefont {Osterkryger}},
  \bibinfo {author} {\bibfnamefont {N.}~\bibnamefont {Gregersen}}, \ and\
  \bibinfo {author} {\bibfnamefont {J.}~\bibnamefont {Mørk}},\ }\href
  {\doibase 10.1103/PhysRevB.98.121306} {\bibfield  {journal} {\bibinfo
  {journal} {Phys. Rev. B}\ }\textbf {\bibinfo {volume} {98}},\ \bibinfo
  {pages} {121306} (\bibinfo {year} {2018})}\BibitemShut {NoStop}%
\bibitem [{\citenamefont {Baba}\ \emph {et~al.}(1991)\citenamefont {Baba},
  \citenamefont {Hamano}, \citenamefont {Koyama},\ and\ \citenamefont
  {Iga}}]{Baba_JQE_1991}%
  \BibitemOpen
  \bibfield  {author} {\bibinfo {author} {\bibfnamefont {T.}~\bibnamefont
  {Baba}}, \bibinfo {author} {\bibfnamefont {T.}~\bibnamefont {Hamano}},
  \bibinfo {author} {\bibfnamefont {F.}~\bibnamefont {Koyama}}, \ and\ \bibinfo
  {author} {\bibfnamefont {K.}~\bibnamefont {Iga}},\ }\href {\doibase
  10.1109/3.89951} {\bibfield  {journal} {\bibinfo  {journal} {IEEE Journal of
  Quantum Electronics}\ }\textbf {\bibinfo {volume} {27}},\ \bibinfo {pages}
  {1347} (\bibinfo {year} {1991})}\BibitemShut {NoStop}%
\bibitem [{\citenamefont {Noda}\ \emph {et~al.}(2007)\citenamefont {Noda},
  \citenamefont {Fujita},\ and\ \citenamefont {Asano}}]{Noda_NatPhot_2007}%
  \BibitemOpen
  \bibfield  {author} {\bibinfo {author} {\bibfnamefont {S.}~\bibnamefont
  {Noda}}, \bibinfo {author} {\bibfnamefont {M.}~\bibnamefont {Fujita}}, \ and\
  \bibinfo {author} {\bibfnamefont {T.}~\bibnamefont {Asano}},\ }\href
  {\doibase 10.1038/nphoton.2007.141} {\bibfield  {journal} {\bibinfo
  {journal} {Nature Photonics}\ }\textbf {\bibinfo {volume} {1}},\ \bibinfo
  {pages} {449} (\bibinfo {year} {2007})}\BibitemShut {NoStop}%
\bibitem [{\citenamefont {Takiguchi}\ \emph {et~al.}(2013)\citenamefont
  {Takiguchi}, \citenamefont {Sumikura}, \citenamefont {Danang~Birowosuto},
  \citenamefont {Kuramochi}, \citenamefont {Sato}, \citenamefont {Takeda},
  \citenamefont {Matsuo},\ and\ \citenamefont {Notomi}}]{Takiguchi_APL_2013}%
  \BibitemOpen
  \bibfield  {author} {\bibinfo {author} {\bibfnamefont {M.}~\bibnamefont
  {Takiguchi}}, \bibinfo {author} {\bibfnamefont {H.}~\bibnamefont {Sumikura}},
  \bibinfo {author} {\bibfnamefont {M.}~\bibnamefont {Danang~Birowosuto}},
  \bibinfo {author} {\bibfnamefont {E.}~\bibnamefont {Kuramochi}}, \bibinfo
  {author} {\bibfnamefont {T.}~\bibnamefont {Sato}}, \bibinfo {author}
  {\bibfnamefont {K.}~\bibnamefont {Takeda}}, \bibinfo {author} {\bibfnamefont
  {S.}~\bibnamefont {Matsuo}}, \ and\ \bibinfo {author} {\bibfnamefont
  {M.}~\bibnamefont {Notomi}},\ }\href {\doibase 10.1063/1.4820194} {\bibfield
  {journal} {\bibinfo  {journal} {Applied Physics Letters}\ }\textbf {\bibinfo
  {volume} {103}},\ \bibinfo {pages} {091113} (\bibinfo {year}
  {2013})}\BibitemShut {NoStop}%
\bibitem [{\citenamefont {Takiguchi}\ \emph {et~al.}(2016)\citenamefont
  {Takiguchi}, \citenamefont {Taniyama}, \citenamefont {Sumikura},
  \citenamefont {Birowosuto}, \citenamefont {Kuramochi}, \citenamefont
  {Shinya}, \citenamefont {Sato}, \citenamefont {Takeda}, \citenamefont
  {Matsuo},\ and\ \citenamefont {Notomi}}]{Takiguchi_OptExpress_2016}%
  \BibitemOpen
  \bibfield  {author} {\bibinfo {author} {\bibfnamefont {M.}~\bibnamefont
  {Takiguchi}}, \bibinfo {author} {\bibfnamefont {H.}~\bibnamefont {Taniyama}},
  \bibinfo {author} {\bibfnamefont {H.}~\bibnamefont {Sumikura}}, \bibinfo
  {author} {\bibfnamefont {M.~D.}\ \bibnamefont {Birowosuto}}, \bibinfo
  {author} {\bibfnamefont {E.}~\bibnamefont {Kuramochi}}, \bibinfo {author}
  {\bibfnamefont {A.}~\bibnamefont {Shinya}}, \bibinfo {author} {\bibfnamefont
  {T.}~\bibnamefont {Sato}}, \bibinfo {author} {\bibfnamefont {K.}~\bibnamefont
  {Takeda}}, \bibinfo {author} {\bibfnamefont {S.}~\bibnamefont {Matsuo}}, \
  and\ \bibinfo {author} {\bibfnamefont {M.}~\bibnamefont {Notomi}},\ }\href
  {\doibase 10.1364/OE.24.003441} {\bibfield  {journal} {\bibinfo  {journal}
  {Opt. Express}\ }\textbf {\bibinfo {volume} {24}},\ \bibinfo {pages} {3441}
  (\bibinfo {year} {2016})}\BibitemShut {NoStop}%
\bibitem [{\citenamefont {Gwo}\ and\ \citenamefont
  {Shih}(2016)}]{Gwo_RepProgrPhys_2016}%
  \BibitemOpen
  \bibfield  {author} {\bibinfo {author} {\bibfnamefont {S.}~\bibnamefont
  {Gwo}}\ and\ \bibinfo {author} {\bibfnamefont {C.-K.}\ \bibnamefont {Shih}},\
  }\href {\doibase 10.1088/0034-4885/79/8/086501} {\bibfield  {journal}
  {\bibinfo  {journal} {Reports on Progress in Physics}\ }\textbf {\bibinfo
  {volume} {79}},\ \bibinfo {pages} {086501} (\bibinfo {year}
  {2016})}\BibitemShut {NoStop}%
\bibitem [{\citenamefont {Takemura}\ \emph {et~al.}(2021)\citenamefont
  {Takemura}, \citenamefont {Takiguchi},\ and\ \citenamefont
  {Notomi}}]{Takemura_PRA_2021}%
  \BibitemOpen
  \bibfield  {author} {\bibinfo {author} {\bibfnamefont {N.}~\bibnamefont
  {Takemura}}, \bibinfo {author} {\bibfnamefont {M.}~\bibnamefont {Takiguchi}},
  \ and\ \bibinfo {author} {\bibfnamefont {M.}~\bibnamefont {Notomi}},\ }\href
  {\doibase 10.1364/JOSAB.413919} {\bibfield  {journal} {\bibinfo  {journal}
  {J. Opt. Soc. Am. B}\ }\textbf {\bibinfo {volume} {38}},\ \bibinfo {pages}
  {699} (\bibinfo {year} {2021})}\BibitemShut {NoStop}%
\bibitem [{\citenamefont {Vyshnevyy}\ and\ \citenamefont
  {Fedyanin}(2022)}]{Vyshnevyy_PhysRevLett_2022}%
  \BibitemOpen
  \bibfield  {author} {\bibinfo {author} {\bibfnamefont {A.~A.}\ \bibnamefont
  {Vyshnevyy}}\ and\ \bibinfo {author} {\bibfnamefont {D.~Y.}\ \bibnamefont
  {Fedyanin}},\ }\href {\doibase 10.1103/PhysRevLett.128.029401} {\bibfield
  {journal} {\bibinfo  {journal} {Phys. Rev. Lett.}\ }\textbf {\bibinfo
  {volume} {128}},\ \bibinfo {pages} {029401} (\bibinfo {year}
  {2022})}\BibitemShut {NoStop}%
\bibitem [{\citenamefont {Carroll}\ \emph {et~al.}(2022)\citenamefont
  {Carroll}, \citenamefont {D'Alessandro}, \citenamefont {Lippi}, \citenamefont
  {Oppo},\ and\ \citenamefont {Papoff}}]{Carroll_PhysRevLett_2022}%
  \BibitemOpen
  \bibfield  {author} {\bibinfo {author} {\bibfnamefont {M.~A.}\ \bibnamefont
  {Carroll}}, \bibinfo {author} {\bibfnamefont {G.}~\bibnamefont
  {D'Alessandro}}, \bibinfo {author} {\bibfnamefont {G.~L.}\ \bibnamefont
  {Lippi}}, \bibinfo {author} {\bibfnamefont {G.-L.}\ \bibnamefont {Oppo}}, \
  and\ \bibinfo {author} {\bibfnamefont {F.}~\bibnamefont {Papoff}},\ }\href
  {\doibase 10.1103/PhysRevLett.128.029402} {\bibfield  {journal} {\bibinfo
  {journal} {Phys. Rev. Lett.}\ }\textbf {\bibinfo {volume} {128}},\ \bibinfo
  {pages} {029402} (\bibinfo {year} {2022})}\BibitemShut {NoStop}%
\bibitem [{\citenamefont {Scully}\ and\ \citenamefont
  {Lamb}(1967)}]{Scully_PhysRev_1967}%
  \BibitemOpen
  \bibfield  {author} {\bibinfo {author} {\bibfnamefont {M.~O.}\ \bibnamefont
  {Scully}}\ and\ \bibinfo {author} {\bibfnamefont {W.~E.}\ \bibnamefont
  {Lamb}},\ }\href {\doibase 10.1103/PhysRev.159.208} {\bibfield  {journal}
  {\bibinfo  {journal} {Phys. Rev.}\ }\textbf {\bibinfo {volume} {159}},\
  \bibinfo {pages} {208} (\bibinfo {year} {1967})}\BibitemShut {NoStop}%
\end{thebibliography}%

\end{document}